\documentclass[11pt,a4paper]{article}
\usepackage[T1]{fontenc}
\usepackage[utf8]{inputenc}
\usepackage{lmodern,amsmath,amssymb,amsthm,mathtools,microtype}
\usepackage[margin=28mm]{geometry}
\usepackage{booktabs,array,longtable,enumitem,xcolor}
\usepackage[colorlinks=true,linkcolor=blue!45!black,citecolor=blue!45!black,urlcolor=blue!45!black]{hyperref}
\hypersetup{pdftitle={Degree profiles of uniform polynomial consequences},pdfauthor={Soren Riis},pdfsubject={Linear degree lower bounds for uniform Nullstellensatz and Polynomial Calculus consequences},pdfkeywords={proof complexity, uniform consequences, degree profiles, linear lower bounds}}
\usepackage{fancyhdr}
\setlist{nosep,leftmargin=*,topsep=5pt}
\newtheorem{theorem}{Theorem}[section]
\newtheorem{lemma}[theorem]{Lemma}
\newtheorem{proposition}[theorem]{Proposition}
\newtheorem{corollary}[theorem]{Corollary}
\theoremstyle{definition}\newtheorem{definition}[theorem]{Definition}
\newtheorem{example}[theorem]{Example}
\theoremstyle{remark}\newtheorem{remark}[theorem]{Remark}
\newcommand{\NS}{\mathrm{NS}}\newcommand{\PC}{\mathrm{PC}}
\newcommand{\NN}{\mathbb N}\newcommand{\Z}{\mathbb Z}\newcommand{\Fp}{\mathbb F_p}
\DeclareMathOperator{\im}{im}\DeclareMathOperator{\Span}{span}
\DeclareMathOperator{\rec}{Rec}\DeclareMathOperator{\Soc}{Soc}\DeclareMathOperator{\Top}{Top}\DeclareMathOperator{\LIS}{LIS}
\newcommand{\join}{\mathbin{\vee_*}}
\newcommand{\Ftwo}{\mathbb F_2}
\newcommand{\D}{\mathcal D}\newcommand{\Del}{\mathcal D}\newcommand{\Lift}{\mathsf L}
\DeclareMathOperator{\Inj}{Inj}
\title{Degree profiles of uniform polynomial consequences\\[3pt]
\large Linear lower bounds for Nullstellensatz and Polynomial Calculus}
\author{S\o ren Riis}
\date{}
\begin{document}
\maketitle
\begin{abstract}
This paper proves asymptotically optimal linear degree lower bounds for Nullstellensatz
and Polynomial Calculus, yielding a four-way polynomial-versus-exponential
proof-size classification for uniform polynomial consequences over fields
of positive characteristic. The size bounds use full, explicit
coefficient lists, including zeros.

Apart from a fixed finite initial range, exactly one of four cases
occurs: $\mathrm T_1$, bounded degree; $\mathrm T_2$, degree at least
linear in the universe size $n$; $\mathrm T_3$, constant degree on some
residue classes and at least linear degree on the others; or
$\mathrm T_4$, infinitely many recurring finite degrees, with
bounded-degree subsequences and finite $\Omega(n)$ peaks at arbitrarily
large sizes. Thus the size patterns are polynomial bounds, exponential
lower bounds, a periodic mixture, or polynomial subsequences with finite
exponential peaks, respectively. Underivable targets have degree
$\infty$, and both proof systems assign the same type.

The result strengthens my 2008--2009 refutation classification and extends
it to uniform consequences. The proof uses degree-preserving lifting
with onset and period bounds linear in the degree. A common $p$-adic
hardness spectrum describes the four types topologically. Distinguishing
any two types is undecidable already for refutations. The linear degree bounds, four-way classification and pairwise
undecidability results have been formalised and verified in Lean. The Lean
formalisation, verification records and an accompanying tutorial are
available in the supporting materials~\cite{RiisSupporting}.
\end{abstract}
\medskip
\section{Introduction}\label{sec:introduction}
An algebraic consequence can be simple to state and difficult to derive.
Given polynomial equations $\Gamma_n$ on a universe of size $n$, we ask
how much degree is needed to prove a prescribed equation $G^{(n)}=0$ from
them. Both the axioms and the target are generated by fixed rules as
$n$ varies. The answer has four mutually exclusive and exhaustive
patterns: bounded degree, linear hardness, a periodic mixture of the
two, and systematic fluctuation with finite linear peaks. This paper
proves the classification for uniform polynomial consequences, with
refutation as the special target $G^{(n)}=1$. It strengthens the earlier
logarithmic Nullstellensatz and double-logarithmic Polynomial Calculus
lower bounds to linear bounds in the universe size.

The NS proof system was introduced by Beame, Impagliazzo,
Kraj\'i\v cek, Pitassi and Pudl\'ak~\cite{NSFoundations}; PC was
introduced by Clegg, Edmonds and Impagliazzo~\cite{PCFoundations}.
To fix the proof-theoretic meaning of a consequence, let
$\Gamma=\{f_1,\ldots,f_m\}$ be a finite set of polynomials over a field
$K$, interpreted as the equations $f_i=0$. A \emph{Nullstellensatz}
(NS) derivation of $G$ is an identity
\[
                         G=\sum_{i=1}^m a_i f_i.
\]
Its degree is the largest total degree of a nonzero product $a_i f_i$.
\emph{Polynomial Calculus} (PC) derives polynomials line by line: each
line is an input axiom, a linear combination of earlier lines, or an
earlier line multiplied by a variable. A derivation of $G$ ends with
that actual polynomial, and its degree is the largest total degree
of any line after collecting coefficients. Thus $x-y$ and $y$ derive
$x$ at degree one, while deriving $1$ certifies that the input equations
have no common zero. Soundness makes every derivation a semantic
consequence. The converse is not imposed as an inference rule:
semantic entailment and the existence of an NS or PC derivation are
distinct questions for a general polynomial system.

Degree measures the complexity of the algebraic calculation rather
than its number of lines. A fixed degree bound confines all lines to
a vector space with polynomially many monomials in the variable count.
An NS certificate uses the span of the allowed monomial multiples of
axioms. PC can also cancel terms and multiply the resulting polynomial
again. Its bounded-degree consequences therefore form an iterated
linear closure, and its minimum degree can be smaller than the NS
minimum. Section~\ref{sec:systems} gives the exact characterisations
used throughout the proof.

A \emph{uniform family} $(\Gamma_n)_n$ is a family of systems of
polynomial equations, generated by a fixed finite list of bounded-degree
polynomial patterns using the same coefficient rules at every universe
size. At each size $n$, all equations in $\Gamma_n$ are imposed
simultaneously. Each pattern can generate many equations as its indices
range over the universe, so the number of equations can grow with $n$
even though the list of patterns is fixed.
Variable and axiom indices each carry a bounded number of labels,
and relabelling the universe permutes the equations. Each axiom is fixed
by permutations fixing its distinguished index labels, even when it
contains sums over the entire universe. A \emph{fully invariant uniform
target} has no distinguished labels and is unchanged by every
permutation; $G^{(n)}=\sum_{i<n}x_i$ is a simple example. Its logical
antecedents include the Paris--Wilkie translation~\cite{ParisWilkie},
treated systematically by Kraj\'{i}\v{c}ek~\cite[Chapter~9]{KrajBounded},
and my early work on independence and finitisation in bounded
arithmetic~\cite{RiisThesis,RiisFinitisation}, followed by the
complexity-gap results~\cite{RiisTree,Riis}. My joint journal papers
with Meera Sitharam develop the uniform translation and its
permutation-module viewpoint~\cite{RiisSitharamTautologies,RiisSitharamModules}.
Here the hypotheses are stated entirely in algebraic terms. Definitions~\ref{def:template}
and~\ref{def:uniform-target} make the coefficient rules and support
conditions precise.

\paragraph{Examples of uniform families.}
For these examples label the universe by $1,\ldots,n$. With three
unary variable types $x_i,y_i,z_i$, consider the generating polynomial
\[
 P_{1,n}=x_1+y_1+\sum_{j,k=1}^{n}x_jx_k+
                       \sum_{i,j,k=1}^{n}z_i z_j z_k.
\]
Its total degree is at most three, independent of $n$, although the
number of summands and variables grows with $n$. A permutation fixing
$1$ fixes this whole polynomial. Its distinguished index therefore has
support $\{1\}$. Taking its permutation orbit gives the symmetric axiom
system $P_{a,n}=0$ for all $a\in\{1,\ldots,n\}$, where only the first
two terms become $x_a+y_a$: both global sums are unchanged.
The degree bound concerns each monomial, not the number of terms or
the union of labels mentioned by the sum. All sums are evaluated in
the fixed coefficient field, including any coefficient cancellations.

Binary and higher-arity variable types are equally admissible. For
binary variables $u_{ij}$, the row axioms
\[
               F_{i,n}=\sum_{j=1}^{n}u_{ij}-1=0
\]
have degree one and index support $\{i\}$, while each variable has
support contained in $\{i,j\}$. A cubic polynomial such as
\[
       H_n=\sum_{i,j,k=1}^{n}u_{ij}u_{jk}u_{ki}
\]
is fully invariant, with one empty-support axiom or target index.
A single system can combine several such patterns. For example, it can
impose all $n$ row equations $F_{i,n}=0$, all $n^2$ Boolean equations
$u_{ij}^2-u_{ij}=0$, and the global equation $H_n=0$ simultaneously.
These three patterns define a system of $n+n^2+1$ indexed equations.
There can be $n^2$ binary variables, or $n^a$ variables of any fixed
arity $a$, without violating the bounded-support hypothesis. These
examples are instances of the rank-independent coefficient rules in
Definition~\ref{def:template}, rather than arbitrary symmetric
polynomials chosen afresh at each rank.

\paragraph{Familiar examples from the literature.}
Algebraic encodings of pigeonhole and counting principles provide
important examples of this uniform structure. Razborov studied
Polynomial Calculus degree lower bounds for pigeonhole
principles~\cite{RazborovPC}; Beame and Riis studied Nullstellensatz
bounds for onto-pigeonhole principles~\cite{BeameRiis}. Fixed-block-size
partition principles also underlie the Nullstellensatz lower bounds of
Beame, Impagliazzo, Kraj\'{i}\v{c}ek, Pitassi and
Pudl\'{a}k~\cite{NSFoundations}. In their bounded-degree encodings,
Boolean constraints, pairwise incompatibility constraints and sums
expressing that each element is assigned exactly once are generated by
fixed patterns invariant under relabelling. Fixed-offset pigeonhole
families fit the present single-universe formulation by treating the
extra elements as permanent parameters, as described below. Thus these bounded-degree, fixed-pattern encodings are instances of
the uniformly generated polynomial systems considered here.
The encoding matters: growing-degree axioms are outside the
bounded-degree hypothesis used here.

\paragraph{Orbit generators and permanent parameters.}
A label in an orbit generator is moved when the orbit is taken. Thus
the orbit of $x_1y_2=0$, with $1\ne2$, consists of $x_ay_b=0$ for
all $a\ne b$. It does not include the diagonal equations. This differs
from permanently naming two distinct domain elements: the single
constraint involving those named elements need not impose the analogous
constraint on every pair.

Permanent parameters can be included by using separate variable types.
For one named element $c$, a binary family $z_{ij}$ splits into the
scalar variable $z_{cc}$, the unary families $z_{ci}$ and $z_{ic}$,
and the binary family $z_{ij}$ for $i,j\ne c$. The letter $c$ is part
of the type name; only the other labels are permuted. Correspondingly,
\[
 \sum_{i,j\in\{c\}\cup U}z_{ij}
 =z_{cc}+\sum_{i\in U}z_{ci}+\sum_{i\in U}z_{ic}
                                      +\sum_{i,j\in U}z_{ij}.
\]
The scalar type has empty support and is fixed by every permutation.
Such zero-arity variable types are allowed by
Definition~\ref{def:template}: the positive parameter $k$ is an upper
bound on support size, not a lower bound on arity. More generally,
a fixed set of $\ell$ distinct named elements splits each finite-arity
variable family into finitely many types, according to which coordinates
are named and which remain free. Sums split in the same way. In a
universe of total size $n$, the remaining action is $S_{n-\ell}$;
reindexing the free universe puts the system in the usual $S_m$ form
with $m=n-\ell$. This fixed shift preserves bounded degree and the
asymptotic linear bounds; statements about residues and initial universes
are read with the same shift. For example, with permanent parameters
$c_1,c_2$, the equation $x_{c_1}y_{c_2}=0$ is a single equation in two
scalar variable types. Its orbit under permutations of the remaining
labels contains only itself.
Here the parameters are preselected distinct domain elements. An
uninterpreted constant symbol, whose value is chosen as part of a model,
is instead represented by its function-graph variables in the logical
translation below.

The proof is organised around one question. If an invariant uniform
target $G^{(n)}$ has a degree-at-most-$d$ derivation from $\Gamma_n$, when
can the corresponding target $G^{(n+q)}$ be derived within the same
degree bound from $\Gamma_{n+q}$? Here $q=p^a$ is the number of additional
elements in the finite universe, whose size increases from $n$ to $n+q$. Once such a lift is available and the universe is sufficiently large
relative to the degree bound, it can be repeated:
\[
 \Gamma_{n+\ell q}\vdash_{\mathrm S,\,\le d}G^{(n+\ell q)}
                    \qquad(\ell=0,1,2,\ldots).
\]
The selected numerical degree $d$ is constant on this progression;
the proof lines and their coefficients may change. The essential
quantitative question is how large the initial universe must be relative
to $d$. For indexed targets the same question is asked for the whole
prescribed target family, using its coefficient map rather than a
literal lift of each polynomial.

\paragraph{First-order logic and finite structures.}
The framework includes polynomial encodings of arbitrary fixed
first-order sentences over a finite vocabulary of uninterpreted
relations and functions. A sentence can first be Skolemised and
normalised to universal clauses, with auxiliary relation or function
symbols where needed. The new functions record existential witnesses:
for example, $\forall u\,\exists v\,R(u,v)$ can be represented by
$\forall u\,R(u,f(u))$. This preserves existence of models on each
nonempty finite universe. A model may have several witness expansions,
so it does not assert a one-to-one correspondence with models of the
original sentence before those witnesses are chosen.

Relation variables record truth values, and function-graph variables
record chosen outputs. Boolean equations, totality equations and
pairwise functionality collisions give the translation of
Section~\ref{sec:systems}. Its polynomial system has a common root
at size $n$ if and only if the original sentence has an $n$-element
model; roots correspond to models of the chosen normalised expansion.
The vocabulary, arities, clause lengths and hence polynomial degrees
are fixed while $n$ grows. This is the translation setting used in
LICS~2008~\cite[Section~1.2, p.~273]{Riis} and the expanded
2009 report~\cite[Section~1.2, pp.~3--4]{RiisLong}, following the joint work
with Meera Sitharam~\cite{RiisSitharamTautologies}; a precise translation
statement is Theorem~1 of its BRICS version~\cite[Section~II]{RiisSitharamTranslationPreprint}. The joint work gives the
translation for existential second-order sentences, including these
first-order existence problems, and its permutation-module setting is
developed in~\cite{RiisSitharamModules}.

Existential quantification is not simply replaced by a sum. For a
witness graph, $\sum_j x_{ij}=1$ is used together with Booleanity and
$x_{ij}x_{ij'}=0$ for $j\ne j'$. Only together do these require exactly
one chosen output; in characteristic $p$, the sum equation alone also
allows $p+1$ Boolean ones. The example of an injective function omitting
one named point is obtained in this way and has no finite nonempty
model, although it has an infinite model.

The domain symbols here are uninterpreted: order, arithmetic and
other relations are themselves part of the structure being sought.
One may axiomatise a linear order and then quantify over its possible
interpretations. Fixing an external order or arithmetic on the labels
in advance is a different translation and need not satisfy the full
permutation and rank-independent coefficient hypotheses. The theorem
does not automatically cover such built-in structures.

A fixed coefficient rule can likewise prescribe the polynomial
expressions to be derived. At universe size $n$, write these targets as
$G_b^{(n)}$, where $b$ is the target index. The superscript records the
universe size; it does not name a distinguished element. For a fully
invariant target we omit $b$ and write $G^{(n)}$. An indexed target family
may request $G_b^{(n)}=0$ for every $b$, or the representatives of one
fixed index orbit. No arbitrary choice of a different orbit at each
size is made.

The governing hypotheses are fixed bounds on axiom and target degrees,
bounded supports of the
variable and axiom indices, and coefficients prescribed consistently
across universe sizes and under relabelling. An index supported on a
few labels can specify a polynomial involving the entire universe:
the totality equation $\sum_j x_{ij}-1=0$ has index $i$, even though
the sum uses every $j$. The same distinction applies to targets.
Uniformity does not require all terms of a target polynomial to use
one bounded set of labels. The parameter $n$ is universe size, not
variable count; variables indexed by $k$-tuples may number on the
order of $n^k$. The coefficient field has one fixed positive
characteristic throughout.

There is a proof-size interpretation when each polynomial is represented
by its \emph{full, explicit coefficient list}. For a polynomial of
total degree $d$, this lists all monomials of degree at most $d$ in
the entire ambient variable set, including zero coefficients. The
convention applies to PC lines and NS multipliers as well as to the
input. Bounded-degree derivations can then be written in polynomial
size, whereas linear degree lower bounds force size at least
$2^{\Omega(n)}$ on the relevant instances. These are lower bounds;
no universal matching exponential upper bound is asserted. The size
conclusions concern finite derivations of the prescribed targets.
An underivable target has infinite degree and proof size. Over a fixed
finite field, coefficient counts also give the corresponding bit-size
bounds. If the input length is $L(n)=O(n^{c_0})$ for a fixed $c_0>0$, the exponential bound in
$n$ becomes a stretched-exponential bound
$2^{\Omega(L(n)^{1/c_0})}$. Corollary~\ref{cor:dense-types} supplies the
counting argument and the precise scope; sparse representations that
omit zero coefficients use a different size measure.

\paragraph{Complexity gaps and the earlier classification.}
The restriction to a fixed uniform description belongs to a broader
programme of complexity gap theorems. Resolution combines
$A\lor x$ and $B\lor\neg x$ to derive $A\lor B$. In tree-like
resolution, a derived clause cannot be shared between branches;
general resolution permits this reuse. Riis's tree-resolution gap
theorem~\cite{RiisTree} gives, for the standard translations of a
first-order sentence with no finite models, either polynomial-size
refutations or a $2^{\Omega(n)}$ size lower bound. The hard case occurs
exactly when the sentence has an infinite model; if it has no model
at all, the polynomial case holds. Distinguishing the two cases is
undecidable.

Dantchev and Riis~\cite{DantchevRiis} proved a size gap between
polynomial upper bounds and $2^{\Omega(n^\varepsilon)}$ lower bounds,
for some fixed $\varepsilon>0$, for \emph{full resolution} on
unary-relativised families. A unary predicate selects the domain on which the sentence
is interpreted. When the original sentence has no finite models, its
relativised translations have polynomial-size refutations, even
tree-like ones, if it has no model at all. If it has an infinite model,
every full-resolution refutation has size
$2^{\Omega(n^\varepsilon)}$ for some $\varepsilon>0$. That work also
treats tree-like resolution with bounded-size conjunctions and shows
that a built-in order destroys the gap phenomenon.

Dantchev, Martin and Szeider~\cite{DMS} refine the tree-resolution
classification in the parameterised setting. Dantchev and
Martin~\cite{DMRank} obtain a constant-versus-polynomial rank gap for
the Lov\'asz--Schrijver and Sherali--Adams systems, again governed by
infinite models. Beyersdorff, Clymo, Dantchev and Martin~\cite{QBFGap}
establish tree-like QBF-resolution analogues, with explicit dependence
on the translation. These results concern different resources---size,
parameterised size or rank---and do not themselves yield an NS or PC
degree bound.

Positive characteristic allows more than a dichotomy for algebraic
degree. Easy and hard sizes can occupy different residue classes, and
one family can have infinitely many recurring finite degree levels.
My LICS 2008 paper~\cite[Theorems~1 and~10]{Riis}, with fuller treatment
in the 53-page 2009 report~\cite{RiisLong}, established four refutation types,
their periodic mixed case and agreement of the NS and PC type labels.
Its general degree lower bounds were $\Omega(\log n)$ for NS and
$\Omega(\log\log n)$ for PC. Their periodicity and degree-dependent
onsets came from Kraj\'i\v cek's Theorem~3.5~\cite{Krajicek}, building
on Ajtai's analysis of symmetric linear systems~\cite{Ajtai}.
Here the linear bounds are proved for
the more general task of deriving uniform targets. The constant
target $1$ recovers the refutation classification and its strengthened
bounds.

In the final remarks of the LICS paper and the longer report, I
conjectured that the classification should admit a degree lower bound
of the form $\Omega(n^\varepsilon)$, for sufficiently small fixed $\varepsilon>0$,
at least for NS and possibly for PC~\cite[Section~8]{Riis}\cite[Section~9]{RiisLong}.
I recorded the considerable effort already spent on this question.
Both the LICS abstract and the longer report explicitly asked about a
linear degree lower bound~\cite{Riis,RiisLong}.
The present theorem proves that bound for both systems, with the
hard-residue and finite-peak qualifications below, and extends the
setting from refutations to uniform polynomial consequences.

The journal work on uniformly generated submodules concerns fields of
characteristic zero~\cite{RiisSitharamModules}. In a separate 1997
preprint with Meera Sitharam~\cite{RiisSitharam1997}, we announced a
nonconstant-to-linear degree gap for uniform NS and PC families.
In my 2009 report, I explicitly acknowledged an unresolved gap in the
inducing-down proof and stated that the main claim remained conjectured
\cite[p.~50, footnote~3; p.~53, reference~33]{RiisLong}.
That qualification concerns the proof; it does not refute the asserted
gap theorem. The present argument gives independent finite algebraic
constructions for the hypotheses and four profiles stated here.

The preliminary fresh-label construction---a construction introducing
additional labels---in Section~\ref{sec:quadratic-route} already yields an $\Omega(\sqrt n)$
degree gap in characteristic two, answering the positive-power question
in that setting. The linear construction makes a structural change:
corrections at successive tuple lengths reuse one retained pool of labels.
The odd-characteristic argument supplies the additional nilpotent-chain
structure needed to obtain the same linear degree lower bound in every
positive characteristic.

The linear degree lower bound is asymptotically optimal in general. The
injection example in Section~\ref{sec:examples} has minimum NS and PC
degrees $\Theta(n)$, while the fluctuating example has finite
$\Theta(n)$ peaks. These examples show that a superlinear degree lower
bound cannot replace the linear bound throughout the classification.

\paragraph{The four degree profiles.}
For either proof system, let $h(n)$ be the least derivation degree of
the uniform target; for an indexed target family, take the maximum of
the least degrees over its finite rank-$n$ index set. Infinity records
an underivable target, and an empty index set has degree zero. Let
$\rec(h)$ be the set of finite degrees occurring at infinitely many
sizes. The four alternatives and their coefficient-list size readings
are as follows.
\begin{center}
\begin{tabular}{@{}p{0.065\textwidth}p{0.46\textwidth}p{0.39\textwidth}@{}}
\toprule
Type & Degree behaviour & Derivation-size consequence\\
\midrule
$\mathrm T_1$ & Bounded degree beyond a fixed initial range
 & Polynomial upper bounds.\\
$\mathrm T_2$ & No recurring finite degree; linear lower bounds
 & Exponential lower bounds.\\
$\mathrm T_3$ & Finitely many recurring levels without eventual boundedness;
fixed low degrees on a nonempty proper set of residues, linear lower
bounds on the others
 & Periodic polynomial upper bounds and exponential lower bounds.\\
$\mathrm T_4$ & Infinitely many recurring levels; finite liminf
and finite linear peaks
 & Polynomial subsequences and finite exponential peaks at arbitrarily
large sizes.\\
\bottomrule
\end{tabular}
\end{center}
Finite initial exceptions are allowed. In $\mathrm T_3$, easy residues
need not share the same degree. In $\mathrm T_4$, a linear degree bound
at every sufficiently large size is false: the least recurring finite
degree gives arbitrarily late low-degree derivations. Its infinitely many recurring finite degrees distinguish it from
$\mathrm T_3$, which has only finitely many. Thus arbitrarily large
sizes have finite linear peaks, while one fixed finite degree also
occurs arbitrarily late.
Section~\ref{sec:classification} proves exhaustiveness and exclusivity
and states the exact quantitative bounds.

\paragraph{From unchanged-degree lifting to a linear bound.}
Fix a proof-degree budget $d$. The central statement says that
derivability within this budget is periodic in universe size after
an onset linear in $d$, with a period also linear in $d$. Consequently,
a finite degree $d$ attained beyond its own onset must recur on an
infinite arithmetic progression. A degree which does not recur can
only occur before that onset. Inverting this restriction gives the
linear lower bound in $\mathrm T_2$ and on the hard residues of
$\mathrm T_3$. In $\mathrm T_4$, the bound on the period locates
representatives of unbounded recurring degrees at sizes proportional
to those degrees, producing finite linear peaks. Both the onset and
period are needed.

The relevant budget counts labels. If variable indices have support
at most $k$ and axiom indices at most $s$, a degree-$d$ monomial
paired with an axiom index uses at most $kd+s$ labels. The fully
invariant target uses this budget. For targets indexed by supports
of size at most $t$, the general proof uses the conservative budget
\[
                             R(d)=2kd+s+t.
\]
The extra factor of two comes from detecting the target image outside
the consequence space by a pairing; both factors of that pairing
must be transported. The fixed target degree affects the initial
rank threshold and excludes budgets below the target's degree.
Above that fixed threshold, a sufficient periodic onset is
$n>2R(d)$, and a period is the least power of the characteristic
exceeding $R(d)$. Thus the extension to indexed targets retains
the linear degree lower bounds.

There are three transport obligations. First, the operation must take
the actual axiom and multiplication maps to those at the smaller
rank. Second, it must preserve the sums, images and intersections
which compute all bounded-degree consequences. This is the difficult
PC requirement: a cancellation can create a new polynomial eligible
for multiplication. Third, it must preserve and reflect whether the
prescribed target belongs to the resulting space. Exact deletion and
a common group-algebra projector meet the first two obligations;
transport of the target coefficient map and nonzero-subquotient
detection meet the third. An abstract isomorphism of spaces alone
would supply none of these map compatibilities.

The construction is developed in stages. In characteristic two, a
first lift uses separate fresh labels at successive tuple lengths;
its quadratic label cost already explains a square-root degree gap.
The linear construction reuses one pool. Its projector is constructed
through tuple lengths $b\le R(d)$,
repairing the current length while leaving the shorter lengths
unchanged. The same retained labels serve all repairs, so the
allocation is controlled by $\max_{b\le R(d)}(2b-1)$ rather than
a sum of separate fresh-label allocations. The proof develops the
square-zero construction in characteristic two, normalisation and
the zero-marginal decomposition, the full rook-forcing induction,
and the top-power construction in arbitrary positive characteristic.
For odd $p$, square-zero identities are replaced by full length-$p$
nilpotent chains and an ordered projector, with a separate transpose
argument. These repairs retain the same linear label allocation.
Section~\ref{sec:transfer} states the exact map requirements and
the resulting transfer, and
Sections~\ref{sec:block-algebra}--\ref{sec:odd} establish its algebraic
and combinatorial ingredients.

The comparison between the two systems is separate. Compressing
invariant-subspace chains to a bounded universe gives
$h_{\PC}(n)\le h_{\NS}(n)\le F(h_{\PC}(n))$ whenever the PC
degree is finite, with a monotone function $F$ independent of
universe size. It applies to
the entire consequence space and hence to each prescribed target.
This establishes common type without asserting equality of exact
degrees or of recurring degree sets. The linear PC bound is proved
directly, rather than inferred by inverting the potentially large
function $F$.

\paragraph{Examples, semantics and undecidability.}
The first-order examples illustrate the four refutation types. A
fresh-variable construction also produces a nonconstant target
whose finite derivation degree has linear peaks; this shows that
the consequence theorem includes actual polynomial targets as well
as the constant $1$. The motivating Boolean translation uses
totality sums and functionality collisions, with no automatic
reduction modulo Boolean equations. Section~\ref{sec:model-theory} gives a $p$-adic hardness spectrum for
all uniform requests: the four types correspond to an empty spectrum,
the whole space, a nonempty proper clopen set, and a closed set that is
not open. This makes explicit the residue-class structure already
implicit in my earlier report~\cite[Section~4]{RiisLong}, now in the
setting of uniform consequences. The same section distinguishes
linear witnesses from actual solutions and recalls Kraj\'i\v cek's
Euler criterion. The topological argument needs no Boolean semantic
hypotheses and does not assert a converse to that criterion.

The undecidability theorem already holds in the refutation special
case. The no-finite-model promise and undecidability for every
nonempty proper union of types were stated in LICS~2008
\cite[Theorem~1]{Riis}. Here explicit reductions separate any two
types under that same promise, refining the published every-union
formulation. On normalised first-order inputs, an explicit finite
computation prefix gives a uniform degree bound when its machine
halts, and a controlled disjunction supplies the pairwise reductions.
The effective reduction and its finite-term certificate are included.

\paragraph{Algebraic antecedents.}
Kraj\'i\v cek's bounded-degree NS--PC comparison
\cite[Theorem~5.5]{Krajicek}
supplied an input to my earlier classification. The present proof
retains the NS-span stabilisation argument and supplies its uniform
chain bound independently. The behaviour of fixed-degree data as the
number of labels grows is a central theme of representation stability
and FI-modules, developed by Church, Ellenberg and Farb~\cite{CEF}.
An FI-module assigns vector spaces and compatible linear maps to finite
sets and injections; it organises comparisons as the label set grows. In positive
characteristic, Harman~\cite{Harman} proves periodic equivalences for
bounded-weight symmetric-group representations with a linear stable
range. His equivalences already respect induction, restriction, tensor
products and duality~\cite[Propositions~2.7 and~2.12]{Harman}.
The deletion functor $\ker z/(\ker z\cap\im z)$ used here is also
one of the explicit functors in Hone and coauthors~\cite[Theorem~A]{HKPSYZ}.
Its group-algebra operator comes from a generic shifted cyclic subgroup.
Our argument proves exactness on its stated bounded-support category,
without invoking a categorical equivalence. These antecedents place
the linear-onset mechanism in an established setting. The
present proof independently constructs exact deletion on the actual
polynomial diagrams, preserves their consequence closures and detects
the prescribed targets, with the quantitative bounds used here. The rook
realisation step is related to the integral tensor-space centraliser
results of Bowman, Doty and Martin~\cite{BDM}; the required realisation
is obtained by the forcing and elimination argument in the paper.
Uniform target membership already belongs to Kraj\'i\v cek's
framework~\cite{KrajEuler,Krajicek}. The contributions here are the
linear quantitative bounds for uniform consequences, the independent
algebraic proof and explicit reductions for every pair of types under
the earlier no-finite-model promise.

\paragraph{Notation and proof guide.}\label{sec:specialist-tools}
The principal notation is collected here; local auxiliary symbols are
defined in their respective constructions.
\begin{center}
\begin{tabular}{@{}ll@{}}
\toprule
Notation & Meaning\\
\midrule
$n$, $N=n+q$ & Retained and larger universe sizes\\
$d$; $D$, $D_G$ & Proof-degree budget; fixed axiom and target degree bounds\\
$k,s,t$ & Variable, axiom-index and target-index support bounds\\
$R(d)=\kappa d+\sigma$ & Support budget used in periodicity\\
$N(d)$, $Q(d)$ & Degree-dependent onset and prime-power period\\
$I_d(n)$, $C_d(n)$ & NS and ordinary-PC consequence spaces\\
$P_b(N)$, $\mathcal C_r(N)$ & Injective-tuple modules and their subquotient category\\
$\mathcal Z_b(N)$, $J_r(N)$ & Zero-marginal space and tuple annihilator ideal\\
$z$, $w$, $e$, $\Del$ & Block operator, common inner inverse, projector, deletion\\
\bottomrule
\end{tabular}
\end{center}
The central dependency chain is
\[
 \begin{gathered}
 \text{rook realisation and layer splitting}\Longrightarrow
 \text{common inner inverse}\Longrightarrow\text{exact deletion},\\
 \text{exact deletion and target detection}\Longrightarrow
 \text{periodicity with linear onset}\Longrightarrow\text{four profiles}.
 \end{gathered}
\]
The NS--PC comparison additionally uses a uniform subspace-chain bound.
The matching obstruction supplies the concrete examples needed for
pairwise undecidability. Standard background is in Lang~\cite{LangAlgebra},
Alperin~\cite{AlperinLocal}, the Stacks Project~\cite{StacksExact} and
Cutland~\cite{Cutland}; the specific algebraic and combinatorial
arguments used here are proved below.

\paragraph{Outline of the paper.}
Section~\ref{sec:systems} defines the proof systems, uniform axiom
and target templates, first-order encoding and bounded-degree
consequence spaces. Section~\ref{sec:classification} derives the
four profiles and their size consequences from quantitative transfer.
Sections~\ref{sec:transfer}--\ref{sec:odd} construct the deletion and
lifting operations, first in characteristic two and then in arbitrary
positive characteristic. Section~\ref{sec:comparison} proves uniform
NS--PC type agreement. Section~\ref{sec:model-theory} gives the topological
classification and discusses algebraic witnesses and Euler structures.
Section~\ref{sec:target-examples} develops nonconstant targets and the
fresh-variable construction. Sections~\ref{sec:examples}--\ref{app:matching}
give the four first-order refutation examples and prove their matching
obstruction.
Section~\ref{sec:undecidability} completes the computation-prefix
reduction. Section~\ref{sec:conclusion} closes with the mathematical
consequences and acknowledgements.

\section{Uniform systems and degree}\label{sec:systems}
Fix a field $K$ of prime characteristic $p$. We study the degree needed
to derive a specified polynomial from a system of equations. Refutation
is the special case in which the specified polynomial is $1$. All rings
are ordinary commutative polynomial rings; Boolean equations, when
present, are axioms rather than a quotient convention.

\subsection{Consequences and their degree}
Let $\Gamma\subseteq K[x_v:v\in V]$ be a finite system and let $G$ be a
polynomial in the same ring. We write $\Gamma\vdash_S G$, for
$S\in\{\NS,\PC\}$, for an algebraic derivation of $G$ from $\Gamma$.
Since the input equations are $f=0$, this means deriving the equation
$G=0$. In particular, deriving $1$ gives a contradiction.

Following the NS proof system of Beame, Impagliazzo, Kraj\'i\v cek,
Pitassi and Pudl\'ak~\cite{NSFoundations}, an NS derivation of $G$ is
an identity
\[
                  G=\sum_{f\in\Gamma}a_f f.
\]
Its degree is the largest total degree of a nonzero product $a_f f$.
Over a field this is $\deg a_f+\deg f$ for each such product. Zero
summands impose no restriction and may be omitted; the empty sum is a
certificate of $G=0$ of degree zero. We use $\deg0=0$ throughout.

In the system introduced by Clegg, Edmonds and
Impagliazzo~\cite{PCFoundations}, an ordinary-PC derivation is a finite
sequence $P_1,\ldots,P_M$ whose
last line is $G$. Each line is an input axiom, a linear combination
$\alpha P_i+\beta P_j$ of earlier lines, or a variable multiple $x_vP_i$
of an earlier line. The zero polynomial is also allowed as the empty
linear combination. The proof degree is the largest total degree of any
line, computed after collecting coefficients. Allowing a linear
combination of any finite number of earlier lines gives the same
degree: it can be expanded into successive two-line combinations. Thus a cancellation may
lower the degree of a derived polynomial and permit a later
multiplication within the same degree budget. A Boolean simplification
is not a further rule.

Write $h_S(\Gamma\vdash G)\in\NN\cup\{\infty\}$ for the least such
degree, assigning $\infty$ when there is no derivation. Every finite
value is at least $\deg G$. For refutations we retain the notation
$d_S(\Gamma)=h_S(\Gamma\vdash1)$. Expanding an NS certificate into
monomial multiples, deriving each product by successive variable
multiplication, and adding gives
\begin{equation}\label{eq:pcns}
 h_{\PC}(\Gamma\vdash G)\le h_{\NS}(\Gamma\vdash G),
 \qquad d_{\PC}(\Gamma)\le d_{\NS}(\Gamma).
\end{equation}
The inequality concerns actual derivations of the specified target;
the target is not added to the input axioms.

\paragraph{A short consequence proof.}
Over $\mathbb F_2$, take $f_1=xy+x$, $f_2=xy+1$ and $f_3=xz$.
The following proof derives the nonconstant target $G=z$ in degree two:
\[
 f_1+f_2=x+1,\qquad z(x+1)=xz+z,\qquad (xz+z)+f_3=z.
\]
We are keeping every line within the degree budget two. The quadratic
terms cancel to give the degree-one line $x+1$; multiplying that line
by $z$ is therefore a permitted step within this budget. Expanding
these same steps into the NS identity $z=zf_1+zf_2+f_3$ gives product
degree three. The input system itself is consistent: $x=y=1$, $z=0$
satisfies it. Thus consequences need not arise from refutations.

\subsection{Uniform axioms and uniform targets}
The running first-order example asks for an injective function whose
image omits a distinguished point. If $x_{ij}$ encodes the graph atom
$g(i)=j$, totality and functionality contribute
\[
                 f_{i,n}=\sum_{j<n}x_{ij}-1,
                 \qquad x_{ij}x_{ij'}\quad(j\ne j').
\]
The same rules are used at every size. A permutation sends $x_{ij}$ to
$x_{\sigma(i),\sigma(j)}$ and sends $f_{i,n}$ to $f_{\sigma(i),n}$.
Every permutation fixing $i$ therefore fixes $f_{i,n}$, although the
sum contains variables involving all labels. This distinguishes the
\emph{support of an axiom index} from the union of labels in its terms.
The complete inconsistent system is given in \eqref{eq:G-witness}.

We call a support assignment \emph{equivariant} when relabelling an
object also relabels its support. For linear maps the same word means
commutation with relabelling: $f(\sigma v)=\sigma f(v)$. Uniformity
below specifies both this symmetry and one coefficient rule used at
every rank.

\begin{definition}[Uniform template]\label{def:template}
A bounded basis system consists of labelled objects, with an action of
permutations of $\NN$, and a finite support for each object, whose size
has a fixed bound independent of rank. Support is
equivariant, and a permutation fixing an object's support pointwise
fixes that object. At rank $n$ the basis consists exactly of the objects
supported in $[n]=\{0,\ldots,n-1\}$, and is finite.

A uniform polynomial template $T$ has a variable basis $\mathcal V$ of
support at most $k\ge1$, an axiom-index basis $\mathcal A$ of support at
most $s$, and coefficients $c(a,\mu)\in K$ indexed by an axiom index and
a monomial in the labelled variables. The coefficients are invariant
under simultaneous relabelling, independent of rank, and vanish when
$\deg\mu>D$ for some fixed $D$. The rank-$n$ instance is
\[
 \Gamma_n=\{f_{a,n}:a\in\mathcal A_n\},\qquad
 f_{a,n}=\sum_{\mu\text{ supported in }[n]}c(a,\mu)\mu.
\]
Choose $n_0\ge\max\{1,kD+s\}$. The parameters $k,s,D,n_0$ belong to
the fixed template.
\end{definition}

Support zero is allowed for variable and index bases. A zero-arity
variable type is a single scalar variable fixed by every permutation;
a finite number of such types can be included. The parameter $k\ge1$
is only an upper bound on variable support, so it does not exclude these
types. This also explains the permanent-parameter encoding in the
introduction without enlarging the template hypotheses.

Fixed finite types of tuples, with specified equality patterns, give
familiar basis systems. A degree-$d$ monomial uses at most $kd$ labels,
even if the polynomial containing it mentions every label in the
universe. Products of bases add support bounds; restriction to an
invariant subset does not increase them. A monomial--axiom pair has
support at most
\begin{equation}\label{eq:safe-budget}
                         r=kd+s.
\end{equation}
Overlaps can reduce this bound. For the injection system one may take
$k=2$ and $s=3$: row totality has index support one, while an
injectivity collision $x_{ij}x_{hj}$ has support at most three. The
multiplication domain at degree $d$ has basis pairs $(x_v,\mu)$ with
$\deg\mu\le d-1$, of support at most $k+k(d-1)=kd$.

\begin{definition}[Uniform target family]\label{def:uniform-target}
For a fixed axiom template, a uniform indexed target family consists of
a bounded basis $\mathcal B$ of support at most $t$ and coefficients
$\gamma(b,\mu)\in K$, invariant under simultaneous relabelling of
$b$ and $\mu$, independent of rank, and zero when $\deg\mu>D_G$.
It requests the targets
\[
 G_b^{(n)}=\sum_{\mu\text{ supported in }[n]}\gamma(b,\mu)\mu,
                         \qquad b\in\mathcal B_n.
\]
Here $n$ is the universe size and $b$ is the target index; its support
is a bounded set of elements of $[n]$.
The degree of the request is the worst of the individual least degrees:
\begin{equation}\label{eq:target-profile}
 h_S(n)=\max_{b\in\mathcal B_n}h_S(\Gamma_n\vdash G_b^{(n)}),
                         \qquad S\in\{\NS,\PC\}.
\end{equation}
The empty maximum is zero; a single underivable target makes the
maximum $\infty$. Thus $h_S(n)\le d$ means that \emph{every} requested
target has a degree-$d$ derivation, with no restriction on the total
number of proof lines used.
\end{definition}

There are three useful forms of this definition. A \emph{fully invariant}
target $G^{(n)}$ is the case of one target index with empty support, so
$\sigma G^{(n)}=G^{(n)}$ for every permutation of $[n]$. For example, with
variables $x_i$, the polynomial $\sum_{i<n}x_i$ is fully invariant:
each monomial uses one label, although the full sum mentions all labels.
An \emph{indexed} family can instead request $G_i^{(n)}=x_i$ for each
$i<n$. Its index has support one; its members are permuted rather than
individually fixed.

One may restrict $\mathcal B$ to one fixed global orbit of indices.
At every rank $n\ge t$ that orbit has a representative; all its
representatives at that rank are related by a permutation of $[n]$ and
have the same least degree. Hence one representative per rank gives the
same profile. Indeed, restrict a global relabelling to the finite support of the index; its image is also in $[n]$. Extend this finite injection to a permutation of $[n]$, which has the same action on the supported index. There are only finitely many global orbits, since each
meets the finite set $\mathcal B_t$. The aggregate request
\eqref{eq:target-profile} takes their maximum. An arbitrary selection
that switches between different orbits with $n$ is a different problem
and is not covered by the theorem.

Rank-independent coefficient rules are essential for both axioms and
targets. Symmetry of each individual instance is insufficient: choosing
$\{1=0\}$ at sizes in an arbitrary set and the empty system elsewhere
produces an arbitrary degree-zero predicate, but is not a fixed
template. Likewise, the target cannot be chosen by an arbitrary
size-dependent rule.

Beyond $n_0$, every axiom-index orbit has a fixed polynomial degree.
A monomial witnessing its formal degree uses at most $kD$ labels;
a relabelling fixing the index places these labels inside $[n]$ once
$n\ge kD+s$. The same argument for targets gives the common stable
starting rank
\begin{equation}\label{eq:stable-rank}
                        n_*=\max\{n_0,1,kD_G+t\}.
\end{equation}
Thus each target orbit has a fixed degree there, and their maximum is
a degree floor for both profiles. The all-target degree-zero predicate
is constant in this range: a nonconstant requested target fails it,
whereas constant targets are consequences exactly when they are zero
or the inputs span a nonzero constant. For general indexed requests the
transport argument uses the larger support budget $2kd+s+t$; the
extra factor of two detects a possible nonzero quotient, as explained
in Section~\ref{sec:transfer}.

\subsection{The normalised first-order input class}
Any fixed first-order sentence in a finite uninterpreted vocabulary can
be converted, by Skolemisation, flattening of terms and clausification,
to a normalised expansion with the same existence of models at each
nonempty finite size. Witnesses are retained as function symbols rather
than translated as unrestricted sums. The resulting roots correspond
to models of this expansion; several expansions may have the same
reduct to the original vocabulary. This is the translation approach of
\cite[Section~1.2]{Riis}\cite[Section~1.2]{RiisLong} and
\cite{RiisSitharamTautologies}; see the explicit translation theorem in
\cite[Section~II, Theorem~1]{RiisSitharamTranslationPreprint}.
We fix one such normalisation.

A normalised sentence is a finite collection of universally quantified clauses over relation symbols and function symbols. Its atoms are $R(x_{i_1},\ldots,x_{i_a})$, flattened graph atoms $x_j=f(x_{i_1},\ldots,x_{i_a})$, and equalities $x_i=x_j$. Constants are functions of arity zero. Symbols are uninterpreted; in particular there is no built-in order or arithmetic. Well-formedness requires the declared arities and variable references to match.

At size $n$ introduce relation variables $r^R_{\bar i}$ and graph variables $x^f_{\bar i,j}$. Include
\begin{align}
 z^2-z&=0 &&\text{for every variable},\label{eq:boolean}\\
 \sum_{j<n}x^f_{\bar i,j}-1&=0,\qquad
 x^f_{\bar i,j}x^f_{\bar i,j'}=0 &&(j\ne j'),\label{eq:function}\\
 \prod_{L\in C}(1-\operatorname{val}(L))&=0 &&\text{for every grounded clause }C.\label{eq:clause}
\end{align}
An equality atom has value $\delta_{ij}$; negation replaces a value by its complement. The collisions in \eqref{eq:function} are indispensable in characteristic $p$: a sum of Boolean values equal to $1$ need not contain exactly one $1$ without them.

\begin{proposition}[Compiler and semantics]\label{prop:compiler}
The displayed translation is an actual uniform template. Its parameters can be computed from the normalised syntax. For every $n\ge1$, its common roots over $K$ correspond to models of the sentence on $[n]$.
\end{proposition}
\begin{proof}
Use the atom tuples for variable indices and clause-grounding tuples for clause indices; totality and collision indices are also fixed tuples. Expansion of \eqref{eq:clause} has coefficients depending only on equalities among these labels. Repeated coordinates and coefficient cancellations are included in this expansion. Safe bounds, writing $a_R,a_f$ for the maximum relation and function arities, are
\[
 k=a_R+a_f+1,\quad s=k+1+\max_C\operatorname{nvars}(C),\quad
 D=2+\max_C|C|,\quad n_0=kD+s+1.
\]
Here $\operatorname{nvars}(C)$ is one more than the largest variable index
occurring in $C$, and is $1$ when none occurs (the compiler reserves an unused variable slot). At a root every variable is $0$ or $1$. Totality and collisions choose exactly one output of each function. Clause products vanish precisely when a literal is true. Conversely, a model supplies these $0/1$ values and satisfies every equation.
\end{proof}

We use this explicit normalised input class for undecidability. There is no assertion that different choices of normalisation or Skolemisation preserve exact degree. Keeping the chosen encoding visible is especially important for the examples.

\subsection{Nullstellensatz completeness and degree upper bounds}\label{sec:booleanrole}
Let $I=(f_1,\ldots,f_m)\subseteq K[x_1,\ldots,x_v]$ be the ideal
generated by a finite polynomial system. Hilbert's Nullstellensatz
relates its common roots over an algebraic closure $\overline K$ to
ideal membership: the system has no common root if and only if
$1\in I$. More generally, a polynomial $G$ vanishes on every common
root if and only if $G\in\sqrt I$, equivalently $G^r\in I$ for some
$r\ge1$; see \cite{LangAlgebra}. The ideal-membership statements may be
made over $K$ itself, since solvability of the finite coefficient
systems descends from field extensions (Lemma~\ref{lem:field}).
An NS derivation of the actual target $G$ requires $G\in I$, rather
than merely $G\in\sqrt I$. Thus semantic consequence and algebraic
derivability need not coincide for a general system.

For Boolean encodings, the equations $x_i^2-x_i=0$ force every variable
to be $0$ or $1$, over $K$ and over its algebraic closure. The quotient
by these equations is a finite product of copies of $K$; hence every
ideal containing them is radical. In this setting a target vanishes
on all solutions if and only if it has an NS derivation. In particular,
absence of a Boolean solution is equivalent to an NS refutation
\cite{NSFoundations}. Without these equations, absence of roots in $K$
alone is insufficient: $x^2+1$ has no root in $\mathbb F_3$ but has
roots over its algebraic closure and admits no refutation.

The quantitative issue is the degree of a certificate. For an
inconsistent Boolean system in $v$ variables, including each equation
$x_i^2-x_i=0$ among its inputs, with input degree at most $D\ge2$, the standard Boolean interpolation argument gives the safe bound
\begin{equation}\label{eq:booleancomplete}
                 d_{\PC}\le d_{\NS}\le v+D.
\end{equation}
Indeed, Boolean point indicators have degree at most $v$; multiplying
each by an input nonzero at that point and reducing modulo the Boolean
equations yields a certificate whose products have degree at most
$v+D$. See~\cite{NSFoundations} for the standard completeness argument.

For comparison, Razborov~\cite{RazborovPC} gives modified pigeonhole
systems with PC degree $\Omega(v)$, allowing logarithmic input degree.
This does not establish sharpness for our uniform class, which requires
a constant input-degree bound $D$ independent of $n$.

Here $v$ counts variables, whereas $n$ denotes the universe size. For
variables indexed by tuples of arity at most $k$, typically $v=O(n^k)$,
so this argument does not give a general $O(n)$ bound. A bound linear
in universe size requires additional structure: for the injection
system, Section~\ref{sec:examples} gives an explicit certificate of
degree at most $n+1$. Together with its linear lower bound, that
example establishes the asymptotic sharpness claimed in this paper.

All degree bounds use the ordinary polynomial ring. Multilinearisation
must be justified by Boolean-equation multiples with their degrees
counted: multiplying a squarefree degree-$d$ monomial by a variable
already in it first gives degree $d+1$. The general uniform-family
theorem does not require Boolean equations. In the first-order
encoding, the pairwise collision equations are also needed for
uniqueness: in characteristic $p$, a totality sum equal to one can
otherwise have $p+1$ nonzero summands.

\subsection{The space of all bounded-degree consequences}\label{sec:pc-space}
Fix a rank-$n$ instance, write $V=\mathcal V_n$ for its variable-index
set, and choose a degree budget $d\ge1$. Let $P_{\le d}$ be the
vector space with the monomials of degree at most $d$ as basis. Zero
axioms may be omitted. For NS define
\begin{equation}\label{eq:nsspan}
 I_d=\Span_K\{\mu f:f\in\Gamma,\ \mu\text{ a monomial},\quad
                       \deg\mu+\deg f\le d\}\subseteq P_{\le d}.
\end{equation}
NS uses one span of permitted monomial--axiom products. PC requires an
additional closure:
\begin{equation}\label{eq:pcclosure}
 \begin{gathered}
 V_0=\Span_K\{f\in\Gamma:\deg f\le d\},\\
 V_{j+1}=V_j+\sum_{v\in V}x_v(V_j\cap P_{\le d-1}).
 \end{gathered}
\end{equation}
For $d=0$ omit the multiplication term, so $V_\infty=V_0$.

\begin{proposition}[Exact consequence-space characterisation]\label{prop:pc-closure}
For every polynomial $G$ and every degree budget $d\ge0$,
\[
 \begin{aligned}
 h_{\NS}(\Gamma\vdash G)\le d&\quad\Longleftrightarrow\quad G\in I_d,\\
 h_{\PC}(\Gamma\vdash G)\le d&\quad\Longleftrightarrow\quad
                       G\in V_\infty:=\bigcup_jV_j.
 \end{aligned}
\]
The PC chain stabilises after finitely many stages and
$I_d\subseteq V_\infty\subseteq P_{\le d}$. In particular, the least
refutation degrees test membership of $1$ in these spaces.
\end{proposition}
\begin{proof}
For NS, expand each multiplier into monomials. Every resulting product
$\mu f$ satisfies the degree bound and belongs to the spanning set in
\eqref{eq:nsspan}. Conversely, group a finite linear combination of
these products by axiom. Its multiplier $a_f$ has degree at most
$d-\deg f$, giving the required identity with right-hand side $G$.
The empty combination gives $G=0$. Each permitted product $\mu f$ is
also a PC consequence: multiply $f$ successively by the variable
factors of $\mu$, with every intermediate degree at most $d$, and add.

For PC, induct on a proof line. A usable axiom lies in $V_0$. A linear
combination of vectors in $V_a$ and $V_b$ lies in $V_{\max(a,b)}$.
If a nonzero line $f$ is multiplied by a variable and the result has
degree at most $d$, then $\deg f\le d-1$. If $f\in V_j$, its multiple
therefore lies in $V_{j+1}$. The zero line causes no difficulty.

Conversely, every vector of $V_0$ is derived by linear combinations of
usable axioms. A vector of $V_{j+1}$ is a finite sum of a vector of
$V_j$ and terms $x_vf_v$, with $f_v\in V_j\cap P_{\le d-1}$.
Concatenate finite derivations of the $f_v$, append the permitted
multiplication steps, and add until the last line is the requested
vector. All partial linear combinations remain in $P_{\le d}$.

The ascending chain lies in the finite-dimensional space $P_{\le d}$.
Every strict inclusion increases dimension, and equality of two
successive stages makes all subsequent stages equal. Hence the chain
stabilises. This argument places no length bound on candidate proofs;
it shows that all their consequences are captured by a finite closure.
\end{proof}

The cancellation in our earlier example occurs in
$V_0\cap P_{\le1}$: $f_1+f_2=x+1$ becomes eligible for multiplication
although neither of its two inputs was linear. A transport argument
for ordinary PC must preserve the intersection in \eqref{eq:pcclosure},
including such vectors, rather than just transport a list of low-degree
input axioms.

Extending scalars from $K$ to a field $E$ keeps the coordinate equations
and allows coefficients in $E$. If $U\subseteq K^M$ has a $K$-basis
$u_1,\ldots,u_a$, its extension is
$U_E=\Span_E\{u_1,\ldots,u_a\}\subseteq E^M$, also denoted
$E\otimes_K U$. This is the finite-dimensional scalar-extension
operation used in the next lemma; see \cite{LangAlgebra} for its
standard linear-algebra background.

\begin{lemma}[Field extension]\label{lem:field}
For a finite system $\Gamma$ and target $G$ over $K$, extending the
coefficient field changes neither least NS nor least ordinary-PC
degree of $\Gamma\vdash G$.
\end{lemma}
\begin{proof}
At fixed degree, NS derivability is membership of $G$ in a finite span;
Gaussian elimination gives the same answer after extending scalars.
For PC, scalar extension commutes with finite-dimensional sums,
intersections and images, hence with every stage of
\eqref{eq:pcclosure}. Membership of the original vector $G$ in the
stable space is therefore preserved and reflected.
\end{proof}

\subsection{Explicit coefficient lists and proof size}\label{sec:dense-size}
Write $v(n)=|\mathcal V_n|$ for the number of polynomial variables,
distinguishing it from the universe size $n$ and the periodicity onset
$N(d)$ used later. Variables, axiom indices and target indices all have
polynomial counts in $n$: each bounded-support object is a relabelling
of one of finitely many objects at its support bound.

\begin{definition}[Size with explicit coefficient lists]\label{def:dense-size}
In an ambient ring with $v$ variables, a nonzero polynomial of total
degree $d$ is stored as the full coefficient table of all monomials of
total degree at most $d$, including zero coefficients. Its length is
\[
                       B_v(d)=\binom{v+d}{d}.
\]
The zero polynomial has a one-entry representation. A PC proof of $G$
has size equal to the sum of these lengths over its lines, including
the final line $G$. An NS identity $G=\sum_fa_ff$ has size equal to
the sum over its nonzero multipliers; the input polynomials and target
are supplied separately. The empty certificate for $G=0$ has size zero.

Let $S^{\mathrm{coeff}}_S(\Gamma\vdash G)$ be the minimum certificate
size in system $S$, with value $\infty$ if the target is underivable.
For a uniform indexed request put
\[
 S^{\mathrm{coeff}}_S(n)=\max_{b\in\mathcal B_n}
                   S^{\mathrm{coeff}}_S(\Gamma_n\vdash G_b^{(n)}),
\]
with empty maximum zero. For the single target $1$, this is the usual
minimum refutation size.
\end{definition}

A bundle of proofs for every requested target has total size between
this maximum and $|\mathcal B_n|$ times the maximum, when the request
is nonempty. Since $|\mathcal B_n|$ is polynomial in $n$, the two
conventions give the same polynomial and exponential size cases.
Size counts coefficient entries over $K$. For a fixed finite field,
coefficients have bounded bit length; variable, axiom and line indices
add only polynomial overhead to the bounds below. The uniformly
bounded-degree inputs and targets themselves have polynomial-size
tables. Every table uses \emph{all ambient variables}, even those
absent from its nonzero terms. Filling in missing powers in
$x_1^d+5$ alone would require only $d+1$ entries, a different measure.

\begin{lemma}[Degree and coefficient-list size]\label{lem:dense-size}
For $v$ variables and input degree at most $D$, if $G$ has a degree-$d$
NS derivation, it has an NS certificate of size at most $B_v(d)^2$.
If it has a degree-$d$ ordinary-PC derivation, it has a PC proof of
size $O(B_v(d)^3)$. Conversely,
writing $h_S=h_S(\Gamma\vdash G)$, finite least degrees give
\[
 S^{\mathrm{coeff}}_{\PC}(\Gamma\vdash G)\ge B_v(h_{\PC}),\qquad
 S^{\mathrm{coeff}}_{\NS}(\Gamma\vdash G)
                \ge B_v(\max\{0,h_{\NS}-D\})\quad(G\ne0).
\]
Also $B_v(d)\ge2^{\min\{v,d\}}$. For an indexed request the same upper
bounds hold when $h_S(n)\le d$. For a nonempty request, the lower
bounds apply to a target attaining the finite maximum, with the NS
zero-target exception just stated.
\end{lemma}
\begin{proof}
Put $M=\dim P_{\le d}=B_v(d)$. For NS, membership of $G$ in the span
$I_d$ has a witness using at most $M$ monomial--axiom products.
Grouping by axiom gives at most $M$ nonzero multipliers, each with at
most $M$ coefficient entries.

For PC, retain a basis of the usable input axioms. Maintain a basis of
the span $W$ of the retained derivable polynomials. Unless $W$ is
closed under permitted multiplication, choose
$f\in W\cap P_{\le d-1}$ with $x_if\notin W$. Derive $f$ as a linear
combination of at most $M$ retained polynomials, using $O(M)$ two-line
steps, then append $x_if$ to
the basis. At most $M$ extensions occur. Once $W$ is closed, it contains
every degree-$d$ consequence, including $G$. Append a linear
combination giving $G$ as the last line. This constructs $O(M^2)$
lines, each with at most $M$ entries; all intermediate linear
combinations stay in $P_{\le d}$. If $G=0$, the one-line proof $0$
suffices.

Every PC proof has a line of degree at least its least possible proof
degree. If $G\ne0$, every NS certificate has a nonzero product of
degree at least $h_{\NS}$, whose multiplier consequently has degree at
least $h_{\NS}-D$. Finally, the table contains every squarefree
monomial in any $\min\{v,d\}$ variables, proving the exponential
estimate. For a finite indexed request, a finite maximum is attained
unless the request is empty, in which case its size is zero.
\end{proof}

This coefficient convention makes the size consequence elementary.
It is a different resource from sparse monomial size: zero entries may
dominate the table. We make no sparse-size lower-bound claim here.

\section{Quantitative transfer and the four degree profiles}\label{sec:classification}
Fix a uniform axiom template and a stable uniform target family
$G_b^{(n)}$ as in Section~\ref{sec:systems}. The parameters $k,s,t$ bound
variable, axiom-index and target-index supports, respectively, and $D_G$
bounds the target degree. Put
\[
                 n_* = \max\{n_0,1,kD_G+t\}.
\]
For $\mathrm S\in\{\NS,\PC\}$, the aggregate profile $h_{\mathrm S}(n)$
is the maximum of the least degrees needed to derive the requested
polynomials $G_b^{(n)}$, $b\in\mathcal B_n$. Zero targets have degree $0$;
an underivable target gives value $\infty$, and the maximum of an empty
request is $0$. We may instead
restrict $\mathcal B$ to one fixed global orbit. At every $n\ge t$ that
orbit has a representative, and equivariance makes its profile the least
degree of any representative. These are two precise choices of a uniform
request. Choosing different global orbits according to $n$ is not covered.

Use the conservative parameters
\begin{equation}\label{eq:profile-parameters}
                      \kappa=2k,\qquad \sigma=s+t.
\end{equation}
The general transfer budget is then $\kappa d+\sigma=2kd+s+t$.
For a fully invariant target, and in particular for refutation, a sharper
version with $\kappa=k$ and $\sigma=s$ is given below. All the profile
deductions are stated in the same notation so that both versions have a
common presentation.

The next theorem is the quantitative lifting statement from which
the classification follows. It fixes the degree budget and varies the
universe size. Its two directions are important: an upward lift keeps
a derivation available on one progression, while the equivalence also
rules out a new derivation appearing later in that residue class.

\begin{theorem}[Periodicity with linear onset]\label{thm:periodicity}
Over any field of prime characteristic $p$, let $h$ be either the NS or
ordinary-PC aggregate profile, or either profile of one fixed target
orbit. For $d\ge1$ put $R(d)=\kappa d+\sigma$. If
\[
 n,n'\ge n_*,\qquad n,n'>2R(d),\qquad
 n\equiv n'\pmod q,\qquad q=p^a>R(d),
\]
then $h(n)\le d$ if and only if $h(n')\le d$. The degree-zero predicate
is constant for $n\ge n_*$.
\end{theorem}
The proof in Section~\ref{sec:target-transfer} transports the actual
coefficient map of the targets and the actual consequence spaces. It does
not require invariant proofs, and it does not infer the PC assertion from
an NS lower bound. The same argument works whether or not the axioms are
consistent.

Define
\begin{equation}\label{eq:NQ}
 \begin{gathered}
 N(0)=n_*,\qquad Q(0)=1,\\
 N(d)=\max\{n_*,2(\kappa d+\sigma)+1\},\qquad
 Q(d)=\min\{p^a:p^a>\kappa d+\sigma\}\quad(d\ge1).
 \end{gathered}
\end{equation}
Here $N(d)$ denotes the onset for a degree threshold; it is not the
larger rank $N=n+q$ used in the deletion construction. $Q(d)$ is the
chosen block size and period. Both are bounded linearly in $d$ beyond
the fixed template threshold.
The onsets are nondecreasing, and $Q(e)$ divides $Q(d)$ for $e\le d$.
Applying Theorem~\ref{thm:periodicity} at every threshold up to $d$ gives
\begin{equation}\label{eq:periodic}
 n,m\ge N(d),\quad n\equiv m\pmod {Q(d)}
 \quad\Longrightarrow\quad
 \min\{h(n),d+1\}=\min\{h(m),d+1\}.
\end{equation}
Here $\min\{\infty,d+1\}=d+1$. For $d\ge1$,
\begin{equation}\label{eq:periodbound}
                         Q(d)\le p(\kappa d+\sigma).
\end{equation}
Indeed, the preceding power of $p$ is at most $\kappa d+\sigma$.
The fixed initial threshold includes the stabilisation of target degrees.
A budget below a target's stabilised degree cannot derive that target;
thus such thresholds are not replaced by tests on a truncated target.

\subsection{The classification theorem}
Define the set of recurring finite degrees by
\[
 E(h)=\{d\in\NN:\text{some }n\ge N(d)\text{ has }h(n)=d\}.
\]
Equation~\eqref{eq:periodic} shows that every member $d$ occurs along an
infinite arithmetic progression:
\[
                  h(n+\ell Q(d))=d\qquad(\ell\ge0).
\]
The degree remains $d$ throughout this progression: lifting does not lower the absolute degree. Its ratio to the universe size tends to zero. Nor does one such progression make the whole profile bounded; other residue classes may remain hard.

Conversely, a value occurring at arbitrarily large sizes occurs beyond
$N(d)$. Thus $E(h)=\rec(h)$, independently of the admissible onset.
The key restriction is that a nonrecurring finite degree $d$ can occur
only before its own onset $N(d)$.

To include the small ranks in an exact formula, write $x_+=\max\{x,0\}$
and put
\[
             \beta(n)=\left\lfloor
                       \frac{(n-2\sigma-1)_+}{2\kappa}
                       \right\rfloor.
\]
The positive part agrees with natural-number subtraction and is irrelevant
once $n\ge2\sigma+1$.

\begin{theorem}[Four-way classification]\label{thm:classification}
For each profile considered in Theorem~\ref{thm:periodicity}, exactly one
of the following four alternatives holds.
\begin{enumerate}[label=\textup{T\arabic*.}]
\item The profile is bounded by a finite constant for all sufficiently
large $n$.
\item $E(h)$ is empty. For every $n\ge n_*$,
\begin{equation}\label{eq:linear}
                              h(n)>\beta(n),
\end{equation}
including $h(n)=\infty$ in the extended order. In particular the finite
values have an $\Omega(n)$ lower bound.
\item $E(h)$ is finite and nonempty, and $h$ is not eventually bounded.
Put $D=\max E(h)$. There is a nonempty proper set
$U\subseteq\{0,\ldots,Q(D)-1\}$ such that, for each $c\in U$, one
finite value $v_c\le D$ satisfies
\[
        h(n)=v_c\quad\text{whenever }n\ge N(D),\quad n\bmod Q(D)=c.
\]
For $n\ge N(D)$ with $n\bmod Q(D)\notin U$, the lower bound
\eqref{eq:linear} holds.
\item $E(h)$ is infinite. For every bound $B$ there are $n,d\ge B$,
with $d\in E(h)$ and $h(n)=d<\infty$, such that
\begin{equation}\label{eq:peaks}
 n\le(p+2)(\kappa d+\sigma),\qquad
 h(n)\ge\frac{n}{(p+2)\kappa}-\frac{\sigma}{\kappa}.
\end{equation}
Also $\liminf_{n\to\infty}h(n)=\min E(h)<\infty$.
\end{enumerate}
In T1 and T3 the liminf likewise equals the least recurring finite value;
in T2 it is $\infty$. All statements leave the fixed finite initial
range unconstrained. NS and PC assign each fixed uniform request the
same type, by Theorem~\ref{thm:comparison}.
\end{theorem}
\begin{proof}
For $n\ge n_*$, the inverse threshold satisfies $N(\beta(n))\le n$.
If $\beta(n)=0$ this follows from $N(0)=n_*$. Otherwise
$2\kappa\beta(n)\le n-2\sigma-1$, which gives the assertion from
\eqref{eq:NQ}. Consequently, if $h(n)\le\beta(n)$, its finite value
belongs to $E(h)$: its own onset is at most $N(\beta(n))$.
If $E(h)=\varnothing$, this is impossible and proves T2, including the
small ranks for which $\beta(n)=0$.

Suppose $E(h)$ is finite and nonempty and let $D=\max E(h)$.
By truncated periodicity, on each class modulo $Q(D)$ the profile has
one fixed value at most $D$, or is always greater than $D$, once
$n\ge N(D)$. The classes of the first kind form $U$. Recurrence puts
some value from $E(h)$ beyond $N(D)$, so $U$ is nonempty. If all classes
belong to $U$, the profile is eventually bounded. Otherwise, at a size
off $U$, the inequality $h(n)\le\beta(n)$ would give a recurring value,
hence a value at most $D$, contrary to the definition of $U$.
This proves the complete T3 split whenever the profile is not bounded.

If $E(h)$ is infinite, choose its elements $d$ arbitrarily large so that
$N(d)=2(\kappa d+\sigma)+1$. A recurring class for $d$ has a
representative in the interval
\[
              N(d)\le n<N(d)+Q(d),\qquad h(n)=d.
\]
Using integrality and \eqref{eq:periodbound} gives
$n\le(p+2)(\kappa d+\sigma)$, hence \eqref{eq:peaks}. The chosen $d$
and $n$ can both exceed any prescribed bound. These are finite degree
values, rather than occurrences of $\infty$.

Whenever $E(h)$ is nonempty, let $e=\min E(h)$. Each of the finitely
many values below $e$ occurs only finitely often, so eventually
$h(n)\ge e$. The value $e$ recurs, proving the liminf formula. If $E(h)$
is empty, the same argument for every finite bound shows $h(n)\to\infty$.
Finally, eventual boundedness forces $E(h)$ to be finite and nonempty:
a recurring value cannot exceed an eventual bound, and at least one of
the finitely many bounded values occurs infinitely often. This proves
exhaustiveness and mutual exclusivity of the four cases.
\end{proof}

The finite liminf in T4 is one number, not an assertion that the profile
is eventually constant. Its infinitely many recurring levels persist
arbitrarily far out. The bound on $N(d)$ provides the degree gap; the
bound on $Q(d)$ places finite representatives of large recurring levels
at sizes linear in their degrees. Both estimates are needed for the peak
statement. The easy classes in T3 may have different finite degrees.
Neither the theorem nor the comparison identifies the NS and PC sets of
recurring degrees.

\begin{corollary}[Sharper invariant-target and refutation bounds]\label{cor:invariant-budget}
For one fully $S_n$-invariant target given by a stable coefficient rule,
Theorems~\ref{thm:periodicity} and~\ref{thm:classification} hold with
$\kappa=k$, $\sigma=s$ and
$n_*=\max\{n_0,1,kD_G\}$. For refutation, take the single target $1$;
then $D_G=0$, $n_*=n_0$, and the transfer budget is $kd+s$.
\end{corollary}
\begin{proof}
The target map has the trivial module $K$ as source. Its image in a
consequence-space quotient is zero or a copy of $K$, and deletion fixes
$K$. Lemma~\ref{lem:constant} therefore detects membership without the
doubled ambient support needed for indexed targets. The actual target
matrix is still transported. The unchanged axiom and PC-closure budget
$kd+s$ gives periodicity, and the preceding profile argument applies
with the smaller parameters.
\end{proof}

\begin{proposition}[The fluctuating case has no fixed periodic split]\label{prop:no-periodic-split}
In T4 there is no period $q_0$, set of its residue classes $U$, finite
bound $B$, and constant $c>0$ such that, at every sufficiently large
size, $h(n)\le B$ on $U$ and $h(n)\ge cn$ off $U$.
\end{proposition}
\begin{proof}
Choose a recurring finite value $e>B$. Its sufficiently late occurrences
cannot lie in $U$, and for $n>e/c$ cannot lie off $U$. This contradicts
recurrence. The argument uses unbounded recurring finite levels, not
merely an unbounded profile.
\end{proof}

\subsection{The polynomial--exponential consequence}\label{sec:dense-types}
With the full explicit coefficient lists of
Definition~\ref{def:dense-size}, let $S(n)$ be the maximum of the minimum
sizes of proofs of the requested targets. Use $S(n)=0$ for an empty
request and $S(n)=\infty$ if any target is underivable. A collection of
proofs for every target has minimum total size between $S(n)$ and
$|\mathcal B_n|S(n)$ when the request is nonempty. Since
$|\mathcal B_n|=n^{O(1)}$, these two size conventions give the same
conclusions below. For one fixed orbit, any representative has the same
minimum size by relabelling.

\begin{corollary}[Polynomial--exponential size bounds]\label{cor:dense-types}
For either NS identities or ordinary-PC derivations of a fixed uniform
request, the type in Theorem~\ref{thm:classification} has the following
size consequences.
\begin{enumerate}[label=\textup{T\arabic*.}]
\item $S(n)\le n^C$ for all sufficiently large $n$, for some constant $C$.
\item $S(n)\ge2^{cn}$ for all sufficiently large $n$, for some $c>0$.
\item On the easy residue classes, $S(n)\le n^C$ for all sufficiently
large $n$; on their complement, $S(n)\ge2^{cn}$, for fixed $C,c>0$.
\item $S(n)\le n^C$ at infinitely many sizes for some fixed $C$, and
$2^{cn}\le S(n)<\infty$ at arbitrarily large sizes for some fixed $c>0$.
Moreover, for every $e>0$ there is $b$ such that
$n^e<S(n)\le n^b$ at infinitely many sizes. There is no eventual periodic
partition with one uniformly polynomial part and one part satisfying an
exponential lower bound.
\end{enumerate}
The exponential statements are lower bounds. They give no universal
matching exponential upper bound. Infinite sizes are interpreted in the
extended order; finiteness is asserted explicitly on the T4 peak
subsequence.
\end{corollary}
\begin{proof}
Write $D_A$ for a fixed upper bound on axiom degree. If the variable
basis contains an object with support size $j>0$, relabelling supplies
a distinct object supported on each $j$-subset of $[n]$. Thus
$v(n)\ge\binom nj\ge n$ for sufficiently large $n$, while
$v(n)=n^{O(1)}$. The target and axiom index counts are polynomial as well.
Lemma~\ref{lem:dense-size} gives polynomial-size proofs for every target
whose derivation degree is bounded by a fixed constant, uniformly over
the request.

If the aggregate degree is a finite value $d>D_A$, some target attains
that maximum. Its minimum proof size is at least
$B_{v(n)}(d-D_A)$ in either system: for NS, a degree-$d$ identity must
contain a multiplier of degree at least $d-D_A$; for PC, any proof must
contain a line of degree at least $d$. Such a target is nonzero, so the
empty NS certificate creates no exception. A linear degree lower bound
therefore yields $2^{\Omega(n)}$. If a target is underivable, its infinite
size satisfies the same lower inequality. This proves T1--T3 and the two
subsequences in T4.

For the additional T4 assertion, choose a recurring degree $d$ with
$\ell=d-D_A>e$. At its infinitely many sufficiently large occurrences,
\[
                   S(n)\ge B_{v(n)}(\ell)
                        \ge\frac{n^{\ell}}{\ell!}>n^e,
\]
while the fixed degree $d$ gives $S(n)\le n^b$ for some $b$.
A periodic polynomial/exponential partition would be contradicted by
choosing $e$ above its polynomial exponent: these occurrences are too
large for its polynomial part and too small for its exponential part.

If every variable support is empty, the entire variable basis is finite
and every permutation fixes it. Each axiom-index orbit consequently
gives one rank-independent polynomial; each target-index orbit does too.
There are finitely many such orbits, and all appear at sufficiently
large rank. The actual axiom and target polynomial sets therefore
stabilise. The aggregate degree is then a finite constant, or is
$\infty$, giving respectively T1 or T2 with infinite size. This covers
the degenerate case without a growing variable count.
\end{proof}

The coefficient-list estimate is \emph{at least} exponential in universe
size. For instance, $v(n)=\Theta(n^2)$ and a line of degree $\Theta(n)$
require $2^{\Theta(n\log n)}$ coefficient entries. Over a fixed finite
field, encoding the axiom and target polynomials by full lists has length
$L(n)=n^{O(1)}$, since their degrees and index supports are fixed.
If $L(n)=O(n^{c_0})$ for a fixed $c_0>0$, the hard-instance bound gives
$2^{\Omega(L(n)^{1/c_0})}$. This is a stretched-exponential lower bound in
input length, with an exponent depending on the template.

\section{Deleting labels while preserving uniform consequences}\label{sec:transfer}
The closure in Section~\ref{sec:pc-space} reduces a question about
arbitrarily long proofs to one about subspaces and linear maps. The
proof works with these spaces of consequences rather than with invariant certificates. Even an invariant target can belong to the image of an equivariant map without having an invariant preimage in characteristic $p$. Averaging over $S_n$ would therefore miss the issue. The construction below transports the whole image and, for PC, its iterated closure.

A representation, or module, here is a vector space with a linear action
of label permutations. An invariant subspace is preserved by every
such permutation; a quotient by it inherits the action. A
\emph{subquotient} is a quotient of an invariant subspace. The trivial
module is $K$ with every permutation acting as the identity. In
characteristic $p$, the usual averaging operator
$(n!)^{-1}\sum_{g\in S_n}g$ is unavailable once $p\mid n!$.
We therefore construct compatible maps directly, rather than obtain
them by averaging.

There is also a reason not to delete variables by substitution. Setting
all $x_{ij}$ with a deleted row label $i$ to zero turns the totality axiom
$\sum_jx_{ij}-1$ into $-1$. The resulting system is immediately refutable;
it is not the intended smaller instance. Instead we need an operation
which deletes labels on permutation bases while preserving the linear
relations among their subspaces.

The proof separates three requirements: transport the actual coefficient
and multiplication maps, preserve the linear operations forming the
consequence spaces, and reflect target membership. We establish them
below using one deletion operation. A common projector then realises
the reverse direction as an explicit lifting operator. Sections~\ref{sec:block-algebra}--\ref{sec:odd}
construct that projector; the present section explains the contracts
it must satisfy and how they imply the desired transfer.

Here \emph{exactness} means preservation of kernels and images, and hence
of the operations in the PC recurrence. For example, $U+V$ is the image
of $(u,v)\mapsto u+v$, while $U\cap V$ is identified with the kernel of
$(u,v)\mapsto u-v$ on $U\oplus V$. Preserving only dimensions, or only
the spaces generated by the input axioms, would not suffice.

\begin{remark}[Why prime-power blocks are natural]\label{rem:prime-power-counts}
For $q=p^a$ and $0\le j<q$, the images of the integers
$\binom{n+q}{j}$ and $\binom n j$ agree in any field of characteristic
$p$. Indeed, in $K[T]$ repeated use of the characteristic-$p$ binomial
identity gives
\[
 (1+T)^{n+q}=(1+T)^n(1+T^q).
\]
Comparing coefficients below degree $q$ proves the assertion. At
$j=q$ an additional coefficient $1$ appears, so the strict inequality
matters. This explains the persistence of small counting expressions
after a prime-power extension. It is only a counting observation:
ordinary-PC transfer also has to preserve intersections and the
restricted multiplication maps, which require the construction below.
\end{remark}

\subsection{The exact deletion lemma}
For this representation-theoretic construction, $K$ is algebraically
closed of characteristic $p$. Lemma~\ref{lem:field} will return the
conclusions about proof degree to an arbitrary coefficient field.
For a finite set $X$, $K[X]$ denotes the vector space with basis
$e_x$ for $x\in X$. A permutation action on $X$ gives the permutation
module $g e_x=e_{gx}$. The group algebra $K[S_N]$ consists of linear
combinations of label permutations, with composition extended
bilinearly as multiplication. Its elements act by the same expressions
on every representation; an equivariant map commutes with these
expressions. This is the standard module language used below
\cite{AlperinLocal}.

Let $P_b(N)=K[\operatorname{Inj}([b],[N])]$ be the permutation module on injective tuples. Write $\mathcal C_r(N)$ for the finite-dimensional $K[S_N]$-modules annihilated by every group-algebra element which annihilates all $P_b(N)$, $0\le b\le r$. Equivalently, it consists of subquotients of finite direct sums of the $P_b(N)$, $b\le r$. Indeed, their sum is a faithful module for the finite-dimensional quotient algebra; the regular module embeds into a finite sum of copies by evaluation on a basis. This class is closed under submodules, quotients and finite direct sums. It contains the permutation module of every bounded basis system of support at most $r$: each orbit is a quotient of a tuple orbit obtained by choosing an ordering of its support.

Thus $\mathcal C_r(N)$ imposes the operator relations shared by the
tuple spaces up to length $r$. It does not require each vector to
mention at most $r$ labels: a vector in $P_1(N)$ may have nonzero
coefficients at all $N$ labels.

The regular action on the block means that exactly one translation
takes any chosen block label to any other. Every retained label is
fixed by those translations.

Take $q=p^a>r$, write $N=n+q$, and let $A\cong(\Z/p\Z)^a$ act regularly on the last $q$ labels. Its coordinate translations
$\sigma_1,\ldots,\sigma_a$ have order $p$ and fix the retained labels
$[n]$. Choose $c_1,\ldots,c_a\in K$ linearly independent over $\Fp$ and put
\[
 z=\sum_{i=1}^a c_i(\sigma_i-1),\qquad z^p=0,\qquad
 \mathcal D(M)=\ker(z:M\to M)/(\ker z\cap\im z).
\]
Here $1$ is the identity permutation, and multiplication of operators
means composition. The relation $z^p=0$ expresses nilpotence of a
linear operator; it does not assert that $z$ itself is zero.
Permutations of the first $n$ labels commute with $z$, so $\mathcal D(M)$ is an $S_n$-module. Equivariant maps induce the maps of $\mathcal D$.

A nilpotent Jordan chain has basis $v_0,\ldots,v_{\ell-1}$ with
$zv_0=0$ and $zv_j=v_{j-1}$ for $j>0$. Its kernel is $Kv_0$;
for $\ell>1$, this line is already in the image. A length-one chain
has zero image instead. These are statements about the operator $z$;
the chain decomposition need not respect the full label action.

On a Jordan decomposition for the nilpotent operator $z$, this quotient
retains the one-dimensional blocks and discards the longer ones. That
operation is generally not exact. The point of the following theorem is
that the bounded-support representation class, with enough retained labels,
has additional structure making it exact. On permutation bases it then
has the concrete effect of deleting the last block of labels.

All tensor products below are over the working field. Concretely,
$K[X]\otimes Y$ is a list of vectors $y_x\in Y$, written
$\sum_x e_x\otimes y_x$. Labels act on both factors by
$g(e_x\otimes y)=e_{gx}\otimes gy$. A natural identification is one
which commutes with the applicable equivariant maps, rather than merely
matching their dimensions.

\begin{theorem}[Exact deletion]\label{thm:deletion}
For $q>r$ and retained rank $n\ge\max\{0,2r-1\}$ there is a choice of $c_1,\ldots,c_a$, linearly independent over $\mathbb F_p$ in the algebraically closed working field, for which $\mathcal D$ is exact on $\mathcal C_r(n+q)$. It fixes the trivial module. For a support-$u$ basis system $\mathcal U$ with $u<q$, there are natural identifications
\begin{align}
 \mathcal D(K[\mathcal U_{n+q}])&\cong K[\mathcal U_n],\label{eq:permdelete}\\
 \mathcal D(K[\mathcal U_{n+q}]\otimes Y)&\cong
 K[\mathcal U_n]\otimes\mathcal D(Y)\label{eq:tensordelete}
\end{align}
for arbitrary $Y$. For rank-independent invariant coefficient matrices, the first identification transports the actual matrix to its realisation at rank $n$.
\end{theorem}

The algebraic certificate is a single group-algebra element $w$ satisfying
$z^{p-1}wz^{p-1}=z^{p-1}$ on all tuple modules of lengths at most $r$.
Sections~\ref{sec:block-algebra}--\ref{sec:odd} construct the common witness and prove this theorem. The natural matrix and tensor identifications are part of that proof.

The witness is one shared inner-inverse expression; it is not an
inverse to the nilpotent operator on the whole space. Exactness is
asserted for the stated class and retained-rank range, with a suitable
coefficient choice, and does not follow from $z^p=0$ alone.

The tensor identification is required only with a permutation module in its first factor. When applying exactness to a tensor product with an arbitrary subspace, we place that product inside a known bounded-support ambient tensor product. Thus no unstated tensor closure of $\mathcal C_r$ is needed.

\subsection{Transporting the entire consequence space}
Fix a positive degree budget $d$ and write $W_N=P_{N,\le d}$
and $C_d(N)=V_\infty(N)$ for the stabilised PC consequence space.
We first transport the spaces $I_d(N)$ and $C_d(N)$, without choosing
any target. This separates the proof-system calculation from the
question of whether a particular uniform expression belongs to it.

For NS the certificate map has source basis the pairs $(\mu,a)$ with
\[
 \deg\mu+\deg f_{a,N}\le d,
 \qquad (\mu,a)\longmapsto\mu f_{a,N}\in W_N.
\]
Its source has support at most $kd+s$, and $W_N$ has support at most
$kd$. The matrix entry at a target monomial $\nu$ is zero unless
$\mu\mid\nu$, and otherwise is $c(a,\nu/\mu)$. These are the actual
rank-independent invariant entries of the certificate map. The
stability threshold $n_0$ makes each axiom degree independent of rank.
Exactness and matrix naturality therefore identify the embedded image
$\Del(I_d(N))$ with $I_d(n)$.

For PC let $X_N=K[\mathcal V_N]$ and $L_N=P_{N,\le d-1}$.
The multiplication map
\[
       m_N:X_N\otimes L_N\longrightarrow W_N,\qquad
                 x_v\otimes f\longmapsto x_vf
\]
has stable entries. Its source is supported on at most
$k+k(d-1)=kd$ labels. The usable axioms give the image $V_0(N)$
of their coefficient map. The recurrence is
\[
 V_{j+1}(N)=V_j(N)+\im\bigl(m_N|_{X_N\otimes(V_j(N)\cap L_N)}\bigr).
\]
Every stage is a submodule of $W_N$, and every multiplication domain
is a submodule of the fixed support-bounded space $X_N\otimes L_N$.
The intersection is the kernel of $(v,l)\mapsto v-l$ on
$V_j(N)\oplus L_N$. Exactness transports that kernel, the multiplication
image and the sum. The natural tensor identification
\eqref{eq:tensordelete} supplies compatibility with the restricted
multiplication map. Induction thus identifies the embedded
$\Del(V_j(N))$ with $V_j(n)$ for every $j$.

Both ascending chains stabilise because their ambient spaces are finite
dimensional. Choose a stage beyond both stabilisation points. We obtain
\begin{equation}\label{eq:consequence-transport}
 \Del(I_d(N))=I_d(n),\qquad \Del(C_d(N))=C_d(n),
\end{equation}
where equality means equality after the natural identification
$\Del(W_N)\simeq W_n$. This transports every cancellation in a PC
proof, including a low-degree result subsequently multiplied by a
variable. No bound on proof length and no symmetry of individual proof
lines was used.

\subsection{Why deletion also reflects target membership}
Preserving an image is not by itself enough: an exact operation can
annihilate a nonzero quotient. The next lemma supplies the reflection
needed for a general target source.

The detector uses the bilinear coefficient pairing on a permutation
basis, $\langle v,w\rangle=\sum_x v_xw_x$. Its orthogonal space
$S^\perp$ consists of vectors pairing to zero with $S$. This pairing
is nondegenerate over any field; it involves no conjugation or
positivity. A vector represents the functional given by pairing
against it.

\begin{lemma}[Nonzero subquotients survive]\label{lem:detect}
Let $P=K[\mathcal U_N]$ have support at most $u$. Suppose deletion is
exact on $\mathcal C_{2u}(N)$, its block size exceeds $2u$, and the
natural identifications \eqref{eq:permdelete}--\eqref{eq:tensordelete}
hold. If $S\subsetneq S'\subseteq P$ are invariant subspaces, then
$\Del(S'/S)\ne0$.
\end{lemma}
\begin{proof}
The permutation basis gives a nondegenerate dot product on $P$,
equivariant under $S_N$. It induces a pairing
\[
                         S^\perp\otimes(S'/S)\longrightarrow K.
\]
This pairing is surjective: for $v\in S'\setminus S$, a linear
functional vanishing on $S$ can take value one at $v$; the dot product
represents that functional by an element of $S^\perp$.
Let $Y=S'/S$. Tensoring over a field is exact, so the displayed
source lies in $P\otimes Y$, itself a subquotient of $P\otimes P$.
The last permutation module has support at most $2u$, justifying
every use of exactness. If $\Del(Y)=0$, permutation-first-factor tensor
compatibility gives $\Del(P\otimes Y)=0$. Exactness then takes the
surjective pairing to a surjection from zero onto
$\Del(K)\simeq K\ne0$, a contradiction.

We use this surjective pairing, not a trace or an averaged inner
product. In characteristic $p$, composing a pairing with coevaluation
can yield a vanishing dimension scalar and would not prove the lemma.
\end{proof}

The invariant-target case has a simpler detector: the trivial module
$K$ is one-dimensional, so a nonzero map from it has a one-dimensional
image. It has no proper nonzero submodule, which is the meaning of
\emph{simple} in the following argument.

\begin{lemma}[An invariant target needs no doubled budget]\label{lem:constant}
Let $j:K\to W$ and $H\subseteq W$ be equivariant and belong to the
exactness class. Then $j(1)\in H$ if and only if its deleted image
belongs to $\Del(H)\subseteq\Del(W)$.
\end{lemma}
\begin{proof}
Membership is the vanishing of $K\xrightarrow{j}W\to W/H$.
Its image is either zero or isomorphic to the simple trivial module
$K$. Exactness preserves this image, and $\Del(K)\simeq K\ne0$.
Thus the composite vanishes exactly when its deleted composite does.
The vector $j(1)$ may be any invariant target, not just the constant
polynomial one.
\end{proof}

\subsection{The actual uniform target map}\label{sec:target-transfer}
For indexed targets, let $\mathcal B$ be the fixed index basis and
let $\gamma(b,\mu)$ be its stable coefficient rule. Write $\delta$ for the largest stabilised target degree, taking
$\delta=0$ for an empty request. At a rank where
all target degrees are at most $d$, define
\[
 J_N:K[\mathcal B_N]\longrightarrow W_N,\qquad
                [b]\longmapsto G_b^{(N)}.
\]
Its entry in row $\mu$, column $b$ is precisely $\gamma(b,\mu)$.
Equivariance and rank independence are the hypotheses of the natural
matrix identification in Theorem~\ref{thm:deletion}. Consequently
$\Del(J_N)$ is the actual smaller target map $J_n$.
This statement concerns the prescribed full polynomials $G_b^{(n)}$;
we do not replace a target by its degree-$d$ truncation.

Let $H_N$ be either $I_d(N)$ or $C_d(N)$. Every target belongs to
$H_N$ exactly when $\im J_N\subseteq H_N$. Put
\[
          S_N=H_N+\im J_N,\qquad Y_N=S_N/H_N.
\]
The condition is $Y_N=0$. Exactness, target-map naturality and
\eqref{eq:consequence-transport} identify its deletion with
\[
          Y_n=(H_n+\im J_n)/H_n.
\]
The polynomial permutation basis of $W_N$ has support at most $kd$.
Lemma~\ref{lem:detect}, applied with this ORIGINAL support bound,
reflects vanishing of $Y_N$ once the budget covers $2kd$.
The affine budget
\[
                            R(d)=2kd+s+t
\]
also covers the certificate source ($kd+s$), the target source ($t$)
and every multiplication domain. It therefore proves membership in
both directions. The target source need not be simple; reflection
comes from a subquotient of the TARGET ambient polynomial space.
Doubling $R(d)$ once again would be unnecessary: the pairing uses
$W_N\otimes W_N$, whose support is $2kd$.

\begin{proof}[Proof of Theorem~\ref{thm:periodicity}]
First extend scalars to an algebraic closure, as permitted by
Lemma~\ref{lem:field}. The degree of every indexed target stabilises
beyond $n_*=\max\{n_0,1,kD_G+t\}$. If $d$ is smaller than the
largest stable target degree, the all-target predicate is false at both
ranks. Otherwise the full map $J$ just described applies.
Take $N=n+q$, $q=p^a>R(d)$ and retained rank
$n\ge\max\{n_*,2R(d)-1\}$. Exact deletion transports $H_N$, $J_N$
and their sum and quotient. The detection lemma yields
\[
 \bigl(\forall b\in\mathcal B_N,\ G_b^{(N)}\in H_N\bigr)
 \quad\Longleftrightarrow\quad
 \bigl(\forall b\in\mathcal B_n,\ G_b^{(n)}\in H_n\bigr).
\]
Field descent holds for each actual target separately, so the equivalence
holds over the original field. Repeated steps connect any eligible
ranks in a common residue class and prove the stated periodicity.
For one fixed global target orbit, replace $\mathcal B$ by that orbit;
at eligible ranks, equivariance makes every representative equivalent.

At degree zero the stabilised targets are either all zero, or are
scalars with at least one nonzero value. In the latter case every target
is derivable in degree zero exactly when the original axioms include a
nonzero constant. If a target has positive stable degree, degree-zero
all-target membership fails. These are fixed predicates beyond $n_*$
and give the separate zero-level assertion.

For a fully invariant target use $j_N:K\to W_N$, $j_N(1)=G^{(N)}$,
in place of $J_N$. Lemma~\ref{lem:constant} reflects membership without
the tensor pairing. The budget reduces to $kd+s$, with stable onset
$\max\{n_0,1,kD_G\}$. Taking $G^{(N)}=1$ gives the sharper refutation
special case. No target has been added to the axiom system.
\end{proof}

\subsection{A lifting operator for bounded-degree consequences}
Fix $d\ge1$, write $W_n=P_{n,\le d}$ and use a block
$q=p^a>r=kd+s$ in the deletion range. Let
$\iota:W_n\to W_N$ include retained monomials and
$\rho:W_N\to W_n$ discard all others, where $N=n+q$.
We seek a common projector lift satisfying
\begin{equation}\label{eq:newcontract}
 \Lift:W_n\longrightarrow W_N,\qquad
 \deg\Lift(f)\le\deg f,\qquad\Lift(1)=1,\qquad
 \Lift(V_j(n))\subseteq V_j(N)\quad\text{for every }j.
\end{equation}
It will also satisfy $\rho\Lift=\mathrm{id}$. The previously proved
transfer gives
\begin{equation}\label{eq:newiff}
 1\in C_d(n)\quad\Longleftrightarrow\quad1\in C_d(N).
\end{equation}
The operator below realises the upward direction while preserving the
whole PC closure. Indexed targets still require the target-map
argument above; their projector lifts need not equal the prescribed
larger targets individually.

\subsection{The lift from a common projector}\label{lift-sec:formula}
We now give the map, leaving the construction of its certificate to the
next sections. Temporarily allow a coefficient extension $E/K$ and use
the same notation for the extended polynomial spaces. Section~\ref{lift-sec:descent}
returns the lifting statement to $K$.

Identify the added block $B$ with $A=\Fp^a$, acting on it by translations
and fixing $[n]$. Let $\sigma_1,\ldots,\sigma_a$ be the coordinate
translations. Choose $c_1,\ldots,c_a\in E$ linearly independent over
$\Fp$ and set
\[
 z=\sum_{i=1}^a c_i(\sigma_i-1)\in E[S_N],\qquad z^p=0.
\]
The group algebra $E[S_N]$ consists of linear combinations of label
permutations; it acts on polynomials by the corresponding linear
combinations of variable renamings. The permutations commute on the
block, and $\sigma_i^p=1$, which proves $z^p=0$.

The finite algebraic target is one element $w\in E[S_N]$ for which
\begin{equation}\label{lift-eq:witness}
 u w u=u\quad\text{on every }P_b(N)=E[\Inj([b],[N])],\quad b\le r,
 \qquad u=z^{p-1}.
\end{equation}
Thus the same expression must work on every tuple length up to the
support budget. This inner-inverse identity does not make the
nilpotent operator $u$ invertible. For characteristic two put
\[
                         e=(1-zw)(1-wz).
\]
For general $p$, put $m=p-1$ and
\begin{equation}\label{lift-eq:projector}
 S_0=1,\qquad S_{j+1}=(1-z^jwz^{m-j})S_j\ (0\le j<m),
 \qquad e=(1-z^mw)S_m.
\end{equation}
An idempotent, or projector, is an operator satisfying $e^2=e$.
For $v\in\ker z$, the constructed $e$ will supply the representative
$ev$ of its deletion class. Its idempotence and this identification
are proved below from the common witness and the block action.

The two formulas agree at $p=2$. The general formula requires a further
property of the block action: every nilpotent Jordan chain on a tuple
module has length one or $p$. Nilpotence $z^p=0$ alone would also allow
intermediate lengths, for which this projector need not work. The
block decomposition establishes this chain property; together with
the common witness it gives the projector identities in
Section~\ref{sec:odd}. Once these inputs are
proved, the formulas are explicit finite algebra expressions. The lifting operator is
\begin{equation}\label{lift-eq:actual-lift}
 \boxed{\quad\Lift_{n,N,d}(f)=e\,\iota(f).\quad}
\end{equation}
Equivalently, writing $e=\sum_{g\in S_N}a_g g$,
\[
                    \Lift(f)=\sum_{g\in S_N}a_g\,g\bigl(\iota(f)\bigr).
\]
This explains why degree does not increase: inclusion and every
renaming preserve degree. The difficult point is to choose the
\emph{common} coefficients $a_g$ so that all consequence spaces are
preserved. An arbitrary sum of renamings does not have that property.

\subsection{Compatibility with the entire PC closure}\label{lift-sec:verify}
The common projector constructed in Sections~\ref{sec:block-algebra}--\ref{sec:odd}
satisfies $e^2=e$, $ze=ez=0$ and, on $\ker z$,
\begin{equation}\label{lift-eq:onker}
                       ev=v-z^{p-1}wv.
\end{equation}
It identifies $\Del(M)$ with $eM$ naturally as vector spaces.
Together with the retained-coordinate description of deletion,
\eqref{eq:consequence-transport} and its stagewise proof give
\begin{equation}\label{lift-eq:stage-transport}
 \rho\bigl(V_j(N)\cap\ker z_{W_N}\bigr)=V_j(n)\quad\text{for every }j.
\end{equation}
Thus the exactness and multiplication-map work need not be repeated.

\begin{proposition}[The lifting formula preserves and reflects consequences]\label{lift-prop:lift}
Assume the two facts in Section~\ref{lift-sec:formula}. Then
$\Lift=e\iota$ preserves degree, $\rho\Lift=\mathrm{id}$,
$\Lift(1)=1$, and
\[
                  \Lift(f)\in V_j(N)\quad\Longleftrightarrow\quad
                  f\in V_j(n).
\]
The same equivalence holds for the stable spaces $C_d$.
\end{proposition}
\begin{proof}
Every retained monomial is fixed by the block action, so $z\iota(f)=0$.
Equation~\eqref{lift-eq:onker} gives
\[
                  \Lift(f)=\iota(f)-z^{p-1}w\iota(f).
\]
The image of $z$ has no retained component: $z$ kills retained basis
vectors and preserves the span of moved ones. Hence $\rho\Lift(f)=f$.
On constants, every permutation acts trivially, so the correction term
vanishes and $\Lift(1)=1$.

If $f\in V_j(n)$, equation~\eqref{lift-eq:stage-transport} supplies
$v\in V_j(N)\cap\ker z$ with $\rho v=f$. The permutation-space
identification says that $v-\iota(f)\in\ker z\cap\im z$. Since
$e\im z=0$, we have $\Lift(f)=ev$. The space $V_j(N)$ is invariant
under all label permutations, hence under $e$, and $ev\in V_j(N)$.
Conversely, $\Lift(f)\in V_j(N)$ is killed by $z$, so
\eqref{lift-eq:stage-transport} and $\rho\Lift(f)=f$ give $f\in V_j(n)$.
Degree preservation follows from the renaming formula. Pass to the
unions for the stable statement.
\end{proof}

This proves both the forward lifting implication and
\eqref{eq:newiff}. The image of the lift need not be all of $C_d(N)$:
the larger universe has many additional consequences. Every smaller consequence has this projector lift. In the transfer range, for $n\ge n_*$ and a fully
invariant target $G$ whose stabilised degree is at most $d$, naturality of its map from $K$ gives
$\Lift(G^{(n)})=G^{(N)}$: the target map commutes with the common projector,
which acts as the identity on $K$. More concretely, invariance gives
$zG^{(N)}=0$ and $eG^{(N)}=G^{(N)}$, while stable coefficients give
$\rho G^{(N)}=G^{(n)}$. The retained-vector identification and the argument
in Proposition~\ref{lift-prop:lift} therefore yield $e\iota(G^{(n)})=G^{(N)}$. Hence invariant targets, including
$1$, are exactly preserved. For indexed targets this equality need
not hold. The image-subquotient argument above proves transfer of
prescribed indexed targets without assuming it.

The common action also explains how individual inference steps are
realised. For a retained axiom index $a$, naturality of its coefficient
map $F_N$ gives
\begin{equation}\label{lift-eq:axiomlift}
                 \Lift(f_{a,n})=F_N(e_A\iota_A(a)).
\end{equation}
For $f\in V_j(n)\cap P_{n,\le d-1}$, act on the joint tensor input:
\begin{equation}\label{lift-eq:jointlift}
 e_{X_N\otimes L_N}(\iota(x_v)\otimes\iota(f))
       =\sum_y x_y\otimes g_y,
 \qquad X_N=K[\mathcal V_N],\quad L_N=P_{N,\le d-1}.
\end{equation}
The tensor identification in \eqref{eq:tensordelete} places each
$g_y$ in $V_j(N)\cap L_N$. Equivariance of multiplication then gives
\begin{equation}\label{lift-eq:liftmul}
                         \Lift(x_vf)=\sum_y x_yg_y.
\end{equation}
The same permutation acts on both tensor factors; replacing this joint
action by $e_X\otimes e_L$ would lose that compatibility. Storing
closure-basis derivations and expressing the $g_y$ in those bases
recovers a literal PC proof, with no proof-length claim.

\subsection{A one-label lifting example}\label{lift-sec:small}
Work in characteristic two. Retain the label $a$ and add $b,c$.
Let $z=1+(bc)$ and $w=(ab)$ act on the linear variables
$x_a,x_b,x_c$. On their three-dimensional span,
\[
 z=\begin{pmatrix}0&0&0\\0&1&1\\0&1&1\end{pmatrix},\qquad
 w=\begin{pmatrix}0&1&0\\1&0&0\\0&0&1\end{pmatrix}.
\]
Direct multiplication gives $zwz=z$ and
\[
                       e=(1-zw)(1-wz)
                         =\begin{pmatrix}1&1&1\\1&1&1\\1&1&1\end{pmatrix}.
\]
Consequently
\begin{equation}\label{lift-eq:smalllift}
                  \Lift(\alpha+\beta x_a)
                      =\alpha+\beta(x_a+x_b+x_c).
\end{equation}
Its retained coordinate is the original polynomial, and $\Lift(1)=1$.

Take the uniform degree-one axiom $\sum_{i<m}x_i=0$ at every universe size $m$.
At size one the axiom is $x_a$; at size three it is $x_a+x_b+x_c$.
Formula~\eqref{lift-eq:smalllift} sends the former exactly to the latter and
lifts every degree-one consequence. Plain inclusion would leave $x_a$,
which is not a consequence of the larger system: the assignment
$(x_a,x_b,x_c)=(1,1,0)$ satisfies its sum but not $x_a=0$.

Here the invariant target $G^{(m)}=\sum_{i<m}x_i$ is transported to
the prescribed larger target. In contrast, the indexed expression
$G_a^{(m)}=x_a$ has projector lift $x_a+x_b+x_c$, not $x_a$. The map
is linear but is not a ring homomorphism and does not preserve every
indexed expression literally. This calculation illustrates the distinction but lies outside the
indexed transfer range: for $k=d=t=1$ and $s=0$, that theorem uses
$R(d)=3$, whereas here $q=2$ and $n=1$. It is therefore no
counterexample to indexed periodicity.

\subsection{Descent to the coefficient field}\label{lift-sec:descent}
Lemma~\ref{lem:field} returns every target-membership equivalence to
the coefficient field $K$. A concrete forward lift can also be descended:
all its coefficients lie in a finite extension $E/K$. Choose a $K$-linear
map $\lambda:E\to K$ with $\lambda(1)=1$ and apply it coefficientwise:
\begin{equation}\label{lift-eq:fieldlift}
                  \widehat\Lift(f)=\lambda_*(\Lift_E(f)).
\end{equation}
Since $V_j(N;E)=E\otimes_K V_j(N;K)$, this preserves membership in each
closure stage, fixes $1$, preserves degree and satisfies
$\rho\widehat\Lift=\mathrm{id}$. Recover a proof from the $K$-closure
bases; coefficientwise projection of individual proof steps need not
preserve their scalar inferences. Reflection of prescribed target
membership follows from Lemma~\ref{lem:field}, not from a claim that
$\widehat\Lift$ reflects every individual consequence.

\section{Nilpotent blocks and a common projector}\label{sec:block-algebra}
The transfer proof reduces the whole theorem to Exact Deletion
(Theorem~\ref{thm:deletion}). We now prove that finite-dimensional input,
using a support budget $r$ as an abstract parameter. For general indexed
targets its application is $r=2kd+s+t$; for invariant targets it is
$r=kd+s$. The construction is independent of the equations and targets.
We begin with the simpler square-zero operator in characteristic two,
then build a witness with a quadratic retained-label cost, replace that
cost by the linear threshold, and finally handle odd characteristic.

There are two obligations. On permutation bases, deletion must retain
precisely the objects using the $n$ retained labels and transport the
actual coefficient maps. On their subspaces and quotients it must also
preserve kernels and images, so that it transports PC intersections
after cancellation. The orbit calculation supplies the first
obligation; one common group-algebra witness will supply the second.

\subsection{A block of labels and a square-zero operator}\label{sec:block2}
For this part $\operatorname{char}K=2$. Choose $q=2^a>r$, put $N=n+q$, and split the labels into retained labels $R=[n]$ and a block $B$ with $q$ labels. Identify $B$ with the vector space $A=\Ftwo^a$. Translation by the $i$th unit vector gives a permutation $\sigma_i$ of $B$, extended by the identity on $R$. These permutations commute and square to the identity.

Choose $c_1,\ldots,c_a$ linearly independent over $\Ftwo$ in a suitable extension field, and put
\begin{equation}\label{eq:z2}z=\sum_{i=1}^ac_i(\sigma_i-1).\end{equation}
Here $z$ is a linear operator, not a scalar, and its products mean composition. Thus $z^2=0$ can hold although $z\ne0$. Since the summands commute, cross terms cancel in characteristic two, and
\[z^2=\sum_i c_i^2(\sigma_i-1)^2=0.\]
For $q=2$, $z$ can be the swap of the two block labels plus the identity. For $q=4$, work over $\mathbb F_4=\Ftwo[\omega]/(\omega^2+\omega+1)$ and use $(c_1,c_2)=(1,\omega)$.
During this finite-algebra construction, $K$ denotes the chosen
working extension field in all tuple spaces and group algebras.
The extension field is used to construct the block operator and its
witness; Lemma~\ref{lem:field} preserves derivability and minimum
degree over the original coefficient field.

The independence is essential. On a two-point orbit where $\sigma_1$ and $\sigma_2$ induce the same swap, choosing $c_1=c_2=1$ makes $z=0$. The orbit is moved, yet would survive deletion. Independent coefficients ensure that every nonzero binary combination of the $c_i$ remains nonzero.

For a space $M$ with label action define
\begin{equation}\label{eq:D2}\D(M)=\ker z/\im z.\end{equation}
This quotient is legal because $z^2=0$ implies $\im z\subseteq\ker z$. Two vectors killed by $z$ represent the same class if their difference is $zw$ for some $w$. A map commuting with label permutations commutes with $z$ and induces a map between these quotients. Permutations of the retained labels commute with $z$, so they still act on the quotient.

\subsubsection{The two-point orbit calculation}
For the illustrative choice $z=\sigma-I$, on basis vectors $e_\beta,e_\gamma$ swapped by $\sigma$,
\[ze_\beta=ze_\gamma=e_\beta+e_\gamma.\]
The kernel and image are both the one-dimensional span of $e_\beta+e_\gamma$, so the quotient is zero. On a fixed basis vector $e_i$, $z=0$, and its line survives. This is the prototype of deletion: moved directions disappear, fixed directions remain.

\subsubsection{The calculation for every moved orbit}
An orbit comprises the objects reached from one object by the block
translations. Its stabiliser consists of translations fixing that
object. Since $A$ is abelian, an orbit with stabiliser $H$ can be
coordinatised by the cosets $A/H$; the coordinate function below uses
that description.

Let $X$ be a finite $A$-set with no point fixed by all of $A$. On each orbit choose $x_0$. Its stabiliser $H$ is a proper binary subspace. Choose a nonzero linear functional $\ell:A\to\Ftwo$ vanishing on $H$. The function
\[\pi(vx_0)=\ell(v)\]
is well defined: two choices of $v$ differ by an element of $H$. Let $D$ multiply $e_x$ by $\pi(x)$, and put $e_i=\ell(\varepsilon_i)$. These are constants on that orbit. If $S_i$ denotes the action of $\sigma_i$, direct evaluation at a basis vector gives
\[DS_i-S_iD=e_iS_i,\qquad Dz-zD=W:=\sum_i c_ie_iS_i.\]
The difference $DS_i-S_iD$ is a \emph{commutator}, measuring the
failure of the two operators to commute. Here $D$ is diagonal in the
orbit basis, and $e_i\in\Ftwo$ is a scalar, distinct from a basis
vector $e_x$.

Moreover $Wz=zW$ and
\[W^2=\Bigl(\sum_i c_ie_i\Bigr)^2I=\alpha^2I,\qquad\alpha\ne0.\]
Thus $\Theta=W^{-1}D$ satisfies $\Theta z-z\Theta=I$. For $zm=0$ this gives $m=-z\Theta m$, hence $\ker z\subseteq\im z$. The reverse inclusion was automatic, so the quotient vanishes.

This calculation also works on $K[X]\otimes Y$ for \emph{any} space $Y$ carrying the label action. Let $D$ inspect only the $X$ coordinate and let $S_i$ act on both factors. We still have $S_i^2=I$ and the same commutator calculation. This stronger statement will justify multiplication, whose domain contains a variable factor tensored with a consequence subspace.

\subsection{Natural deletion on permutation bases}
Let $\mathcal U$ be a basis system whose objects have support less than $q$. An object fixed by all of $A$ cannot have support meeting $B$. Otherwise invariance of its support under the regular action would force its support to contain all $q$ block labels. Thus its fixed objects are precisely $\mathcal U_n$.

Split the basis into fixed and moved objects. The preceding calculation kills the moved summand and fixes the retained one. Hence
\begin{align}
\D(K[\mathcal U_{n+q}])&\cong K[\mathcal U_n],\label{eq:basis}\\
\D(K[\mathcal U_{n+q}]\otimes Y)&\cong K[\mathcal U_n]\otimes\D(Y).\label{eq:tensor}
\end{align}
The second identification sends $e_u\otimes[y]$ to $[e_u\otimes y]$ for retained $u$. It is natural in $Y$: applying a map to $y$ before or after taking its class gives the same result. For a product of two permutation bases, it simply keeps pairs of retained objects. There is no choice of unrelated isomorphisms here.

Suppose a matrix from one permutation basis to another has stable relabelling-invariant coefficients. On a retained input vector, its image is killed by $z$. Its retained component has exactly the smaller-rank coefficients; its moved component is in $\ker z=\im z$ on the moved summand, so vanishes in the quotient. Thus the \emph{actual matrix}, not merely its rank, is transported correctly.

\subsubsection{A small failure of exactness}
Take the two-point permutation space $E=Ke_\beta\oplus Ke_\gamma$ and its invariant line $T=K(e_\beta+e_\gamma)$. Inclusion $T\hookrightarrow E$ is injective. But $\D(T)=T$ whereas $\D(E)=0$. Deletion takes an injection to the zero map on a nonzero space. Therefore the basis calculation alone does not prove that deletion preserves kernels, images or intersections.

This failure is precisely why a proof cannot stop at ``the monomials delete correctly.'' We must show that, with enough retained labels, every subspace produced by the PC calculation lies in a class where the quotient operation is exact.

\subsection{A common group-algebra witness}\label{sec:common}
The group algebra $K[S_N]$ consists of finite formal sums $\sum_ga_gg$ of label permutations. It acts on every space with an $S_N$ action by the corresponding linear combination of operators. Its multiplication is composition. This provides a common language for certificates valid on many spaces at once.

For $b\ge0$ let
\[P_b(N)=K[\Inj([b],[N])],\]
the space on ordered injective $b$-tuples. This is a tuple space, distinct from the polynomial ambient $P_{N,\le d}$; $b$ counts positions rather than polynomial degree. Label permutations act on entries; a second group $S_b$ permutes tuple positions. These actions commute. Define
\[J_r(N)=\{a\in K[S_N]:aP_b(N)=0\text{ for every }b\le r\}.\]
It is a two-sided ideal. Write $\mathcal C_r(N)$ for the finite-dimensional spaces with label action annihilated by this ideal. Subspaces invariant under labels, quotients and finite direct sums remain in this class.
The ideal $J_r(N)$ is unrelated to the prescribed-target map $J_N$.

A two-sided ideal is closed under multiplication on either side by
any group-algebra element. Here this follows because an operator
annihilating every tuple space still does so after either composition.
Its relations therefore hold on the subspaces and quotients in
$\mathcal C_r(N)$.

Every bounded-support permutation basis of support at most $r$ belongs to the class. For an orbit of an object, order its support of size $b\le r$. Sending an injective tuple to the corresponding relabelling of that object gives a surjective equivariant map from $P_b(N)$ onto the orbit space. The support condition makes this well defined. Elements of $J_r$ kill that quotient as well.

\begin{definition}
A support-$r$ witness is one element $w\in K[S_N]$ satisfying
\begin{equation}\label{eq:witness2}zwz-z\in J_r(N).
\end{equation}
\end{definition}
Such a $w$ gives $zwz=z$ on \emph{every} object of $\mathcal C_r(N)$. It is a inner inverse: for $v\in\im z$, applying $w$ produces a preimage under $z$. It does not invert the nilpotent operator on the whole space. Choosing a separate matrix inverse on each space would not give this compatibility. A module map $f$, meaning a linear map commuting with label permutations, satisfies $f(wv)=w f(v)$ for this common group-algebra element.

\subsubsection{The projector calculation, without an exactness assumption}
Recall that a projector is an idempotent linear operator. Its image
can model a quotient without being invariant under the original label
action. What matters for exactness here is that the same formula for
$e$ commutes with every equivariant map between the spaces concerned.

Suppose $z^2=0$ and $zwz=z$. Set
\[e=(I-zw)(I-wz).\]
Then
\[ze=z(I-wz)=z-zwz=0,\qquad ez=(I-zw)z=0.\]
If $zv=0$, then $ev=v-zwv$. Also
\[(zw)e=(zw-zwzw)(I-wz)=0,\]
since $zwzw=zw$. Applying the formula on $\ker z$ to $ev$ gives $e^2v=ev$. Thus $e$ is a projector.

The map $[v]\mapsto ev$ is an isomorphism $\ker z/\im z\to eM$: it is well defined because $ez=0$, its image lies in $\ker z$, and $ev-v=-zwv$ is in $\im z$ when $zv=0$. Its inverse sends a vector of $eM$ to its quotient class.

A short exact sequence $0\to U\to V\to W\to0$ consists of an
injection, followed by a surjection whose kernel is the image of that
injection. Preservation means that these three properties still hold
after applying deletion. The common projector verifies each property
directly.

Now let $0\to U\to V\to W\to0$ be an exact sequence in $\mathcal C_r(N)$. All maps commute with the same $e$. Restricting to $eU,eV,eW$ preserves injectivity. A vector of $eW$ has a lift $v\in V$; $ev$ is a lift in $eV$, proving surjectivity. A vector $ev$ mapping to zero lies in $U$ and equals its own projection, so lies in $eU$. This proves exactness explicitly. By the natural isomorphism, $\D$ is exact.

The projector need not commute with permutations of the retained labels. We use its image as a vector-space model to prove exactness, and keep the original quotient for the residual label action. Nothing here asserts an unwarranted equivariant direct-sum decomposition for that residual group.

\begin{example}[One retained label]
Let the retained label be $a$ and the two block labels be $b,c$. Put $z=I+(bc)$ and $w=(ab)$. On $P_1(3)$,
\[zwz(e_b)=zw(e_b+e_c)=z(e_a+e_c)=e_b+e_c=ze_b,\]
and similarly for $e_c$; both sides vanish on $e_a$. The projector is the all-ones $3\times3$ matrix. Its image is the constant line, naturally representing the retained one-dimensional quotient. With no retained label, every element of $K[S_2]$ satisfies $zwz=0$, so there is no such witness. The spare label makes a real difference.
\end{example}

\section{A preliminary construction with fresh labels}\label{sec:quadratic-route}
The first construction exhibits all the algebraic steps with a
generous label budget. In characteristic two, it repairs the common
identity $zwz=z$ at tuple length $b$ by allocating $b$ new retained
labels. The cumulative cost is quadratic in $r$; the next section
improves this cost by using the residual's special shape on one pool.

\subsection{Fresh-label construction}\label{sec:quadratic}
We now prove that a support-$r$ witness exists in characteristic two when
\[n\ge T(r):=1+2+\cdots+r=\frac{r(r+1)}2,\qquad r<q.\]
This proof is independent of the rook theorem used later. Its ingredients are finite algebra, a position-compatible auxiliary inverse, and an explicit way to simulate that inverse using fresh labels.

Choose the generic coefficients in a finite field $F$ of characteristic two, large enough to contain $a$ independent elements. We may work inside the algebraic closure of the original field; the degree question descends by Lemma~\ref{lem:field}. The finiteness of $F$ will be used only to normalise an element of the finite unital group algebra on the current label set.
If that label set is $\Omega$, the algebra $F[S_\Omega]$ has
$|F|^{|S_\Omega|}$ elements, one coefficient in $F$ for each
permutation. This finite construction field imposes no finiteness
restriction on the original coefficient field.

\subsubsection{Normalisation of the residual}
Suppose $w_0$ works on shorter tuples but not yet on length $b$. Put $y=z-zw_0z$. An auxiliary linear map $h$ may satisfy $zhz=z$ on the larger tuple space. It is tempting to conclude $yhy=y$. In general the expansion contains an extra term $zw_0zw_0z$, and this need not collapse. Before correcting the error we arrange the additional identity $w_0zw_0=w_0$ in the actual group algebra. This normalisation was a genuine mathematical step in the earlier construction.

Here normalisation replaces a ring element to impose an additional
algebraic identity. A ring representation interprets ring elements as operators,
respecting addition, multiplication and the identity. Thus a relation
proved in the group algebra holds simultaneously in all its
representations. This is why the next lemma preserves every previously
successful tuple length at once.

\begin{lemma}[Normalisation in a finite ring]\label{lem:normal}
Let $a,z$ belong to a finite ring. There is an element $b$ in the same ring with $bzb=b$, such that every ring representation in which $zaz=z$ also satisfies $zbz=z$.
\end{lemma}
\begin{proof}
Put $x=az$. The sequence of powers of $x$ eventually repeats, so $x^{i+t}=x^i$ for some $i\ge0,t>0$. Multiplication by further powers gives $x^{m+t}=x^m$ for $m\ge i$. Choose $k\ge i$ with $k+1$ divisible by $t$. Then $2k+1$ differs from $k$ by a multiple of $t$, and $x^{2k+1}=x^k$.

Set $b=x^ka$. Direct multiplication gives
\[bzb=x^k(az)x^ka=x^{2k+1}a=x^ka=b.\]
In a representation where $zaz=z$, induction gives $z(az)^m=z$ for every $m\ge0$. Hence $zbz=z(az)^{k+1}=z$. One exponent chosen in the original finite ring works simultaneously in all such representations.
\end{proof}
The notation $b$ in this lemma is an element, not a tuple length; in the construction below we denote this normalised element by $b_0$.

\begin{lemma}[Normalised residual]\label{lem:residual}
If $b_0zb_0=b_0$ and $zhz=z$, then, for $y=z-zb_0z$,
\[yhy=y.\]
\end{lemma}
\begin{proof}
Expand without commuting any factors:
\begin{align*}
yhy&=zhz-zhzb_0z-zb_0zhz+zb_0zhzb_0z\\
&=z-2zb_0z+zb_0zb_0z\\
&=z-zb_0z=y.
\end{align*}
The last equality uses the normalisation. This identity works in any ring; characteristic two is not needed for this particular calculation.
\end{proof}

\subsubsection{An auxiliary inverse compatible with tuple positions}
Position permutations reorder the entries of a tuple; label
permutations rename those entries. These actions commute. In this
step we require compatibility with positions, since the subsequent
simulation must work on all orderings of each underlying label set.

Let $\Omega$ consist of the block $B$ and the retained labels already allocated. On $Q=P_b(\Omega)$, partition the tuples according to the $A$-orbit of their \emph{underlying label set}. A set meeting $B$ has proper stabiliser because it has size $b<q$. Choose the functional $\ell$ and coordinate $\pi$ as in Section~\ref{sec:block2}, now on these sets. Let $D$ read $\pi$ from the underlying set, and use the same $W$ and nonzero $\alpha$ as before, defined separately on each orbit. Then on each moved orbit,
\[Dz-zD=W,\quad Wz=zW,\quad W^2=\alpha^2I,\quad z^2=0.\]
Multiplying the commutator by $z$ shows $zDz=zW$. Thus
\[h=\alpha^{-2}WD\quad\text{satisfies}\quad zhz=z.\]
On tuples wholly outside $B$, put $h=0$, since $z=0$ there. Because $D$ depends on the underlying set and translations act on labels, $h$ commutes with every permutation of tuple positions. We have not claimed that $h$ is a group-algebra element. It is just an explicitly constructed linear operator.

\subsubsection{Fresh labels simulate any position-compatible operator}
Adjoin $b$ new labels $\kappa_1,\ldots,\kappa_b$ to $\Omega$. Let $i:P_b(\Omega)\to P_b(\Omega\cup C)$ include the old tuples and let $p$ discard every tuple using a fresh label. We prove that any operator $h$ commuting with position permutations has the form
\begin{equation}\label{eq:compression}h=pci\end{equation}
for some $c\in F[S_{\Omega\cup C}]$.

The inclusion and projection satisfy $pi=I$. The expression $pci$
is the old-to-old part of the larger operator $c$: it makes no claim
about inputs or outputs containing fresh labels. The residual
localisation in the induction below is what makes this part sufficient.

Fix a $b$-element set $U=\{u_1,\ldots,u_b\}\subseteq\Omega$, with one chosen ordering. Form
\[E_U=\prod_{j=1}^b\bigl(1-(u_j\ \kappa_j)\bigr).\]
The transpositions are disjoint. On an old tuple missing some $u_j$, the corresponding factor kills the tuple, and the other factors cannot introduce $u_j$. On a tuple using all the $u_j$, the identity term leaves the tuple old, while every nonidentity term introduces a fresh label. Therefore $pE_Ui$ is precisely the projector onto the orderings of $U$.

For an ordered output tuple $v$, choose a permutation $g\in S_\Omega$ taking $(u_1,\ldots,u_b)$ to $v$. Extend $g$ by fixing fresh labels. Then $pgE_Ui$ takes that input ordering to $v$, takes its position permutations to the corresponding permutations of $v$, and kills all other underlying sets. The orderings of each $U$ form a free orbit for $S_b$. Consequently an operator commuting with $S_b$ is completely determined by its value on one ordering for each $U$. Expanding those values as linear combinations of output tuples and using the operators $pgE_Ui$ proves \eqref{eq:compression}. There is no division by $b!$.

A free position orbit means that no nonidentity position permutation
fixes an injective tuple. Its $b!$ orderings are distinct basis vectors.
Consequently position compatibility determines the remaining values
from one ordering without averaging over $S_b$.

\begin{example}[Two registers]
For $U=\{u_1,u_2\}$,
\[E_U=1-(u_1\kappa_1)-(u_2\kappa_2)+(u_1\kappa_1)(u_2\kappa_2).
\]
On $(u_1,u_2)$, projection $p$ kills the last three terms. On $(u_1,v)$ with $v\ne u_2$, the identity contribution is cancelled by the second swap factor, which fixes that tuple. The detector recognises the set $U$, while leaving the two possible orderings available for a position-compatible output rule.
\end{example}

\subsubsection{The induction, including localisation of the error}
For support zero take $w_0=0$: $z$ is zero on constants. Suppose a witness exists on the block and $T(b-1)$ retained labels, collectively $\Omega$. Normalise it in the finite ring $F[S_\Omega]$ to $b_0$. It still works on all tuple lengths below $b$ and satisfies $b_0zb_0=b_0$ in the ring itself. Set $y=z-zb_0z$.

Adjoin $b$ fresh labels. Decompose $P_b(\Omega\cup C)$ according to the positions occupied by fresh labels and their actual values. The group $S_\Omega$ fixes each fresh label, so every summand with at least one fresh label is a copy of $P_j(\Omega)$ for $j<b$. On it $y=0$ by the old witness relation. The summand with no fresh labels is $Q=P_b(\Omega)$. Thus, as actual maps,
\[y=i y_Q p.\]
This localisation is what allows a compressed inverse to suffice. Construct the auxiliary $h$ on $Q$ above. Lemma~\ref{lem:residual} gives $y_Qhy_Q=y_Q$. Choose $c$ with $pci=h$. Then
\[ycy=i y_Q(pci)y_Qp=i y_Qp=y.\]
Define the new group-algebra element
\[w_b=b_0+(1-b_0z)c(1-zb_0).\]
Since $z(1-b_0z)=(1-zb_0)z=y$,
\[zw_bz=zb_0z+ycy=zb_0z+y=z\]
on length $b$. On every shorter tuple space the same fresh-label decomposition makes $y=0$, so the relation remains valid there too. The new cost is exactly $b$ retained labels, giving
\[T(b)=T(b-1)+b.\]
If still more retained labels are present, extend the constructed permutations by fixing them. Decomposing according to those outside labels reduces every tuple of length at most $b$ to a shorter tuple on the constructed labels. Thus the same witness works for all $n\ge T(b)$.

\begin{theorem}[Quadratic retained-label budget]
For every $r<q=2^a$ and $n\ge r(r+1)/2$, there is a common witness $w$ with $zwz=z$ on all tuple modules of length at most $r$. Therefore deletion is exact on the support-$r$ class. It transports the
entire degree-$d$ PC consequence space with $r=kd+s$, beyond the fixed
template threshold. For prescribed indexed targets use the budget
$r=2kd+s+t$ to include the target map and the doubled-support detector.
\end{theorem}
Every construction needed for this theorem has now been supplied. In particular, the proof is not an inference from witnesses found for small tuple lengths.

\begin{remark}[The square-root degree lower bound]
For either prescribed-target budget, the construction gives an onset
$O(d^2)$ for degree-$d$ transfer.
Inverting that onset gives an $\Omega(\sqrt n)$ lower bound when no
finite degree recurs, and on the hard residues of the periodic case.
The loss is precisely the cumulative allocation $T(r)=r(r+1)/2$.
The following construction replaces this sum by the maximum $2r-1$,
using one retained pool at every layer.
\end{remark}

\section{A common witness with a linear label budget}\label{sec:linear-construction}
Keep $N=n+q$ and the characteristic-two operator $z$, with $q>r$.
The improvement is to repair only the old witness's residual: it
vanishes on incoming shorter-tuple directions and has zero output
marginals. We first construct a correction of this shape, then
realise it by permutations annihilating all shorter tuples. Every
stage uses the same retained pool. Its cost is therefore
$\max_{b\le r}(2b-1)=2r-1$ for $r\ge1$, rather than a sum of
fresh-label allocations.

\subsection{The zero-marginal layer}\label{sec:layers}
We still work in characteristic two. Fix a tuple length $b\ge1$, distinct from the proof-degree budget $d$. Let $V=P_b(N)$. For each position $i$ there is a deletion map
\[\partial_i:P_b(N)\longrightarrow P_{[b]\setminus\{i\}}(N)
\]
which forgets the entry in position $i$. On a vector of coefficients it sums over that entry, so we call it a marginal. The target is a copy of $P_{b-1}(N)$ with the surviving positions named. Give tuple spaces the dot product making their bases orthonormal, and put
\begin{equation}\label{eq:anatomy}
V'=\sum_i\im\partial_i^{\mathsf T},\qquad
\mathcal Z_b(N)=\bigcap_i\ker\partial_i=(V')^\perp,\qquad T_b(N)=V/V'.
\end{equation}
Here $\mathcal Z_b$ is the zero-marginal space; we also write $\Soc=\mathcal Z_b$ and $\Top=T_b$. This notation distinguishes the tuple zero-marginal space from the
polynomial consequence space $C_d(n)$. All statements below use the
displayed definitions.

The dot product is the bilinear coefficient pairing, not a positive
inner product. Thus $U^\perp$ denotes the annihilator of $U$, not an
assumed complement. The pairing identifies $V/V'$ with
$\mathcal Z_b^*$: a quotient class $[v]$ acts on $\zeta\in\mathcal Z_b$
by $\langle v,\zeta\rangle$, independently of its representative.
We use only this elementary finite-dimensional duality; see
\cite{LangAlgebra} for background.

The incoming space $V'$ contains directions lifted from shorter tuples. The zero-marginal space contains directions invisible after deleting any one position. The quotient $T_b$ forgets the incoming directions. Label permutations commute with all marginal maps, and position permutations permute the marginal maps. Thus these spaces carry the two appropriate actions.

\begin{example}[Lengths one and two]
For $b=1$, $\partial$ sums coordinates, $V'$ is the constant line and $\mathcal Z_1(N)$ is the zero-sum space. For $b=2$, a coefficient vector is an off-diagonal array $f(i,j)$, $i\ne j$. Its marginals are its row and column sums. Thus $\mathcal Z_2(N)$ consists of off-diagonal arrays with every row and column sum zero. On three labels it is spanned by
\[e_{12}+e_{23}+e_{31}-e_{21}-e_{32}-e_{13}.
\]
In characteristic two the six coefficients are all one, but every line still sums to zero because it contains two ones.
\end{example}

\subsubsection{The old error already has a special shape}
Suppose $w_0$ is a witness on all lengths below $b$, and set
\[\varepsilon=zw_0z-z\quad\text{on }V.\]
This is the action of an element of $J_{b-1}(N)$. For any $x\in J_{b-1}(N)$,
\[\partial_i x=x\partial_i=0,\qquad x\partial_i^{\mathsf T}=\partial_i^{\mathsf T}x=0.
\]
The same element $x$ acts on both tuple spaces: it commutes with the
marginal maps and acts as zero at length $b-1$.
Therefore $x$ maps $V$ into $\mathcal Z_b(N)$ and kills $V'$. In particular $\varepsilon$ factors through
\[V\longrightarrow T_b(N)\longrightarrow \mathcal Z_b(N)\longrightarrow V.
\]
It also commutes with position permutations. This is much more special than an arbitrary operator on $V$. The improved route exploits exactly this special form.

Factoring through the quotient means that
$\bar\varepsilon([v])=\varepsilon v$ is well defined because
$\varepsilon$ kills $V'$. Its output lies in the actual subspace
$\mathcal Z_b\subseteq V$. The left and right correction maps must
be defined on these two spaces.

There are two tasks. First, construct a position-compatible correction of this shape with $z\varphi z=-\varepsilon$. Second, realise that correction by an element of the actual label group algebra which kills all shorter tuples. The first task needs retained labels; the second will be solved by a rook-placement theorem without adding labels.

\subsection{Partitioning a zero-marginal vector so that one label is avoided}\label{sec:partition}
For a finite position set $S$ and label set $L$, let $P_S(L)=K[\Inj(S,L)]$ and $C_S(L)=\bigcap_{i\in S}\ker\partial_i$. This is the zero-marginal space also denoted $\mathcal Z_S(L)$. If $L'\subseteq L$, regard $C_S(L')$ as the vectors of $C_S(L)$ supported on tuples using only $L'$. Write $\mathcal Z_c$ when there are $c$ positions.

Suppose a block label $\beta$ occupies a new position and $y\in \mathcal Z_c(L)$ occupies the other positions. The vector $e_\beta\otimes y$ has zero marginals in those other positions, but its new-position marginal is $y$. We want a retained correction with the same marginal. If
\[y=\sum_{x\in L}\pi_x(y),\qquad \pi_x(y)\in \mathcal Z_c(L\setminus\{x\}),
\]
then
\begin{equation}\label{eq:standin}
e_\beta\otimes y-\sum_{x\in L}e_x\otimes\pi_x(y)
\end{equation}
has every marginal zero. Avoiding $x$ is essential: it makes each displayed tuple injective. The maps must also commute with permutations of the old positions, so they can be used consistently in all layers.

\begin{example}[One old position]
Take $L=\{1,2,3\}$ and $y=a e_1+b e_2-(a+b)e_3$. Define
\[\pi_1(y)=b(e_2-e_3),\quad \pi_2(y)=a(e_1-e_3),\quad\pi_3(y)=0.
\]
Each summand avoids its subscript and has coordinate sum zero; their sum is $y$. For $y=e_1-e_2$, formula \eqref{eq:standin} is the six-term vector
\[e_{\beta1}-e_{\beta2}+e_{12}-e_{13}-e_{21}+e_{23}.
\]
Check its two marginals by grouping terms with the same surviving label. With only labels $1,2$, every $\mathcal Z_1(L\setminus\{x\})$ is zero, while $e_1-e_2\ne0$. The third label is what permits the construction.
\end{example}

Here a partition of unity means a sum of linear maps equal to the
identity. Its summands may overlap and need not be idempotent.
Equivariance concerns tuple positions; the construction may choose
distinguished labels.

\begin{lemma}[Partition of unity]\label{lem:unity}
If $|L|\ge2c+1$, there are position-equivariant linear maps
\[\pi_x:\mathcal Z_c(L)\longrightarrow \mathcal Z_c(L\setminus\{x\}),\qquad
\sum_{x\in L}\pi_x=I.
\]
This statement and its proof work over any field.
\end{lemma}
The label called $d$ in the following proof is a local distinguished
label, not the polynomial-degree budget.
\begin{proof}
We induct on $c$. At $c=0$ the space is $K$ on the empty tuple; choose one label $x$, make $\pi_x=I$ and all others zero.

Let $c\ge1$. Choose $d\in L$ and put $L'=L\setminus\{d\}$. For $y\in \mathcal Z_c(L)$ and position $i$, let $y^{(i)}$ be the coefficients of those tuples in which $d$ appears in position $i$, with that position then removed. This lies in $\mathcal Z_{[c]\setminus\{i\}}(L')$. To see this, fix $d$ in position $i$ and sum another position: it is a marginal of $y$, hence zero. There cannot be a second occurrence of $d$ in an injective tuple.

By induction, partition $y^{(i)}$ on $L'$ as $\sum_{x\in L'}\pi'_x(y^{(i)})$, each term avoiding $x$. For a vector $v\in \mathcal Z_{[c]\setminus\{i\}}(L'\setminus\{x\})$, partition once more on $L'\setminus\{x\}$:
\[v=\sum_{x'\in L'\setminus\{x\}}\pi''_{x'}(v).
\]
Both applications are justified: $|L'|\ge2c$ and $|L'\setminus\{x\}|\ge2c-1=2(c-1)+1$.

Construct $F_{i,x}(v)$ by inserting $d$ into position $i$ of $v$, then subtracting the sum obtained by inserting $x'$ into position $i$ of $\pi''_{x'}(v)$. Every term avoids $x$. All tuples are injective because the lower-dimensional vector avoids the inserted label. A marginal in a position other than $i$ vanishes term by term. The marginal in position $i$ is $v-\sum_{x'}\pi''_{x'}v=0$. Thus
\[F_{i,x}(v)\in \mathcal Z_c(L\setminus\{x\}).
\]
Its $d$-part in position $i$ is $v$, and its $d$-parts in other positions are zero.

For $x\in L'$ set
\[\pi_x(y)=\sum_{i=1}^cF_{i,x}\bigl(\pi'_x(y^{(i)})\bigr),\qquad
\pi_d(y)=y-\sum_{x\in L'}\pi_x(y).
\]
The $d$-part of the last expression in every position is $y^{(i)}-\sum_x\pi'_x(y^{(i)})=0$. It therefore uses no label $d$, and belongs to $\mathcal Z_c(L')$. The sum identity is built into the definition.

For equivariance, use the same inductively constructed maps on every position set of size $c-1$, transported along a bijection. The transport is independent of the chosen bijection because the inductive maps commute with position permutations. Permuting the $c$ positions simply permutes the terms indexed by $i$ in the displayed sum. No averaging over a symmetric group is used.
\end{proof}

The count $2c+1$ is now visible in an actual induction: reducing one position removes two labels, the label $d$ being cleared and the label $x$ being avoided. This is a sufficient bound proved by the construction, not an appeal to a dimension formula.

\subsection{Adding several positions without losing the zero marginals}\label{sec:cone}
A section of a surjection $f$ is a linear map $\sigma$ with
$f\sigma=I$: it chooses preimages coherently with linear combinations.
It gives a complement to $\ker f$, but compatibility with an operator
or group action requires a further argument. Here the section adds
positions and the surjection sums them out.

Let $R_0$ and $J$ be disjoint position sets of sizes $c$ and $t$. Here $t$ counts new positions, not target-index support. We want to add the positions in $J$ while keeping the marginals in $R_0$ zero. If $|L|\ge2c+t$, the map deleting all positions in $J$ has a position-$R_0$-equivariant section
\begin{equation}\label{eq:cone}
\mathcal Z_{R_0}(L)\longrightarrow
\{\Phi\in P_{R_0\cup J}(L):\partial_i\Phi=0\ (i\in R_0)\}.
\end{equation}
Here is the construction. Suppose some positions $J'\subsetneq J$ have already been added. Split the vector according to their injective labelling $\gamma$. On that component the remaining zero-marginal vector uses the label set $L\setminus\im\gamma$. Its size is at least
\[2c+t-|J'|\ge2c+1.
\]
Partition that vector by Lemma~\ref{lem:unity}. For a new position $j$, insert label $x$ in the summand $\pi_x$. Summing the new position returns the original vector, and the old zero marginals stay zero. Repeat until all $t$ positions have been added. Every step is linear and commutes with the $R_0$ position action, proving \eqref{eq:cone}.

\subsubsection{Sorting a zero-marginal vector by its block positions}
Split the universe as retained labels $[n]$ and block labels $B$. Every vector $f\in P_b(n+q)$ has components $f_D$, for subsets $D\subseteq[b]$, consisting of tuples whose block positions are exactly $D$:
\[f_D\in K[\Inj(D,B)]\otimes P_{[b]\setminus D}(n).
\]
Write $\epsilon_i$ for summing a block entry and $\partial_i^{\rm ret}$ for summing a retained entry. The condition of zero marginal in position $i\notin D$ is
\begin{equation}\label{eq:cycleeq}
\partial_i^{\rm ret}f_D+\epsilon_i f_{D\cup\{i\}}=0.
\end{equation}
It simply separates the possible value of the deleted entry into a retained label or a block label.

Let $\mathcal Z^{\le t}$ be the zero-marginal vectors with no components using more than $t$ block positions. The first space is $\mathcal Z^{\le0}=\mathcal Z_b(n)$. The top components of a vector in $\mathcal Z^{\le t}$ have zero retained marginals by \eqref{eq:cycleeq}. We obtain a map onto the prospective layer
\[G_t=\bigoplus_{|D|=t}K[\Inj(D,B)]\otimes \mathcal Z_{[b]\setminus D}(n).
\]
Its kernel is $\mathcal Z^{\le t-1}$. We next prove surjectivity with a position-compatible section, so this is an actual quotient description rather than a count of dimensions.

The increasing spaces $\mathcal Z^{\le t}$ form a filtration; its
$t$th layer is the quotient by $\mathcal Z^{\le t-1}$. In the
displayed tensor coordinates, $e_\beta\otimes y$ places the retained
vector $y$ at a specified block pattern $\beta$. The direct sum separates
the sets of block positions. This is a tuple layer, not a change in
the polynomial-degree budget.

\subsubsection{An explicit lift of one top component}
Take $J=[t]$, an injective block tuple $\beta:J\to B$, and $y\in \mathcal Z_{R_0}(n)$ with $R_0=[b]\setminus J$. If
\[n\ge2(b-t)+t=2b-t,
\]
the section \eqref{eq:cone} gives $\Phi$ on the $b$ retained positions such that $\partial_J\Phi=y$ and $\partial_i\Phi=0$ for $i\in R_0$. For $J'\subseteq J$, set
\[\phi_{J'}=(-1)^{|J'|}\partial_{J\setminus J'}\Phi.
\]
For $D\subseteq J$ define
\[f_D=e_{\beta|D}\otimes\phi_{J\setminus D},
\]
and put the other components zero. At $D=J$ this is $e_\beta\otimes y$, the required top component. For $i\in R_0$, the retained marginal vanishes. For $i\in J\setminus D$,
\[\partial_i\phi_{J\setminus D}=-\phi_{J\setminus(D\cup\{i\})},
\]
which cancels the block marginal of the adjacent component in \eqref{eq:cycleeq}. Thus the whole lift belongs to $\mathcal Z^{\le t}$.

To lift all components equivariantly, choose an ordering of each $t$-element subset of block labels. Use it in the representative positions $[t]$. Its stabiliser among position permutations consists exactly of permutations of the remaining $b-t$ positions. The constructed lift commutes with this stabiliser, by the cone construction. Transport it along position permutations to every other placement of the same ordered block labels. If two permutations give the same placement, their difference is in the stabiliser, so the transported lift is the same. Extend linearly over all block subsets and retained vectors. This supplies the promised $S_b$-equivariant section without division by any group order.

For $1\le t\le b$, the largest requirement $2b-t$ occurs at $t=1$. Therefore the single hypothesis
\begin{equation}\label{eq:linearlabel}n\ge2b-1\end{equation}
supplies all these sections.

\subsection{Making the splitting compatible with the block operator}\label{sec:blocksplitting}
The sections just constructed commute with position permutations, but need not commute with $z$. We now fix that, without adding labels.

A block pattern is an injective partial tuple with a specified nonempty position domain $D$ of size $t$, and values in $B$. Its image is a nonempty subset of $B$ of size $t<q$. Choose the orbit coordinate $\pi$ from Section~\ref{sec:block2} on these \emph{image subsets}. Let $E$ be the space on these patterns, and give a pattern grade $0$ or $1$ according to $\pi$ of its image. Let $E_0,E_1$ be the spans of the two classes. These spaces are stable under position permutations and split by the position domain, since neither operation changes the image subset.

The diagonal grade operator $D$ (distinct from the position domain) and the invertible $W=[D,z]$ satisfy the same identities as before. For either $e=0,1$, suppose $v_0+zv_1=0$ with $v_0,v_1\in E_e$. Apply $D-eI$. The first term vanishes and the second gives $Wv_1$, so $v_1=0$, and then $v_0=0$. Thus $E_e\oplus zE_e$ embeds in the pattern space, giving $2\dim E_e\le\dim E$. Since $E=E_0\oplus E_1$, both inequalities must be equalities. In particular
\[E=E_0\oplus zE_0.
\]
Tensor each domain component with its retained zero-marginal factor. Because $z$ fixes retained labels, the layer has the analogous decomposition
\[G_t=Q_t\oplus zQ_t
\]
with $Q_t$ position-stable and $z:Q_t\to zQ_t$ injective by the preceding embedding. This injectivity makes the two components used below unique.

\subsubsection{Turning a section into a commuting section}
Let $f:E\to Q$ commute with both $z$ and position permutations,
suppose $Q=Q_0\oplus zQ_0$ with $z|_{Q_0}$ injective, as just
established for the layer, and let $\sigma$ be a position-equivariant
section. Define
\[\tau(q_0+zq_1)=\sigma(q_0)+z\sigma(q_1).
\]
The decomposition is unique. We have $f\tau=I$. Moreover
\[z\tau(q_0+zq_1)=z\sigma(q_0)=\tau(zq_0),
\]
using $z^2=0$. Thus $\tau$ commutes with $z$ as well as with positions. Apply this to each quotient $\mathcal Z^{\le t}\to G_t$ and split the filtration successively. We have proved
\begin{equation}\label{eq:socsplit2}
\mathcal Z_b(n+q)=\mathcal Z_b(n)\oplus C\oplus zC,
\end{equation}
where $C$ is position-stable, $z:C\to zC$ is an isomorphism, and $z$ kills the retained summand. This is a concrete direct-sum decomposition, valid under $b<q$ and $n\ge2b-1$.

\subsubsection{The four consequences used for correction}
The witnesses below reverse the arrow $c\mapsto zc$ on the appropriate
chain directions. They are inner inverses, not inverses of the
nilpotent operator on all of $V$. The notation $z^{-1}(U)$ in the
duality calculation denotes the inverse image $\{v:zv\in U\}$.

Write $V=P_b(N)$. The decomposition implies:
\begin{enumerate}
\item $zV\cap \mathcal Z_b(N)\subseteq z\mathcal Z_b(N)$.
\item $zV\cap V'\subseteq zV'$.
\item There is a position-equivariant $w_C$ on $\mathcal Z_b(N)$ with $zw_Cz=z$ there.
\item There is a position-equivariant $w_T$ on $T_b(N)$ with $zw_Tz=z$ there.
\end{enumerate}
For the third assertion, let $w_C$ send $zc$ back to $c$ and kill $C$ and the retained summand. For the first, the kernel of $z$ in \eqref{eq:socsplit2} is $\mathcal Z_b(n)\oplus zC$. An element of $zV\cap \mathcal Z_b(N)$ is therefore $s+zc$. But $zV$ consists of vectors on tuples meeting the block; so does $zc$. The remaining $s$ is supported entirely on retained tuples, forcing $s=0$.

For the other two assertions use transpose. In characteristic two each $\sigma_i$ is an involution, so $z^{\mathsf T}=z$. The pairing identifies $T_b(N)$ with $\mathcal Z_b(N)^*$, giving $w_T$ by duality. To see the second inclusion explicitly, take orthogonal complements. The first inclusion implies
\[z^{-1}(\mathcal Z_b)=\ker z+\mathcal Z_b.
\]
Indeed, if $zv\in \mathcal Z_b$, it equals $zc$ for some $c\in \mathcal Z_b$, so $v-c\in\ker z$. But
\[(zV')^\perp=z^{-1}(\mathcal Z_b),\qquad
(zV\cap V')^\perp=\ker z+\mathcal Z_b.
\]
The complements agree, hence so do the spaces. This proves the required inclusion, with no appeal to semisimplicity.

\subsection{Correcting the error on one layer}\label{sec:layerstep}
The maps $w_C$ and $w_T$ are auxiliary inverses on $\mathcal Z_b$
and $T_b$, not yet common group-algebra elements. In products involving
$w_T$, the residual is read as its induced map $T_b\to\mathcal Z_b$;
composition with the quotient $V\to T_b$ and inclusion
$\mathcal Z_b\hookrightarrow V$ returns the actual map on $V$.

Return to $\varepsilon=zw_0z-z$. It kills $V'$ and has image in $\mathcal Z_b$. It also has image in $zV$, so the first consequence gives $\im\varepsilon\subseteq z\mathcal Z_b$. Hence
\[zw_C\varepsilon=\varepsilon.
\]
If $zv\in V'$, the second consequence gives $zv=zv'$ for some $v'\in V'$. Since $\varepsilon=(zw_0-I)z$,
\[\varepsilon v=(zw_0-I)zv' =\varepsilon v'=0.
\]
On $T_b$, $zw_Tz=z$ says that $z(w_Tzv-v)$ belongs to $V'$ when representatives in $V$ are used. Applying the previous implication gives $\varepsilon w_Tz=\varepsilon$. This notation is unambiguous: $\varepsilon$ kills $V'$, so its value does not depend on the representative chosen after $w_T$.

Consequently the map
\[\varphi=-w_C\varepsilon w_T:T_b\longrightarrow \mathcal Z_b
\]
satisfies $z\varphi z=-\varepsilon$, after composing with quotient and inclusion. It commutes with position permutations and kills $V'$. If this map is realised by an element $j\in J_{b-1}(N)$, then
\[z(w_0+j)z-z=\varepsilon+zjz=0
\]
on length $b$, while $j$ leaves all shorter tuple relations unchanged.

The remaining step is the realisation theorem. Section~\ref{sec:rooks} proves it through a finite combinatorial argument. The realisation uses the existing labels, so the correction closes the witness induction on one retained pool for all $b\le r$ once $n\ge2r-1$ and $r<q$.

\section{Realising zero-marginal corrections by permutations}\label{sec:rook-realization}
\subsection{Rook coordinates and realisation}\label{sec:rooks}
This section realises a position-equivariant correction with image in $\mathcal Z_b$ and kernel containing $V'$ by one element of $J_{b-1}(N)$. It is independent of the block and works over any commutative ring. The proof translates the map into rook coordinates, proves that selected coordinates force the others, then solves for permutation coefficients by triangular elimination. Ordering the finite board is auxiliary bookkeeping; it adds neither labels nor a built-in order to the uniform input.

The centraliser of the position action is the algebra of linear
operators commuting with every permutation of tuple positions.
Label permutations lie in it because the two actions commute.
The realisation problem here has further conditions: the required
operator must have image in $\mathcal Z_b$ and kill $V'$.
The rook calculation will realise operators of this specified shape.

\subsubsection{From a map to a board of non-attacking rooks}
A matrix entry of a position-equivariant map $\varphi:P_b(N)\to P_b(N)$ is indexed by an output tuple $s$ and input tuple $t$. Permuting their positions together leaves the entry unchanged. Thus it depends only on
\[\rho=\{(s_i,t_i):i=1,\ldots,b\},
\]
a set of $b$ cells with distinct rows and distinct columns on an $N\times N$ board. Call this a $b$-rook. Board rows and columns are \emph{labels}; they are different from the positions of the original tuples.

For instance, output $(3,4)$ and input $(1,2)$ give
$\{(3,1),(4,2)\}$. Reordering both tuples preserves this rook;
reordering only one generally changes it. These coordinates retain
the input--output pairing while forgetting simultaneous position order.

The matrix therefore defines a function $F$ on $b$-rooks. For a $(b-1)$-rook $\sigma$, choose an unused column $y$. A column line through $\sigma$ is the family of rooks $\sigma\cup\{(x,y)\}$ as $x$ ranges over unused rows. Define row lines analogously. The map has image in $\mathcal Z_b$ precisely when its column lines sum to zero. It kills $V'$ precisely when its row lines sum to zero. Write $Z_b$ for this zero-line-sum space of rook functions, distinct from the tuple-vector space $\mathcal Z_b$.

A label permutation $g$ has graph $\{(g(y),y)\}$. Its matrix entry is one exactly when $\rho$ is contained in that graph. Set
\[\chi_g(\rho)=[\rho\subseteq\operatorname{graph}g],\qquad
L_b\Bigl(\sum_ga_gg\Bigr)=\sum_ga_g\chi_g
\]
on $b$-rooks. For a fixed unused row or column, exactly one extension of a rook $\sigma$ lies in the graph of $g$ if $\sigma$ itself does, and none otherwise. Therefore
\begin{equation}\label{eq:lines}
\text{each line sum of }L_b(x)\text{ through }\sigma=L_{b-1}(x)(\sigma).
\end{equation}
If $F\in Z_b$ is realised as $L_b(x)$, this identity forces $L_{b-1}(x)=0$, then every smaller $L_j(x)=0$. Thus $x\in J_{b-1}(N)$ automatically. It remains to prove that every $F\in Z_b$ is realisable.

\subsubsection{A first example: matrices with zero row and column sums}
At $b=1$, $Z_1$ is the space of square matrices with zero row and column sums. If row and column $N$ are designated as last, such a matrix is determined by its upper-left $(N-1)\times(N-1)$ entries and is
\[\sum_{i,j<N}F_{ij}(E_{ij}-E_{iN}-E_{Nj}+E_{NN}).
\]
Each four-entry matrix is a difference of two permutation matrices: choose a permutation taking column $j$ to row $i$ and column $N$ to row $N$, and subtract the permutation with these two outputs swapped. The other entries cancel. This proves realisation at length one over every ring. The general rook argument is a higher-length replacement for this elementary construction.

\subsection{A forcing set of coordinates}\label{sec:forcing}
Work on any two equally large finite sets of integer labels $R,C$,
of size $M$, with their inherited orders. Here $M$ is the board size
and $c$ the rook count; later $M=N$ and $c=b$. Numerical thresholds such as
$y+1$ denote the cut strictly above the integer label $y$. A permutation here is a bijection $g:C\to R$. An increasing chain is a subset of its graph increasing in both coordinates. Let $\LIS(g)$ be the largest chain length. Among the column sets of longest chains choose the lexicographically first one, $I_g$: at the first differing column, the set containing that column is earlier. Define
\[\rho_g=\{(g(y),y):y\notin I_g\}.
\]
If $\LIS(g)=M-c$, this is a $c$-rook. Let $\Psi_c$ consist of these rooks.

A forcing set means that restriction to it is injective on the
zero-line-sum functions. Two such functions agreeing there have a
difference vanishing there, so the theorem below establishes uniqueness.
Realisation by permutation coefficients is the separate existence step.

\begin{theorem}[Forcing]\label{thm:force}
A function in $Z_c$ that vanishes on $\Psi_c$ vanishes on every $c$-rook.
\end{theorem}

Permutation strings use one-line notation: $312$ means $g(1)=3$,
$g(2)=1$, $g(3)=2$.
\begin{example}[The $3\times3$ board]
The permutations with longest increasing subsequence of length two are $312,231,213,132$. Their canonical complement rooks are respectively
\[(3,1),\ (1,3),\ (1,2),\ (2,3).
\]
If a zero-line-sum matrix is zero on these four cells, row one first forces $(1,1)$ to be zero. Column one then forces $(2,1)$; row two forces $(2,2)$; column two forces $(3,2)$; and row three forces $(3,3)$. This is the small model of the forcing argument.
\end{example}

\subsubsection{Strengthened induction statement}
To remove the bottom row and the first column, we must know which canonical longest chains survive contraction. Bounds on longest chains in upper parts of the board keep track of this. These are bookkeeping conditions for the induction, not new mathematical assumptions on $F$.

Let $L=M-c$. A condition is $(s,\theta,d)$, where $s$ is ``row'' or ``column'', $\theta$ is a label threshold, and $d\ge1$. This local $d$ is a chain deficit, not polynomial degree. A bijection satisfies it if the longest increasing chain using only cells with that coordinate at least $\theta$ has length at most $L-d$. Bounds are interpreted as integer inequalities; a negative upper bound is impossible for a nonnegative chain length.

For a collection $\mathcal E$ of conditions let $\Psi(\mathcal E)$ be the canonical complement rooks of permutations with $\LIS=L$ satisfying every condition. Let $Q(\mathcal E)$ be the set of $c$-rooks for which, for at least one condition, at most $d-1$ \emph{unused} labels of the specified side lie below $\theta$.

The two descriptions are paired deliberately. A bound on chains above a threshold will, through a line-sum identity, force a value at a rook with too few unused labels below it.

\begin{lemma}[Strengthened forcing]\label{lem:GP}
If $F\in Z_c$ vanishes on $\Psi(\mathcal E)\cup Q(\mathcal E)$, then $F=0$.
\end{lemma}
Theorem~\ref{thm:force} is its case of no conditions.

\subsubsection{Three facts about adding a corner cell}
Let $r_0=\min R$ and $y_0=\min C$. The following statements describe the only possible new increasing chains.
\begin{enumerate}
\item Remove row $r_0$ and column $y$, and let $\widetilde g$ be a bijection on the smaller board with longest chain length $L$. Add the cell $(r_0,y)$. Every new chain using it continues with a chain in columns strictly above $y$. If that region had longest chain at most $L-1$, the new permutation has longest chain $L$. Its lexicographically first longest chain is still the old one: the old chain must use some column below $y$, whereas every new chain through the corner uses columns at least $y$. Therefore $\rho_g=\rho_{\widetilde g}\cup\{(r_0,y)\}$. For a column threshold $\theta\le y$, the new regional maximum is
\[\max\{\LIS(\widetilde g|_{\mathrm{col}\ge\theta}),\ 1+\LIS(\widetilde g|_{\mathrm{col}>y})\}.
\]
\item Remove row $x$ and column $y_0$, and add $(x,y_0)$. If the old region of rows strictly above $x$ has longest chain at most $L-2$, every chain using the new corner has length at most $L-1$. Thus all longest chains remain old, the canonical one remains unchanged, and $\rho_g=\rho_{\widetilde g}\cup\{(x,y_0)\}$. The analogous maximum formula holds for row thresholds at most $x$. The stronger deficit two eliminates a tie involving the smallest column.
\item Remove both minimal labels and add $(r_0,y_0)$. This corner can be placed before every increasing chain of the smaller board. The longest length increases by one, every new longest chain contains the corner, and $I_g=\{y_0\}\cup I_{\widetilde g}$. Thus $\rho_g=\rho_{\widetilde g}$.
\end{enumerate}
These facts follow just by splitting chains according to whether they use the added cell. They do not rely on a representation-theoretic result.

\subsubsection{Base cases and degenerate conditions}
The outer induction decreases the board size $M$; a decreasing corner
sweep supplies the inner induction. Contracting a fixed cell changes
$(M,c)$ to $(M-1,c-1)$ and keeps $L=M-c$ fixed. Removing a row and
column without fixing a cell leaves $c$ fixed and decreases $L$.
The threshold conditions record these distinct changes.

Prove Lemma~\ref{lem:GP} by induction on $M$. If $c>M$ there are no rooks. If $c=M>0$, each line through an $(M-1)$-rook contains exactly one $M$-rook, so the zero-line-sum equation forces its value to zero.

A condition is degenerate if the entire board has at most $d-1$ labels below its threshold. Then every rook belongs to $Q$, and the conclusion is immediate. Assume no condition is degenerate. Every column threshold is therefore above $y_0$ and every row threshold above $r_0$.

If $c=0$, consider the increasing bijection. It has longest length $M$. For a column condition, at least $d$ columns lie below $\theta$, so the part above $\theta$ has length at most $M-d$; the same holds for rows. Thus its empty complement rook belongs to $\Psi(\mathcal E)$, and the sole value of $F$ is zero. We may now assume $1\le c<M$.

\subsubsection{First sweep: eliminate every rook meeting the bottom row}
For $y\in C$ in decreasing order, contract the cell $(r_0,y)$ and define
\[G(\sigma)=F(\sigma\cup\{(r_0,y)\})
\]
on $(c-1)$-rooks of the board with that row and column removed. Its line sums are line sums of $F$, so $G\in Z_{c-1}$. The smaller board has size $M-1$ and complement length $(M-1)-(c-1)=L$.

Retain the old conditions and add
\begin{align*}
&(\mathrm{column},y+1,1),\\
&(\mathrm{column},y+1,d+1)
\quad\text{for each old }(\mathrm{column},\theta,d)\text{ with }\theta\le y.
\end{align*}
The first new condition bounds the right region by $L-1$, giving corner fact 1. The second bounds it by $L-(d+1)$ whenever needed. An old column condition with $\theta\le y$ is preserved upon adding the corner because
\[\max\{L-d,1+L-(d+1)\}=L-d.
\]
Old column conditions above $y$ and old row conditions are unaffected by the corner. Thus a canonical rook satisfying the new conditions extends to one in $\Psi(\mathcal E)$, and $G$ vanishes on its new $\Psi$ set.

We must also verify every case of its new $Q$ set. An old condition passes directly to the enlarged rook: the deleted row and column are occupied by the added cell, so the count of unused labels below an old threshold is unchanged.

For a new condition $(\mathrm{column},y+1,e)$, let
\[U=\{y'\in C:y'<y,\ y'\text{ unused by }\sigma\}.
\]
On the contracted board this is exactly the unused set below $y+1$, so $|U|\le e-1$. If $e=1$, there are no such columns. If the condition came from an old $(\mathrm{column},\theta,d)$, then $e=d+1$ and $|U|\le d$. For each $y'\in U$, the enlarged rook $\sigma\cup\{(r_0,y')\}$ has at most $d-1$ unused columns below $\theta$. Indeed, if $y'<\theta$, its occupation removes one member from a subset of $U$; if $y'\ge\theta$, $y'$ itself lies outside that subset, again leaving at most $|U|-1$. Thus every such rook lies in the old $Q$ set.

Now use the row line through $\sigma$ in row $r_0$. Its terms at columns greater than $y$ are zero by the decreasing inner induction. Its terms below $y$ are either absent ($e=1$) or zero by the old $Q$ condition. The only remaining term is $G(\sigma)$, so it is zero. This proves vanishing on the new $Q$ set in all cases. The outer induction on $M$ now gives $G=0$.

Completing the decreasing sweep shows that $F$ vanishes on every rook meeting row $r_0$.

\subsubsection{Second sweep: eliminate every rook meeting the first column}
Adjoin the condition $(\mathrm{row},r_0+1,1)$ to obtain $\mathcal E^+$. Its new $Q$ rooks are exactly those using row $r_0$: there must be no unused row below $r_0+1$, and that row is the only candidate. They have just been proved zero. Its $\Psi$ set is a subset of the old one. Thus the vanishing assumptions still hold for $\mathcal E^+$.

Sweep $x\in R\setminus\{r_0\}$ in decreasing order. Contract $(x,y_0)$ and keep every old condition in $\mathcal E^+$. For each row condition $(\mathrm{row},\theta,d)$ with $\theta\le x$, add $(\mathrm{row},x+1,d+1)$. In particular the recently added condition produces $(\mathrm{row},x+1,2)$, because $x>r_0$. Therefore the region above $x$ has maximum at most $L-2$, exactly the hypothesis of corner fact 2.

The verification for $\Psi$ is now explicit: row thresholds at most $x$ use the same maximum calculation as in the first sweep; row thresholds above $x$ are unchanged; every column threshold exceeds $y_0$, so excludes the new corner. For $Q$, old conditions pass unchanged. A new condition from deficit $d$ says that there are at most $d$ unused rows below $x$. When any such row $x'$ is occupied instead of $x$, the number unused below its source threshold is at most $d-1$, by the same two cases $x'<\theta$ and $x'\ge\theta$. These terms lie in $Q(\mathcal E^+)$. The column-line equation at $y_0$ then forces the remaining term to zero, since larger rows were handled earlier. The induction on $M$ proves the contracted function zero.

The case $x=r_0$ was already covered by the first sweep. Thus $F$ now vanishes on every rook meeting either row $r_0$ or column $y_0$.

\subsubsection{The remaining smaller board}
Restrict $F$ to $(R\setminus\{r_0\})\times(C\setminus\{y_0\})$ without changing $c$. A line in this board is an old line minus one term using a removed label; that term is zero, so the restriction has zero line sums.

For every condition in $\mathcal E^+$ with deficit $d\ge2$, keep its side and threshold and replace its deficit by $d-1$. Discard the conditions with deficit one. The smaller complement length is $L-1$.

A permutation on the smaller board satisfying these conditions and having longest length $L-1$ extends by the minimal corner. Corner fact 3 gives longest length $L$ and the same complement rook. Every old threshold excludes that corner. For $d\ge2$ the regional bound is
\[(L-1)-(d-1)=L-d;
\]
for $d=1$ it follows from the whole smaller board having longest length $L-1$. So the restricted function vanishes on the new $\Psi$ set.

If a smaller rook belongs to the new $Q$ set, its unused count below a threshold is at most $d-2$. In the original board the removed minimum label is also unused by this rook and lies below that threshold, increasing the count by exactly one. The resulting count is at most $d-1$, so the rook lies in the old $Q(\mathcal E^+)$. Thus the new $Q$ vanishes too. The induction on $M$ finishes the proof of Lemma~\ref{lem:GP}, and hence of the forcing theorem.

\subsection{Triangular realisation and the common correction}\label{sec:realization}
This is back-substitution in a triangular system with diagonal one:
each new coefficient is obtained by subtraction, with no division.
The integral tensor-space centraliser context is discussed in
\cite{BDM}; the particular realisation required here follows from
the forcing argument and the recursion below over any commutative ring.

The second ingredient is simpler. If $\rho_g$ lies on the graph of $g'\ne g$, then $g'$ agrees with $g$ outside $I_g$ and permutes the remaining values $g(I_g)$. On $I_g$ the values of $g$ are increasing. At the first column where $g'$ differs, it must use a larger remaining value. Thus $g'$ is lexicographically larger than $g$.

Fix $F\in Z_b(N)$ and let $\mathcal T=\{g:\LIS(g)\ge N-b\}$. Its canonical complement rooks have sizes at most $b$. Define coefficients $a_g$ in decreasing lexicographic order by
\begin{equation}\label{eq:triangular}
a_g=\begin{cases}F(\rho_g)&|\rho_g|=b,\\0&|\rho_g|<b\end{cases}
-\sum_{\substack{g'\in\mathcal T,\ g'\ne g\\\rho_g\subseteq\operatorname{graph}g'}}a_{g'}.
\end{equation}
Every coefficient on the right was already defined by the preceding observation. Put $x=\sum_{g\in\mathcal T}a_gg$, and let $H_k=L_k(x)$ for each $k\le b$. The construction makes $H_k$ zero on $\Psi_k$ for $k<b$ and makes $H_b=F$ on $\Psi_b$.

At level zero the only coordinate is the empty rook, and it is prescribed zero when $b>0$. Inductively suppose $H_{k-1}=0$. Equation~\eqref{eq:lines} shows that $H_k$ has zero line sums. For $k<b$, it vanishes on $\Psi_k$, so forcing gives $H_k=0$. At $k=b$, $H_b-F$ has zero line sums and vanishes on $\Psi_b$, so forcing gives $H_b=F$. The case $b=0$ is immediate by a scalar multiple of any permutation.

We have proved that every zero-line-sum rook function is realised by a group-algebra element, and the realising element annihilates all smaller tuple modules. Translating back to matrices proves:
\begin{theorem}[Realisation of the layer correction]\label{thm:realization}
For $b\ge1$, every position-equivariant map $\varphi:P_b(N)\to P_b(N)$ with image in $\mathcal Z_b(N)$ and killing $V'$ is the action of an element $j\in J_{b-1}(N)$. No additional labels are required.
\end{theorem}
If $b>N$, the tuple space is zero and its only endomorphism is
realised by $j=0$. The witness construction uses $b\le r<q\le N$.

\begin{example}[Actual permutation coefficients]
On a $3\times3$ board take
\[F=\begin{pmatrix}0&1&-1\\1&-1&0\\-1&0&1\end{pmatrix}.
\]
The recursion gives coefficients $-1,-1,2,1,-1$ for $312,231,213,132,123$, respectively. Therefore
\[F=-P_{312}-P_{231}+2P_{213}+P_{132}-P_{123}.
\]
Here $P_g$ has its one in row $g(y)$ of column $y$. The coefficients sum to zero, so the action on the empty tuple is zero, as required for $J_0$. Reducing this integer identity modulo two or any odd prime gives the corresponding realisation in that characteristic.
\end{example}

\section{Exact deletion in every positive characteristic}\label{sec:odd}
The characteristic-two construction used $z^2=0$ and
$z^{\mathsf T}=z$. We retain its signed combinatorial sections and
permutation realisation, but replace these operator properties by
$p$-level chains, a top-power projector and conjugacy with the
transpose. The goal remains exact deletion on $\mathcal C_r(n+q)$
using the same retained pool of size at least $2r-1$ for $r\ge1$.

\subsection{The block operator in prime characteristic}
Let now $\operatorname{char}K=p$ for any prime. Take a regular block $A=\Fp^a$ of size $q=p^a>r$, choose $c_1,\ldots,c_a$ independent over $\Fp$ in a suitable extension field, and put
\[z=\sum_i c_i(\sigma_i-1).
\]
The generators commute and have order $p$, so the binomial formula in characteristic $p$ gives
\[z^p=\sum_i c_i^p(\sigma_i^p-1)=0.
\]
As in Section~\ref{sec:block2}, $K$ denotes the working extension
field during the construction; degree statements descend to the
original field by Lemma~\ref{lem:field}.
For $p>2$ we cannot replace this by $z^2=0$. For example, for a three-cycle $\sigma$ in characteristic three,
\[z=\sigma-I,\qquad z^2=\sigma^2+\sigma+I\ne0
\]
on its three-point permutation space.

\subsubsection{The deletion quotient}
Nilpotence means that a positive power of the operator is zero.
Here $z^p=0$ permits chain lengths from one to $p$; it does not
exclude intermediate lengths. The quotient must therefore be read
before any fixed/full-chain decomposition is established.

Since $\im z$ need not be contained in $\ker z$, the characteristic-two expression $\ker z/\im z$ is no longer defined. Use
\begin{equation}\label{eq:Dp}
\D(M)=\ker z/(\ker z\cap\im z).
\end{equation}
A nilpotent chain has basis $v_0,\ldots,v_{j-1}$ with $zv_0=0$ and $zv_i=v_{i-1}$ for $i\ge1$. Its kernel is $Kv_0$. If $j=1$, its image is zero and the chain survives in \eqref{eq:Dp}. If $j\ge2$, $v_0$ is in the image and the chain disappears. Thus deletion keeps the one-dimensional chains and discards every longer chain.

One might instead try $\ker z/\im z^{p-1}$. It is a defined quotient, but it is the wrong one: a chain of length two in characteristic three would survive because $z^2=0$ there, even though it should be discarded by \eqref{eq:Dp}. Keeping the quotient precise is important.

\subsection{The commutator argument in characteristic \texorpdfstring{$p$}{p}}\label{sec:commutatorp}
On a moved orbit $A/H$, choose a nonzero linear functional $\ell:A\to\Fp$ vanishing on the proper stabiliser $H$, and use the coordinate $\pi(vx_0)=\ell(v)$ from Section~\ref{sec:block2}. Let $D$ multiply a basis vector by its coordinate $\pi$, put $e_i=\ell(\varepsilon_i)$ and $\alpha=\sum_i c_ie_i\ne0$. With $S_i$ the coordinate-translation operators, defined orbit by orbit, we obtain
\[Dz-zD=W=\sum_i c_ie_iS_i,\quad Wz=zW,\quad W^p=\alpha^pI.
\]
The commutator $[D,z]=Dz-zD$ measures the difference between the
two orders of application. The displayed power identity gives the
explicit inverse $W^{-1}=\alpha^{-p}W^{p-1}$; it is $W$, rather
than the nilpotent $z$, that is inverted.
Consequently the auxiliary operator $B=W^{-1}D$ satisfies
\begin{equation}\label{eq:commutator}[B,z]=Bz-zB=I.\end{equation}
This proof works also on a moved permutation basis tensored with an arbitrary representation: $D$ reads only the basis coordinate, while the $S_i$ act diagonally and still have order $p$.

If $zm=0$, commuting $z$ past powers of $B$ yields
\[zB^jm=-jB^{j-1}m,
\qquad z^jB^jm=(-1)^jj!\,m.
\]
The first identity follows by induction from $zB=Bz-I$; applying it repeatedly gives the second. At $j=p-1$ the scalar $(p-1)!$ is nonzero in characteristic $p$, because every factor $1,\ldots,p-1$ is nonzero. Therefore
\begin{equation}\label{eq:freep}\ker z=\im z^{p-1}
\end{equation}
on the moved summand. The reverse inclusion uses $z^p=0$.

This is the full replacement for the square-zero freeness calculation. It does not divide by $p$, by $|A|$, or by a symmetric-group order. It divides only by a nonzero orbit coefficient and by $(p-1)!$.

Here freeness concerns the action of $z$: a full chain is one copy of
$K[T]/(T^p)$ with $T$ acting as $z$, and full chains form a free
module over that ring. It is not freeness for the full symmetric group.

On a bounded-support permutation basis the retained summand has $z=0$ and the moved summand satisfies \eqref{eq:freep}. Therefore the identifications with the retained basis and the tensor identifications from Section~\ref{sec:block-algebra} still hold, with quotient \eqref{eq:Dp}. Stable matrices are transported by exactly the same retained-component argument.

\subsection{The top-power witness}\label{sec:topwitness}
Set $u=z^{p-1}$. The common certificate becomes
\begin{equation}\label{eq:topwitness}uwu-u\in J_r(N),\qquad\text{or }uwu=u\text{ on every }P_b(N),\ b\le r.
\end{equation}
This says that $w$ takes the bottom of a full length-$p$ chain back to a preimage at its top. It is enough to build a projector for the actual quotient, but more steps are needed than in characteristic two.

First note the property of the tuple modules supplied by the moved/fixed decomposition:
\begin{equation}\label{eq:purity}\ker z\cap\im z\subseteq\im z^{p-1}.
\end{equation}
The image has no retained component, and a moved kernel vector is in the top-power image by \eqref{eq:freep}. We will use \eqref{eq:purity} on the original tuple modules, then transfer polynomial identities through their common annihilator. We do not assume it for every arbitrary nilpotent operator.

Condition~\eqref{eq:purity} is the purity condition needed here: a
kernel vector already in $\im z$ must have a preimage under the top
power. A nontrivial chain shorter than $p$ violates it at its bottom.
Fixed lines and full length-$p$ chains satisfy it.

\begin{example}[Why the distinction matters]
In characteristic three, a lone length-two chain has $u=z^2=0$. The relation $uwu=u$ is then vacuous for any $w$, but $\ker z\cap\im z$ is nonzero and $\im u=0$. Such a chain does not satisfy \eqref{eq:purity}. The argument must establish its projector identities on the actual tuple modules first; the top-power relation alone on an arbitrary vector space is insufficient.
\end{example}

\subsubsection{A projector which removes one chain level at a time}
Write $m=p-1$, an exponent rather than a support parameter. Suppose $z^{m+1}=0$, $z^mwz^m=z^m$ and \eqref{eq:purity} holds. Define
\begin{equation}\label{eq:projectorp}
S_0=I,\qquad S_{k+1}=(I-z^kwz^{m-k})S_k\quad(0\le k<m),
\qquad e=(I-z^mw)S_m.
\end{equation}
Operator products act from right to left. We verify every required property.

\paragraph{Eliminating high levels.}
Here the local integer $d$ is a chain height, not a proof-degree budget.
If $z^{d+1}R=0$, then for $d\ge1$ the vector $y=z^dRv$ is in $\ker z\cap\im z$, hence in $\im z^m$. The witness gives $z^mwy=y$. Thus
\[z^d(I-z^{m-d}wz^d)Rv=y-z^mwy=0.
\]
Starting with $z^{m+1}S_0=0$, the successive factors of \eqref{eq:projectorp} give
\[z^{m+1-k}S_k=0,
\]
and finally $zS_m=0$. It follows that $ze=0$.

\paragraph{Action on a kernel vector.}
If $zv=0$, every factor in $S_m$ fixes $v$, because its rightmost power of $z$ has exponent at least one. Therefore
\begin{equation}\label{eq:ekernel}ev=v-z^mwv\quad(zv=0).
\end{equation}

\paragraph{Killing the image.}
The first factor $I-wz^m$ fixes $\im z$, since $z^{m+1}=0$. All later factors preserve $\im z$ because their correction terms have a left factor $z^k$ with $k\ge1$. Hence $S_mzv\in\ker z\cap\im z\subseteq\im z^m$. The final factor $I-z^mw$ kills this vector by the witness relation. Thus $ez=0$.

\paragraph{Idempotence.}
The witness gives
\[z^mwe=(z^mw-z^mwz^mw)S_m=0.
\]
Since $ev\in\ker z$, formula \eqref{eq:ekernel} now gives $e^2v=ev$. We have proved $ze=ez=0$ and $e^2=e$.

\subsubsection{Passing from tuples to every relevant space}
Objects of $\mathcal C_r(N)$ are modules over $K[S_N]/J_r(N)$.
An identity holding on every tuple test space of length at most $r$ is an identity
of this quotient algebra. Equivariant maps intertwine its operators;
this does not make the projector central. Its image is used as a
vector-space model of deletion, while the quotient retains the residual
label action.

Construct $e$ as the same expression in the group algebra. Once these identities hold on all tuple modules up to length $r$, the elements $ze$, $ez$ and $e^2-e$ belong to $J_r(N)$. They therefore act as zero on every object of $\mathcal C_r(N)$, whether or not we have a separate chain decomposition for that object. Formula \eqref{eq:ekernel} follows there directly from the displayed product formula.

Thus $[v]\mapsto ev$ identifies \eqref{eq:Dp} with $eM$: $e$ kills $\im z$, its image is in $\ker z$, and $v-ev=z^{p-1}wv\in\im z$ for $v\in\ker z$. The exactness proof using a common projector is now identical to Section~\ref{sec:block-algebra}. All maps commute with the common group-algebra expression.

\begin{example}[A full chain in characteristic three]
Let $zv_2=v_1$, $zv_1=v_0$, $zv_0=0$, and let $wv_0=v_2$, with $w$ zero on the other chain vectors. Then $z^2wz^2=z^2$. The projector is
\[e=(I-z^2w)(I-zwz)(I-wz^2).
\]
The rightmost factor removes $v_2$, the middle factor removes $v_1$, and the leftmost removes $v_0$. If a separate fixed vector $t$ is added and $wt=0$, every factor fixes $t$. Thus the projector removes exactly the full chain and retains the fixed direction. For $p=2$ the same formula has only the two factors $(I-zw)(I-wz)$ from Section~\ref{sec:block-algebra}.
\end{example}

\subsection{The linear construction in odd characteristic}\label{sec:oddlinear}
We now adapt the construction of Section~\ref{sec:linear-construction}.
The partition proof in Section~\ref{sec:partition}, cone and signed
lifts in Section~\ref{sec:cone}, and rook realisation in
Section~\ref{sec:rook-realization} use $1$ and $-1$ without a
characteristic-two cancellation. These combinatorial proofs remain
valid over any field and supply the same label count $2b-1$.
The splitting and duality arguments require the replacements below.

At the current tuple length $b$, recall $V=P_b(N)$,
$V'=\sum_i\im\partial_i^{\mathsf T}$,
$\mathcal Z_b=\bigcap_i\ker\partial_i$ and $T_b=V/V'$.
The layer $G_t$ records exactly $t$ block positions in the filtration
$\mathcal Z^{\le t}$. The tuple bases and marginal maps have the
same definitions; their block action has a different chain structure.

\subsubsection{The block layers have \texorpdfstring{$p$}{p} levels instead of two}
Let $E$ be the space on the current block patterns with $t$ positions, where $1\le t\le b<q$, and give each pattern a grade $d\in\Fp$ using the orbit coordinate of its image set. Here $d$ is a field-valued grade, not polynomial degree, and $D$ is the diagonal operator reading that grade. Let $E_d$ be the corresponding subspace. We claim
\[E=E_0\oplus zE_0\oplus\cdots\oplus z^{p-1}E_0.
\]
For fixed $d$, suppose $\sum_{j=0}^{p-1}z^jv_j=0$ with $v_j\in E_d$. The commutator identity and $Wz=zW$ imply
\[(D-dI)z^jv_j=jWz^{j-1}v_j.
\]
Apply $D-dI$ to the relation and cancel the invertible $W$. Repeat this operation $p-1$ times. It gives $(p-1)!v_{p-1}=0$, hence $v_{p-1}=0$. Descending induction gives every $v_j=0$. Thus $p\dim E_d\le\dim E$ for every $d\in\Fp$. Since the $p$ spaces $E_d$ partition the basis, their dimensions sum to $\dim E$, forcing equality in each inequality. The direct sum for $E_0$ is therefore all of $E$.

The grade depends only on the image set, so $E_0$ is position-stable and graded by the position domain. Tensoring with the retained zero-marginal factors gives
\[G_t=\bigoplus_{j=0}^{p-1}z^jQ_t
\]
with $Q_t$ position-stable and $z^{p-1}$ injective on $Q_t$ by the preceding directness argument. A position-equivariant section $\sigma$ becomes a section commuting with $z$ by
\[\tau\left(\sum_{j=0}^{p-1}z^jq_j\right)=\sum_{j=0}^{p-1}z^j\sigma(q_j).
\]
The last term causes no problem because $z^p=0$. Splitting the filtration gives
\begin{equation}\label{eq:socsplitp}
\mathcal Z_b(n+q)=\mathcal Z_b(n)\oplus\bigoplus_{j=0}^{p-1}z^jC
\end{equation}
for a position-stable $C$, again under $n\ge2b-1$ and $b<q$.

\subsubsection{Top and zero-marginal spaces}
The perfect pairing $T_b\times\mathcal Z_b\to K$ exchanges an
operator on the quotient with its transpose on the zero-marginal
space. Duality reverses composition order. Thus the odd-characteristic
argument must use the actual transpose, not assume the
characteristic-two equality $z^{\mathsf T}=z$.

With $u=z^{p-1}$, send $uc$ back to $c$ and kill the other levels in \eqref{eq:socsplitp}. This defines a position-equivariant $w_C$ with $uw_Cu=u$. Also
\[uV\cap \mathcal Z_b\subseteq u\mathcal Z_b.
\]
Indeed a vector in this intersection is killed by $z$, so its decomposition in \eqref{eq:socsplitp} is a retained vector plus an element of $uC$. Its membership in $uV$ excludes the retained part, just as before.

For odd $p$, $z$ need not be symmetric. Its transpose is
\[z^{\mathsf T}=\sum_i c_i(\sigma_i^{-1}-1).
\]
This is the ordinary bilinear matrix transpose; it does not conjugate
the coefficients $c_i$.
Let $\iota$ fix the retained labels and send a block label $v\in A$ to $-v$. Then $\iota\sigma_i\iota^{-1}=\sigma_i^{-1}$, so $z^{\mathsf T}=\iota z\iota^{-1}$. This is an actual relabelling, and preserves the tuple-layer constructions. The just-proved statements apply equally to $u^{\mathsf T}$. Orthogonal complements give
\[uV\cap V'\subseteq uV',
\]
and duality between $T_b$ and $\mathcal Z_b$ gives a position-equivariant $w_T$ with $uw_Tu=u$. This replaces the characteristic-two shortcut $z^{\mathsf T}=z$.
The relabelling $\iota$ is the block inversion, distinct from the
inclusion of a zero-marginal space into its tuple ambient.

\subsubsection{The same correction, with the correct power}
Suppose $w_0$ works on shorter tuples for the top-power relation. Its error is
\[\varepsilon=uw_0u-u.
\]
It maps into $\mathcal Z_b$ and kills $V'$. The four properties just proved give, step by step as in Section~\ref{sec:layerstep},
\[uw_C\varepsilon=\varepsilon,\qquad\varepsilon w_Tu=\varepsilon.
\]
The correction $\varphi=-w_C\varepsilon w_T$ therefore satisfies $u\varphi u=-\varepsilon$. The characteristic-independent rook theorem realises it by $j\in J_{b-1}(N)$. Adding $j$ repairs the witness at length $b$ without changing any shorter length.

Inducting on $b$ proves \eqref{eq:topwitness} on the same label pool whenever $n\ge2r-1$. The longer projector gives exactness, the moved-orbit calculation gives natural basis and tensor deletion, and the original PC closure transport gives the same period statement.

\begin{proof}[Completion of the proof of Theorem~\ref{thm:deletion}]
Begin at tuple length zero with $w_0=0$, since $z$ acts as zero on
constants. At length $b$, the preceding correction and
Theorem~\ref{thm:realization} replace $w_{b-1}$ by $w_b$ satisfying
$z^{p-1}w_bz^{p-1}=z^{p-1}$ on every length at most $b$. The same
retained labels are used at every step. The largest requirement is
$\max_{1\le b\le r}(2b-1)=2r-1$; length zero needs no label.
The projector identities then imply exactness on $\mathcal C_r(N)$.
The orbit commutator calculation supplies the natural retained-basis
and tensor identifications, including the actual stable coefficient
maps. These arguments prove every assertion of Theorem~\ref{thm:deletion}.
\end{proof}
The block construction is performed over its working extension field.
Lemma~\ref{lem:field} descends the transported polynomial consequence
spaces to the original field without changing the degree budget.

\section{An independent uniform NS--PC comparison}\label{sec:comparison}
The linear PC lower bounds were proved by transporting the PC closure
itself. We now establish a different result: every bounded-degree PC
consequence has an NS identity of degree bounded independently of
universe size. The comparison applies to arbitrary polynomials, not
only invariant targets or refutations. Its qualitative consequence is
that NS and PC assign every fixed uniform request the same type.

The NS-span plateau argument already occurs in Kraj\'i\v cek's
uniform comparison~\cite[Theorem~5.5]{Krajicek}. We give an independent
proof of its required chain bound and carry the comparison through for
the actual consequence spaces. A uniform bound on strict
invariant-subspace chains forces the intersections of two successive
NS certificate spaces with the fixed degree-$d$ polynomial ambient
to agree. Their common
space is then closed under every permitted PC operation. The plateau
may depend on $n$, but the bound on its NS certificate degree does not.

At a degree cutoff, $I_e(n)$ is the linear space of polynomials with
degree-at-most-$e$ NS certificates. It is not an ideal closed under
arbitrary multiplication.
A \emph{strict subspace chain} is a sequence in which each inclusion
is proper. Its bound counts proper increases; consecutive equal
spaces form a plateau. The proof uses deletion to preserve those
increases, then uses a plateau to absorb variable multiplication.
No decomposition into simple representations is required.

\begin{lemma}[Uniform chain bound]\label{lem:chains}
For any bounded basis system $\mathcal U$, strict $S_n$-invariant
subspace chains in $K[\mathcal U_n]$ have a length bound independent of $n$.
\end{lemma}
\begin{proof}
Let $r$ bound the basis support. Over an algebraic closure, choose a
fixed $q=p^a>2r$. At $N=n+q$ with $n\ge4r+1$, deletion is exact at the
doubled budget $2r$. Lemma~\ref{lem:detect} shows that each nonzero
subquotient $B/A$ of $K[\mathcal U_N]$ survives. Thus
$A\subsetneq B$ implies
$\mathcal D(A)\subsetneq\mathcal D(B)$: the whole strict chain becomes
a strict chain of the same length in $K[\mathcal U_n]$.

Repeatedly subtract $q$ while the rank exceeds $M=4r+q$. Every retained
rank in such a step exceeds $4r$, so all steps are eligible. The final
chain has at most $\dim K[\mathcal U_M]$ strict inclusions, because a
smaller-rank basis injects into the rank-$M$ basis. The same bound covers
the initial ranks at most $M$. Each step may choose its own generic
coefficients: the detector applies to arbitrary invariant subspaces,
without a naturality assumption on the chain across ranks. Finally,
scalar extension preserves strict inclusions, giving the result over
the original coefficient field.
\end{proof}

Only a finite exactness threshold at each fixed support is needed for
this argument. In characteristic two, the earlier quadratic construction also suffices. The
bound may be large; its role is uniformity in $n$, not a sharp comparison
of the two minimum degrees.

\begin{theorem}[Uniform comparison for all consequences]\label{thm:comparison}
For every uniform axiom template over a field of prime characteristic
$p$, there is a nondecreasing $F:\NN\to\NN$, independent of $n$ and of
the target, with $F(d)\ge d$ and
\begin{equation}\label{eq:whole-comparison}
                              C_d(n)\subseteq I_{F(d)}(n)
                       \qquad(n,d\ge0).
\end{equation}
Consequently, every polynomial with an ordinary-PC derivation of degree
at most $d$ from $\Gamma_n$ has a total-product NS identity of degree at
most $F(d)$ from the same axioms. For the aggregate profiles of any
stable uniform target family, and for the profiles of any one fixed
target orbit,
\begin{equation}\label{eq:target-comparison}
                      h_{\PC}(n)\le h_{\NS}(n)\le F(h_{\PC}(n)),
                      \qquad F(\infty)=\infty.
\end{equation}
Their NS and PC types in Theorem~\ref{thm:classification} therefore agree.
\end{theorem}
\begin{proof}
Fix $d$. Monomials of degree at most $d$ form a permutation basis of
support at most $kd$. Let $L_d$ bound strict invariant-subspace chains in
$P_{n,\le d}$, uniformly in $n$, by Lemma~\ref{lem:chains}.
Consider the invariant spaces
\[
              W_j=I_{d+j}(n)\cap P_{n,\le d},
                    \qquad 0\le j\le L_d+1.
\]
They form an ascending chain. Some $W_j=W_{j+1}$ with $j\le L_d$,
otherwise there would be more than $L_d$ strict inclusions.

This plateau contains every axiom usable within degree $d$ and is closed
under linear combinations. If $f\in W_j$ and $\deg f\le d-1$, then
multiplication of an NS identity for $f$ by a variable gives
$x_vf\in I_{d+j+1}(n)$. Its degree is at most $d$, hence
$x_vf\in W_{j+1}=W_j$. For $d=0$ there is no permitted multiplication
of a nonzero line; the zero line already lies in $W_j$.
Induction on \emph{any} degree-$d$ PC derivation now places every one of
its lines in $W_j$. In particular,
\[
                          C_d(n)\subseteq I_{d+L_d}(n).
\]
This induction retains cancellations: a lower-degree linear combination
can be multiplied because it lies in the plateau, even if its separate
summands have degree $d$.

Choose
\[
                          F(d)=d+\sum_{u=0}^d L_u.
\]
It is nondecreasing, satisfies $F(d)\ge d$, and gives
\eqref{eq:whole-comparison}. No symmetry of a target was used.
Conversely, expansion of NS identities into monomial--axiom products
gives a same-degree PC derivation, so the first inequality in
\eqref{eq:target-comparison} holds. If the aggregate PC degree is finite,
every requested target belongs to $C_d(n)$ at that budget; the inclusion
puts every target in $I_{F(d)}(n)$. Empty requests have degree zero.
Infinite PC degree forces infinite NS degree by the first inequality.
These observations prove the complete extended-order comparison.

For type agreement write $g=h_{\PC}$ and $h=h_{\NS}$.
The inequalities $g\le h\le F(g)$ transfer eventual boundedness in both
directions. If $g$ takes one finite value $d$ infinitely often, then on
those sizes $h$ lies in the finite interval $[d,F(d)]$; some finite value
of $h$ recurs. A recurring finite value of $h$ likewise forces a recurring
finite value of $g$. Thus the recurring sets are empty together.
If recurring values of $g$ are unbounded, the same finite-interval
argument yields unbounded recurring values of $h$. Conversely, choose a
recurring value of $h$ exceeding $F(B)$. On its infinite level set,
$g>B$ and $g\le h$, so some finite value above $B$ recurs for $g$.
Thus infinitude of the recurring sets agrees as well. Boundedness,
emptiness and infinitude determine exactly the four alternatives.
\end{proof}

\begin{corollary}[Common finite peaks]\label{cor:common-peaks}
If a fixed uniform request has common type T4, there is a nondecreasing
$F$ such that, beyond every bound $B$, some size $n\ge B$ has
\[
 h_{\PC}(n)=d\ge B,\qquad d\in\rec(h_{\PC}),\qquad
 d\le h_{\NS}(n)\le F(d)<\infty,
\]
with $n\le(p+2)(\kappa d+\sigma)$. Both degrees satisfy the rational
linear lower bound in \eqref{eq:peaks} at these same sizes.
\end{corollary}
\begin{proof}
Take the finite PC peaks from Theorem~\ref{thm:classification}.
The pointwise lower comparison gives the NS lower bound, and the uniform
upper comparison gives its finiteness at those sizes.
\end{proof}

The comparison does not identify exact degrees, easy-class values or
recurring degree sets. Its potentially large function $F$ cannot be
inverted to obtain a linear PC lower bound from a linear NS lower bound.
That quantitative result came directly from the PC transfer theorem.

\section{Topological classification and semantic context}
\label{sec:model-theory}
The four degree types have a purely topological description valid for
all the uniform requests of Theorem~\ref{thm:classification}. It uses
only the eventual prime-power periodicity of bounded-degree
sublevel sets. No Boolean encoding or model-existence assumption is
needed. We then recall the distinction between algebraic witnesses
and actual solutions, and the established Euler-characteristic criterion.

\subsection{The \texorpdfstring{$p$}{p}-adic hardness spectrum}
\label{sec:topological-spectrum}
The underlying structure is already implicit in the residue-class
analysis of my 2009 report~\cite[Section~4, pp.~18--19]{RiisLong}:
once a degree threshold is fixed, sufficiently large sizes are classified
by residues modulo a power of $p$. We express that earlier structure
in $p$-adic language and apply it here to uniform consequences as well
as refutations.

Fix any one of the aggregate or fixed-orbit profiles $h$ in
Theorem~\ref{thm:periodicity}, over a field of characteristic $p$.
A $p$-adic integer is a compatible choice of residues modulo all powers
of $p$. The space $\mathbb Z_p$ is compact, and its basic residue
neighbourhoods $a+p^L\mathbb Z_p$ are both open and closed
(\emph{clopen}). Using the onsets $N(d)$ and periods $Q(d)$ from
\eqref{eq:NQ}, let $A_d(h)\subseteq\{0,\ldots,Q(d)-1\}$ consist of
the residues $a$ for which
\[
 h(n)\le d\quad\text{whenever }n\ge N(d),\qquad
                         n\equiv a\pmod{Q(d)}.
\]
Put
\[
 \mathcal E_d(h)=\bigcup_{a\in A_d(h)}
                         \bigl(a+Q(d)\mathbb Z_p\bigr).
\]
Thus $\mathcal E_d(h)$ records the eventually degree-$d$-easy residue
classes. Each is clopen, and $\mathcal E_d(h)\subseteq\mathcal E_e(h)$
for $d\le e$: refine to the common period and use the periodicity on a
sufficiently late tail. Set
\begin{equation}\label{eq:hardness-spectrum}
 \mathcal H_p(h)=\mathbb Z_p\setminus
                    \bigcup_{d\in\mathbb N}\mathcal E_d(h).
\end{equation}
We call this closed set the \emph{hardness spectrum}. It does not depend
on the choice of valid onsets or prime-power periods. Equivalently,
$x\notin\mathcal H_p(h)$ exactly when some residue neighbourhood of
$x$ has a uniform finite degree bound at all sufficiently large integer
sizes in that neighbourhood. Common residue refinements show the
agreement of this description with \eqref{eq:hardness-spectrum}.

\begin{samepage}
\begin{corollary}[The four topological shapes]
\label{cor:topological-spectrum}
For every uniform request considered in this paper, the NS and PC
profiles have the following classification:
\begin{center}
\begin{tabular}{@{}ll@{}}
\toprule
Type & Hardness spectrum $\mathcal H_p(h)$\\
\midrule
$\mathrm T_1$ & Empty\\
$\mathrm T_2$ & All of $\mathbb Z_p$\\
$\mathrm T_3$ & Nonempty, proper and clopen\\
$\mathrm T_4$ & Closed but not open\\
\bottomrule
\end{tabular}
\end{center}
Moreover, for the same request the two systems have the same hardness
spectrum, although their exact degrees and threshold sets can differ.
\end{corollary}
\end{samepage}
\begin{proof}
Write $E_d=\mathcal E_d(h)$ and $H=\mathcal H_p(h)$. If $H$ is empty,
compactness and the increasing open cover $(E_d)_d$ give
$E_D=\mathbb Z_p$ for some $D$. This is eventual boundedness, or T1;
the converse is immediate. Also $H=\mathbb Z_p$ exactly when every
$E_d$ is empty. This says that no finite degree recurs, or T2.

In T3, let $D$ be the largest recurring finite degree.
Theorem~\ref{thm:classification} gives a fixed nonempty proper union
of easy residue classes. It is $E_D$, and no $E_d$ meets a hard class:
otherwise arbitrarily late integers in a common residue refinement
would have both bounded and linearly growing degree. Hence
$\bigcup_dE_d=E_D$, so $H$ is nonempty, proper and clopen.

Conversely, suppose $H$ is nonempty, proper and clopen. Its compact
complement is covered by the increasing $E_d$, hence equals some $E_D$.
If a finite degree $e$ recurs, periodicity gives a residue neighbourhood
on which that exact value occurs on a tail. This neighbourhood lies
in $E_e\subseteq\bigcup_d E_d=E_D$; comparing late integer representatives gives
$e\le D$. There is a recurring finite degree because the complement
of $H$ is nonempty, but $H\ne\varnothing$ excludes T1. This is T3.
Since $H$ is always closed and the four degree types are exhaustive,
the remaining shape, closed but not open, is exactly T4.

Finally, the comparison theorem gives
$h_{\PC}\le h_{\NS}\le F(h_{\PC})$, with $F$ independent of size.
Thus a residue neighbourhood is eventually bounded in one system if
and only if it is eventually bounded in the other. The equivalent
description above proves equality of their hardness spectra.
\end{proof}

In T3 a single finite congruence resolution separates the easy and
hard regions. In T4 the closed spectrum has a nonempty boundary:
every neighbourhood of a boundary point meets the easy region, but
no such neighbourhood has one uniform finite degree bound.
The spectrum records the qualitative organisation of the profile;
the linear lower bounds and finite linear peaks still come from the
quantitative transfer theorem. It is defined from the polynomial
request and its degree profile, not from the ordinary model class
of a first-order sentence alone.

\subsection{Dual witnesses for failure to derive a target}
A linear functional is a scalar-valued linear map on coefficient
vectors. If it vanishes on every permitted consequence but not on
the target, applying it to a proposed derivation gives a contradiction.
The converse is finite-dimensional separation by a quotient basis;
it does not require the functional to be evaluation at a solution.

For either system write $W_d(n)$ for its degree-$d$ consequence space:
$W_d(n)=C_d(n)$ for PC and $W_d(n)=I_d(n)$ for NS. Let
$\delta=D_G$ be the fixed upper bound on target degree. We view
$W_d(n)$ inside $P_{\le\max\{d,\delta\}}(n)$, so the ambient space
also contains every requested target even when its degree exceeds $d$.

\begin{definition}[A degree-bounded target witness]
\label{def:consistency-witness}
For a requested target $G_b^{(n)}$, a degree-$d$ witness is a $K$-linear
functional
\[
 \lambda:P_{\le\max\{d,D_G\}}(n)\longrightarrow K,\qquad
 \lambda(G_b^{(n)})=1,\qquad\lambda(W_d(n))=0.
\]
For an all-target request, a witness consists of an index $b$ and such
a functional. For a single fixed orbit, one may use any available
representative. With target $1$, this reduces to the usual normalisation
$\lambda(1)=1$ for a bounded-degree non-refutability witness.
\end{definition}

\begin{proposition}[Dual form of nonderivability]
\label{prop:dual-witness}
A degree-$d$ target witness exists exactly when the target has no
derivation of degree at most $d$ in the chosen system. For an all-target
request it exists exactly when $h(n)>d$.
\end{proposition}
\begin{proof}
If $G_b^{(n)}\in W_d(n)$, a functional annihilating $W_d(n)$ cannot
take value one on it. Conversely its class in the finite-dimensional
quotient $P_{\le\max\{d,D_G\}}(n)/W_d(n)$ is nonzero whenever
it is not in $W_d(n)$. Extend that class to a basis, assign it value
one and assign the other basis vectors zero. Composing with the
quotient map gives the required functional. If the target degree is
greater than $d$, it automatically lies outside $W_d(n)$; the enlarged
ambient space also covers this degree-floor obstruction. Failure of
all-target derivability is failure for at least one target.
\end{proof}

A $K$-valued solution $a$ of $\Gamma_n$ which violates $G_b^{(n)}=0$ gives a
witness by $\lambda(f)=f(a)/G_b^{(n)}(a)$. For target one this is ordinary
evaluation, with its multiplicative property. A general witness need
not be multiplicative, positive or normalised on one. It records a
linear separation from all permitted bounded-degree consequences,
not necessarily a truth assignment to the original variables.

\begin{example}[Semantic entailment need not be derivation]
The equation $x^2=0$ over a field forces $x=0$ at every solution.
Nevertheless, neither NS nor PC derives $x$ from this axiom: every
consequence belongs to the ideal $(x^2)$. For $d\ge2$, the functional
on $P_{\le d}$ taking value one on $x$ and zero on all other monomials
annihilates the degree-$d$ consequence space. This is a target witness
although no actual solution violates $x=0$. The example explains why
our minimum degree is a syntactic ideal-membership resource.
\end{example}

\begin{example}[Linear consistency without truth-value evaluation]
Over $\mathbb F_3$, take $x^2-x=0$, target one and degree budget two.
The PC consequence space is $K(x^2-x)$: no nonzero vector in that span
has degree at most one, so no multiplication is usable. The functional
\[
 \lambda(1)=1,\qquad\lambda(x)=\lambda(x^2)=2
\]
is a witness but is not evaluation: $\lambda(x^2)=2$ whereas
$\lambda(x)^2=1$. Boolean equations impose the intended truth-value
semantics on actual solutions, without making every linear witness an
actual evaluation.
\end{example}

For target $1$, the NS witness is the design method for lower bounds
surveyed and generalised by Buss~\cite{BussDesign}: a normalised
linear functional annihilates every permitted axiom multiple.
Annihilating only axiom multiples gives an NS witness. For PC the
functional must annihilate the full closure, including polynomials
obtained by cancellation followed by new multiplication. This is the
dual form of the distinction between $I_d$ and $C_d$.

\subsection{What Kraj\'i\v cek's Euler criterion supplies}
A strong Euler characteristic assigns a value $\chi(A)$ to each
permitted parameter-definable set, possibly infinite. It gives a point
value one, is additive on disjoint unions, is multiplicative on products,
and is invariant under definable bijections. For a definable map
$f:A\to B$ whose fibres all have characteristic $r$, it also satisfies
$\chi(A)=r\chi(B)$. This last fibre rule is essential.

Such a rule replaces finite counting in a definable solution of a
uniform polynomial system. In Kraj\'i\v cek's criterion, the values lie
in $\mathbb Z/p^\nu\mathbb Z$, where the required precision $\nu$
depends on the degree budget and the indexing bounds. Reduction modulo
$p$ interprets a sum of a definable $\mathbb F_p$-valued function by
\[
 \sum_{x\in A}^{\chi}a(x)
       =\sum_{b\in\mathbb F_p}b\,\overline\chi(a^{-1}(b)).
\]
With the definability and index-sort hypotheses of
\cite[Theorem~6.1]{KrajEuler}, including the characteristic on the
required index sorts $M^*$, a solution violating the target $g=0$
excludes derivations within that fixed degree budget at sufficiently
large sizes in the residue class specified by $\chi(M)\bmod p^\nu$.
The refutation case is $g=1$.

The mechanism is that bounded-degree calculations involve only bounded
configurations of indices. The counting rules transport the relevant
finite identities to the definable setting. A hypothetical derivation
would make the target vanish there, contradicting the chosen solution.
A characteristic with sufficient precision for each degree therefore
provides a semantic obstruction to bounded-degree proofs. As the degree
increases, the required precision may increase too.

Definable bijections, disjoint-union relations and products can be
organised in the Grothendieck ring of definable sets
\cite{KrajScanlon}. The strong fibre rule is additional structure.
Neither an arbitrary infinite model nor a merely mod-$p$ counting rule
supplies the full hypotheses of Kraj\'i\v cek's criterion. It is a
sufficient criterion; no converse is used or asserted here.

Richer first-order counting theories can also describe bounded-degree
algebraic reasoning under additional hypotheses~\cite{PTT}. Such
model-theoretic representations are not needed for
Corollary~\ref{cor:topological-spectrum}; in particular, its scope is
not restricted to a fixed Boolean compiler. A general converse to the
Euler criterion remains a separate question.

\section{Nonconstant targets with finite linear peaks}\label{sec:target-examples}
The degree of a target need not reflect the degree needed to derive it.
A polynomial of degree one may require arbitrarily high proof degree.
We first give an exact construction, then apply it to a uniform family
whose hard values are finite.

A simple invariant target illustrates the distinction between finite
hardness and failure of derivability. For $G^{(n)}=\sum_{i<n}x_i$, the input
$\{1\}$ gives degree exactly one in both systems for $n\ge1$. With no
inputs the same target is underivable and has degree $\infty$. Both
fit the uniform definition, but neither supplies a finite hard example.
The next construction does.

Two elementary maps on the enlarged ring will be used. Substitution
$y=1$ is a ring homomorphism. Coefficient extraction $[y^1]$ is instead
a linear map: it takes $\sum_j a_j(x)y^j$ to $a_1(x)$.
It lowers the total degree by one on a nonzero extracted term and
commutes with multiplication by a polynomial independent of $y$.
This distinction accounts for the different exact NS and PC degrees.

\begin{proposition}[Adjoining a fresh target]\label{prop:fresh-target}
Let $K$ be any field, let $\Gamma\subseteq K[x_v:v\in V]$ be any axiom
set, and adjoin a fresh variable $y$. Each certificate uses only
finitely many axioms. Let $\Gamma^\uparrow$ be the same axioms viewed
in $K[x_v:v\in V,y]$, with no additional axiom involving $y$. Then
\begin{equation}\label{eq:fresh-identities}
 \begin{aligned}
 h_{\PC}(\Gamma^\uparrow\vdash y)
       &=\max\{1,d_{\PC}(\Gamma)\},\\
 h_{\NS}(\Gamma^\uparrow\vdash y)
       &=d_{\NS}(\Gamma)+1.
 \end{aligned}
\end{equation}
These equalities include degree zero and infinity, with
$\infty+1=\infty$.
\end{proposition}
\begin{proof}
For PC, embed a refutation of $\Gamma$ in the larger ring and multiply
its last line $1$ by $y$. If the refutation had degree $d$, the resulting
proof of $y$ has degree $\max\{1,d\}$: only the new last line is added,
and its degree is one.

Conversely, substitute $y=1$ in every line of a PC proof of $y$.
The old axioms are unchanged and linear combinations commute with the
substitution. Multiplication by an old variable remains a multiplication
step; multiplication by $y$ becomes a repeated line, which the linear
combination rule admits. No degree increases, and the last line becomes
$1$. Every proof of $y$ also has degree at least one. These two lower
bounds prove the first equality, including the case when no refutation
exists.

For NS, multiplying the coefficients of an identity
$1=\sum_fa_ff$ by $y$ gives an identity for $y$ with degree one larger.
For the reverse inequality, suppose
\[
                    y=\sum_f a_f(x,y)f(x)
\]
has total-product degree at most $e$. Take the coefficient of $y^1$.
Since every $f$ is independent of $y$, this gives the exact identity
\[
                    1=\sum_f [y^1]a_f(x,y)\,f(x).
\]
Every nonzero product on the right has degree at most $e-1$: it is the
coefficient of $y^1$ in a product of total degree at most $e$. Zero
products may be discarded. Hence $d_{\NS}(\Gamma)\le e-1$.
In particular a finite proof of $y$ exists exactly when a finite
refutation exists. If the original refutation degree is zero, both
new degrees are one, as asserted.
\end{proof}

\begin{example}[Why the two exact identities differ]
Take $\Gamma=\{x,1-x\}$. Both refutation degrees are one. PC derives
$1=x+(1-x)$ and then multiplies that last constant by $y$; every line
still has degree at most one. An NS identity for the fresh target is
\[
                             y=yx+y(1-x).
\]
Its products have degree two. No degree-one NS identity can produce
$y$, since its multipliers of these degree-one axioms would have to
be constants. Thus the new PC degree is one and the NS degree is two.
PC can finish the refutation before multiplying its last line; an NS
identity carries the extra variable in its coefficient polynomials.
\end{example}

The absence of a $y$-axiom is part of the construction; in particular,
we do not add $y^2-y=0$. For a uniform template, give the new variable
empty support and keep all old axiom indices and coefficients. Set
all coefficients of axiom monomials involving $y$ to zero. The result
is again a uniform template with the same bounds $k,s,D,n_0$, and
$G^{(n)}=y$ is a fully invariant target of degree one at every rank.
Thus \eqref{eq:fresh-identities} holds rank by rank.

This operation preserves each of the four types, separately for NS
and PC. The NS recurring finite degrees are shifted by one. For PC,
the map $d\mapsto\max\{1,d\}$ merges only degrees zero and one;
all larger finite degrees and infinity are unchanged. Eventual
boundedness, absence or unboundedness of recurring finite levels, and
the mixed periodic case are consequently preserved.

\begin{corollary}[A finite hard nonconstant consequence family]\label{cor:fresh-peaks}
Let $K$ have prime characteristic $p$, and let $\widehat B_n$ be the
normalised weak-bijection system defined in
\eqref{eq:bhat-witness}, with all of its function and coverage witnesses
retained. Adjoin a fresh variable $y$ as above, and request the target
$G^{(n)}=y$. Both target-degree profiles are of type T4.

More precisely, write $n=p^a u$, where $p\nmid u$, and put $q=p^a$.
Whenever $n\ge2q$,
\[
 \begin{gathered}
 q\le h_{\PC}(n)\le q+4,\qquad
 q+1\le h_{\NS}(n)\le q+5,\\
                         h_{\PC}(n)\le h_{\NS}(n).
 \end{gathered}
\]
In particular, at every size $n_a=(p+1)p^a$,
\begin{equation}\label{eq:fresh-peaks}
 \frac{n_a}{p+1}\le h_{\PC}(n_a)\le h_{\NS}(n_a)
                              \le\frac{n_a}{p+1}+5<\infty.
\end{equation}
\end{corollary}
\begin{proof}
The refutation bounds \eqref{eq:bhat-window}, proved in the next
section, give
$q\le d_{\PC}(\widehat B_n)\le d_{\NS}(\widehat B_n)\le q+4$.
Apply \eqref{eq:fresh-identities}; since $q\ge1$, the maximum in the
PC identity changes nothing here. At $n_a=(p+1)p^a$ we have
$v_p(n_a)=a$ and $n_a\ge2p^a$, yielding \eqref{eq:fresh-peaks}.

For each fixed $a$, infinitely many sizes
$n=p^a(p(j+1)+1)$ have the same valuation and satisfy the displayed
bounds. Some finite degree in each of the bounded windows must
therefore recur infinitely often. The windows tend to infinity with
$a$, proving T4 in each system. On the other hand, the old upper
bound gives $h_{\PC}(n)\le5$ and $h_{\NS}(n)\le6$ whenever
$p\nmid n$. The linear peaks thus coexist with infinitely many easy
sizes, rather than giving an eventual linear lower bound at all sizes.
\end{proof}

This example has finite hard proofs of a nonconstant target, but its
axioms are inconsistent: the difficulty is inherited from refuting
$\widehat B_n$. It does not establish a family of satisfiable axioms
with the same hard consequence profile. The small consistent example
in Section~\ref{sec:systems} illustrates ordinary-PC consequences in
a different setting; no growing lower bound is asserted for it.

\section{The refutation special case: four first-order examples}
\label{sec:examples}

We now take the constant target $G^{(n)}=1$. This is the refutation special
case of the uniform-consequence theorem; the sharper invariant-target
budget applies. The four cases all occur among fixed first-order sentences with no finite
nonempty models. We give examples over a field of fixed prime characteristic
$p$. Each sentence is translated by the graph-and-sum convention: relation
and function-graph variables are Boolean, and every function has a totality
sum and pairwise functionality equations. The parameter $n\geq1$ is the
universe size. For the matrix notation in this section, we relabel the
universe as $\{1,\ldots,n\}$; this changes no equation or degree.

\paragraph{T1: a local contradiction.}
The sentence $\forall x\,(R(x)\wedge\neg R(x))$ contributes $r_i$ and
$1-r_i$. Their sum refutes the system in degree one, independently of $n$.
The contradiction is already visible at one element.

\paragraph{T2: an injection that misses a point.}
Let $\eta$ say that a unary function $g$ is injective and that a constant
$c$ is outside its image. Write $x_{ij}$ for $g(i)=j$ and $c_j$ for $c=j$.
In addition to the Boolean equations, its system $\Gamma_n^\eta$ consists of
\begin{equation}\label{eq:G-witness}
\begin{gathered}
 \sum_jx_{ij}-1=0,\qquad
 x_{ij}x_{ik}=0\ (j\ne k),\qquad
 x_{ij}x_{hj}=0\ (i\ne h),\\
 \sum_jc_j-1=0,\qquad
 c_jc_k=0\ (j\ne k),\qquad x_{ij}c_j=0.
\end{gathered}
\end{equation}
A solution would inject $n$ elements into the $n-1$ elements other than $c$,
so every instance is unsatisfiable. For either proof system, every
degree-$d$ refutation satisfies
\[
 n<3pd\qquad(n\geq3).
\]
Here is how the internal matching bound gives this estimate. Choose the
largest $p$-power $q\leq n/3$ and put $r=n-q$. Set $c_n=1$, all other
$c_j=0$, and restrict the image of $g$ to $[r]$. After transposition, every
remaining axiom is an axiom of the bijection system $B_{r,n}$.
Here $B_{r,m}$ is the Boolean $r$-by-$m$ matrix system with every row and
column sum equal to one and both row and column collisions equal to zero.
Since $r=n-q\geq2q$, if $d<q$ then $2d\leq r$ and $q\mid n-r$, contradicting the rectangle
obstruction of Lemma~\ref{lem:rectangle-obstruction}. The substitutions do
not increase ordinary-PC or NS degree. Thus $d\geq q$, while maximality of
$q$ gives $n<3pq\leq3pd$. This proves T2 without an external pigeonhole
lower-bound theorem.

\paragraph{A matching linear upper bound.}
For the injection system, put
\[
 R_i=\sum_jx_{ij},\qquad C=\sum_jc_j,\qquad P=\prod_{i=1}^nR_i.
\]
Every monomial $c_j\prod_i x_{i,\lambda(i)}$ in the expansion of $CP$
is divisible by an input equation from \eqref{eq:G-witness}. If
$\lambda$ is not injective, it contains a column collision. If
$\lambda$ is injective, it is a permutation of the finite universe;
for the row $i$ with $\lambda(i)=j$, the monomial contains the omission
factor $c_jx_{ij}$. Thus $CP$ lies in the raw NS span at degree at most
$n+1$. The telescoping identity
\[
 CP-1=(C-1)+C\sum_{i=1}^n(R_i-1)\prod_{h<i}R_h
\]
expresses $CP-1$ using the constant and row totality equations, with
every product of degree at most $n+1$. Subtraction gives a refutation,
so for every $n\ge1$,
\[
 d_{\PC}(\Gamma_n^\eta)\le d_{\NS}(\Gamma_n^\eta)\le n+1.
\]
This argument works over any field and uses no Boolean simplification.
Together with the preceding lower bound, it shows that both degrees are
$\Theta(n)$ for each fixed characteristic. The general linear gap is
therefore sharp even among all-size inconsistent first-order families.

\paragraph{T3: hardness on one residue class.}
Let $\chi_p$ assert
\[
 \forall x\qquad f^p(x)=x\quad\hbox{and}\quad f(x)\ne x.
\]
Its finite models are disjoint unions of directed $p$-cycles. In particular,
it has a $0/1$ graph assignment exactly when $p\mid n$. An explicit
normalisation uses the function equations for $f$, the equations $f_{ii}=0$,
and
\begin{equation}\label{eq:cycle-path}
 f_{i_0i_1}f_{i_1i_2}\cdots f_{i_{p-1}i_p}=0
 \qquad(i_p\ne i_0).
\end{equation}
If $p\nmid n$, summing all length-$p$ walks gives an NS refutation of degree
at most $p$. Indeed, totality derives the sum of their monomials minus $n$
within degree $p$. Nonclosed walks vanish by \eqref{eq:cycle-path}.
A closed walk with a repeated vertex either contains a functionality
collision or follows a deterministic cycle of length $\ell<p$ with
$\ell\mid p$. Since $p$ is prime, $\ell=1$, a forbidden fixed point. Each remaining cycle contributes its
monomial $p$ times, according to its starting vertex, and hence contributes
zero. The resulting nonzero scalar $n$ yields the refutation.

Now take $\psi_3=\eta\wedge\chi_p$ in disjoint nonlogical symbols.
For $p\mid n$, substitute a $p$-cycle assignment into the second block.
This does not increase proof degree, so the conjunction has degree at least
that of $\eta$. The reverse inequality follows because all axioms of
$\eta$ are present. Thus the degrees agree, separately for NS and PC. For $p\nmid n$, the second
block refutes in degree at most $p$. Consequently this all-size
unsatisfiable family is bounded off the multiples of $p$ and has linear
degree on the multiples: it has type T3. No exact degree is needed on the
bounded residue classes.

\paragraph{T4: the valuation of the universe size controls the difficulty.}
For a positive integer $n$, its $p$-adic valuation $v_p(n)$ is the
largest $a$ such that $p^a$ divides $n$. Thus $n=p^a u$ with
$p\nmid u$. We use this integer statistic, not a valuation on the
coefficient field; the corresponding block size is $p^a$.
The weak-bijection example from the earlier classification~\cite[Section~5]{Riis}
is more revealing; its algebraic complexity is closely related to the
bijective pigeonhole principle studied by Beame and Riis~\cite{BeameRiis}.
Let unary functions $f,g$
be injective with disjoint images, and require their images together to
cover the universe. Use a unary witness $h$ for coverage:
\[
 \forall z\qquad f(h(z))=z\ \vee\ g(h(z))=z.
\]
Let $a_{ij},b_{ij},w_{ji}$ represent $f(i)=j$, $g(i)=j$, and $h(j)=i$.
The combined $2n$-by-$n$ matrix $(a;b)$ has row sums equal to one and both
row and column collisions equal to zero. The witness block consists of
\begin{equation}\label{eq:bhat-witness}
 \sum_iw_{ji}-1=0,\qquad
 w_{ji}w_{ji'}=0\ (i\ne i'),\qquad
 w_{ji}(1-a_{ij})(1-b_{ij})=0,
\end{equation}
together with the Boolean equations. Call this complete system
$\widehat B_n$. Its witnesses are retained: it is not identified with the
simpler matrix system $B_{n,2n}$.

Put $q=p^{v_p(n)}$. The proved bounds give
\begin{equation}\label{eq:bhat-window}
 q\leq d_{\mathrm{PC}}(\widehat B_n)
 \leq d_{\mathrm{NS}}(\widehat B_n)\leq q+4
 \qquad\hbox{whenever } n\geq2q.
\end{equation}
The upper bound holds at every positive size, without the displayed size
condition. We give both degree-preserving reduction and certificate
explicitly.

For the lower bound transpose the combined matrix and substitute
$w_{ji}=a_{ij}+b_{ij}$ into every witness variable. Its totality becomes
the corresponding totality of $B_{n,2n}$. The identities
\begin{align*}
(a+b)^2-(a+b)&=(a^2-a)+(b^2-b)+2ab,\\
(a+b)(1-a)(1-b)&=-(1-b)(a^2-a)-(1-a)(b^2-b)
\end{align*}
put the substituted Boolean and coverage equations in the raw NS spans
at degrees two and three respectively. A substituted witness collision
expands into four column collisions of the raw combined matrix.
Thus every substituted input has a derivation within its own degree.
Substitution is affine: a multiplication by an old variable becomes a
sum of permitted multiplications by new variables, with the preceding
line still of degree at most $d-1$. A degree-$d$ ordinary-PC refutation
would therefore refute $B_{n,2n}$ within the same degree. The rectangle
obstruction excludes this when $2d\le n$ and $d<q$, yielding the lower
bound in \eqref{eq:bhat-window}.

For the upper bound fix a column $j$ and abbreviate
\[
 u_i=a_{ij}+b_{ij},\qquad T=\sum_i u_i,\qquad
 Z=\sum_i w_{ji}u_i.
\]
The coverage axiom and disjointness collision give the exact identity
\[
 Z-1=(\sum_iw_{ji}-1)
 -\sum_iw_{ji}(1-a_{ij})(1-b_{ij})
 +\sum_iw_{ji}a_{ij}b_{ij}.
\]
Consequently $Z-1$ lies in the degree-three NS span. If $x$ is one entry
of this column, then
\[
 Tx-x=(x^2-x)+\sum_{y\ne x}yx
\]
lies in the degree-two span, by Booleanity and column collisions.
Adding this for $a_{ij}$ and $b_{ij}$ gives $Tu_i-u_i$ in that span.
Finally
\[
 T-1=\sum_iw_{ji}(Tu_i-u_i)-(T-1)(Z-1)
\]
puts the missing column totality in the degree-four span. This retains
the coverage witnesses throughout the derivation.

For completeness, consider any matrix with row sums $R_i$ and column
sums $C_j$, with their differences from one in the NS span at degree
$e$, and with both collision families in degree two. For a fixed
$k$-element row set $S$, telescoping the product of its row sums gives
$\prod_{i\in S}R_i-1$ in degree at most $e+k$. Summing over $S$ gives
\[
 \sum_{|S|=k}\prod_{i\in S}R_i-\binom{\#\mathrm{rows}}k
\]
in that same span. Expand the products as partial functions from $k$
rows to columns. A noninjective partial function contains a column
collision, so its monomial is already in the degree-$k$ span.
The remaining monomials are exactly those of $k$-edge matchings.
The column calculation gives the same matching sum, since inversion
pairs partial injections with identical monomials. Subtracting derives
\[
 \binom{\#\mathrm{rows}}k-\binom{\#\mathrm{columns}}k
\]
within degree $e+k$. Applied to the $2n$-by-$n$ matrix with $e=4$, this
is the desired counting certificate.

Write $n=qu$ with $q=p^{v_p(n)}$ and $p\nmid u$. In characteristic
$p$, $(1+t)^n=(1+t^q)^u$. The coefficient of $t^q$ is therefore $u$;
for $(1+t)^{2n}$ it is $2u$. Thus
$\binom{2n}q-\binom nq=u\ne0$ in $K$. Rescaling the counting
certificate gives an NS refutation of degree at most $q+4$.

For each fixed $a$, the sizes $n=p^a(p(k+1)+1)$, $k\geq0$, have
$v_p(n)=a$ and satisfy \eqref{eq:bhat-window}. Hence some finite degree in
$[p^a,p^a+4]$ recurs infinitely often. These windows are unbounded as $a$
varies, proving T4 for both proof systems. On the other hand, $p\nmid n$
implies degree at most five. There are explicit simultaneous linear peaks:
at $n_a=(p+1)p^a$,
\[
 \frac{n_a}{p+1}\leq d_{\mathrm{PC}}(\widehat B_{n_a})
 \leq d_{\mathrm{NS}}(\widehat B_{n_a})
 \leq\frac{n_a}{p+1}+4.
\]
For example, in characteristic two every odd $n$ has degree at most
five, whereas at $n=3\cdot2^a$ both degrees lie between $2^a$ and
$2^a+4$. The peaks are linear in their universe size even though the
profile returns to bounded values infinitely often.
Thus linear peaks coexist with infinitely many easy sizes. This is the
reason the T4 conclusion concerns peaks rather than an eventual linear
lower bound at every size.

\section{Matching designs and the concrete lower bounds}
\label{app:matching}

The examples use a lower bound for a concrete rectangular system.  Its
proof illustrates why a functional excluding a Nullstellensatz certificate
does not by itself exclude a Polynomial Calculus derivation.  To handle
degree falls we construct a whole space of functionals, together with an
extension operator between successive degrees. The concrete bijection
system has its antecedent in Beame and Riis~\cite{BeameRiis}; the use of
designs as algebraic obstruction witnesses goes back to Buss~\cite{BussDesign}.
The construction and the extension through ordinary-PC degree falls
are proved explicitly below.

For positive integers $r\leq m$, let $B_{r,m}$ consist of the equations
\[
 \sum_{j=1}^{m}x_{ij}-1=0\quad(i\in[r]),\qquad
 \sum_{i=1}^{r}x_{ij}-1=0\quad(j\in[m]),
\]
the Boolean equations $x_{ij}^{2}-x_{ij}=0$, and the row and column
collision equations
\[
 x_{ij}x_{i\ell}=0\quad(j\ne\ell),\qquad
 x_{ij}x_{hj}=0\quad(i\ne h).
\]
Here $[r]=\{1,\ldots,r\}$.  A Boolean solution would describe a bijection
between the two sides, so a proper rectangle has no solution.  Nevertheless,
in positive characteristic the two linear totality sums can agree because
$m-r$ vanishes in the field.  The following statement identifies a range
in which this obstruction persists through ordinary PC.

\begin{lemma}[Rectangle obstruction]\label{lem:rectangle-obstruction}
Let $K$ be a field of prime characteristic $p$, and let $a,D$ be
nonnegative integers.  If
\[
 1\leq r\leq m,\qquad 2D\leq r,\qquad D<p^a,
 \qquad p^a\mid m-r,
\]
then $B_{r,m}$ has no ordinary-PC refutation of degree at most $D$.
Consequently it has no Nullstellensatz refutation of total-product degree
at most $D$.
\end{lemma}

We give the full construction over the integers before passing to $K$.
An \emph{edge set} is a subset of $[N]\times[M]$.  It is a matching if
no two distinct edges share a row or a column.  Write $\operatorname{rows}P$
and $\operatorname{cols}P$ for its two projections.  A degree-$d$ design
may take values in a coefficient ring: it is a function $\psi$ on edge
sets, zero on nonmatchings, for which
\begin{equation}\label{eq:matching-design}
 \sum_{j=1}^{M}\psi(P\cup\{(i,j)\})=\psi(P),\qquad
 \sum_{i=1}^{N}\psi(P\cup\{(i,j)\})=\psi(P)
 \quad(|P|<d).
\end{equation}
The indices $i\in[N]$ and $j\in[M]$ are arbitrary. For $K$-valued
designs, denote the resulting vector space by $E_d(N,M)$. The signed
coordinate reductions below are constructed integrally and then
interpreted in $K$.  Normalisation, $\psi(\varnothing)=1$, is a separate
condition.  We will construct a normalised member of $E_D(r,m)$ and prove
that, for $D\ge1$, \emph{every} member of $E_{D-1}(r,m)$ extends to $E_D(r,m)$ while
preserving its values through degree $D-1$.

Associate an edge set $P$ with the squarefree monomial
$\prod_{e\in P}x_e$. A design assigns proposed values to these
monomials and satisfies the displayed linear consistency tests.
An actual solution evaluates multiplicatively; a design need not.
Normalisation fixes the value of the constant monomial, without any
probability or positivity assumption.

\subsection{A terminating signed reduction}
The reduction is a formal integer combination of matchings.
Its coefficients may be negative, and are interpreted in $K$ only
after the integral identities have been proved. Termination is
controlled by a decreasing integer rank, rather than by cancellation
of coefficients.

A chain in a matching is a subset whose row coordinates and column
coordinates both increase.  Let $h(P)$ be the maximum chain length, and
write $P_{\leq b}$ for the edges in rows at most $b$.  A matching $P$ is
\emph{terminal} if, for every positive integer $b$,
\[
 h(P_{\leq b})\leq b-|\operatorname{rows}(P_{\leq b})|,
\]
and the corresponding inequality holds for column prefixes.  The right
side counts unused labels in the prefix.  This definition depends only
on the labelled matching, not on the ambient rectangle.  Prefixes beyond
its last occupied row or column cannot introduce a new failure: their
left side stays fixed and their right side increases.

For a nonterminal matching, use the least failed row prefix if one exists;
otherwise use the least failed column prefix.  In the selected prefix
choose a longest chain, breaking ties by the lexicographically greatest
list of its edges, with the edges themselves in row-first lexicographic
order.  In a column step use this rule after transposition.  Thus all
choices are deterministic.

Consider a row step with failed prefix $b$ and chosen chain $A$.  Put
\[
 P_0=P\setminus A,\qquad S=\operatorname{cols}A,\qquad
 B=\{i\leq b:i\notin\operatorname{rows}P\},
\]
\[
 U=\operatorname{rows}A\ \cup\
       \{b<i\leq N:i\notin\operatorname{rows}P\}.
\]
The sets $B$ and $U$ are disjoint and partition all rows unused by $P_0$.
Moreover, $|B|<|A|$, because the selected prefix fails.  A longest chain
in the first failed prefix ends in row $b$.  Indeed, if it ended earlier,
the same chain would belong to the preceding prefix, whereas the number
of unused labels cannot decrease when one adds the next row.  That
preceding prefix would already fail.

For sets of columns $S$ and rows $V$, a full injection from $S$ into $V$
means a matching $J$ with $\operatorname{cols}J=S$ and
$\operatorname{rows}J\subseteq V$.  A partial injection has only
$\operatorname{cols}J\subseteq S$.  Replace $P$ by
\[
 \sum_{\substack{J\text{ partial injection}\\S\to B}}
       (-1)^{|J|}[P_0\cup J]
 \ -\!
 \sum_{\substack{J\text{ full injection}\\S\to U\\J\ne A}}
       [P_0\cup J],
\]
omitting colliding terms.  Square brackets here denote formal basis
elements, not polynomial equivalence classes.  The transposed rule defines
a column step.

To verify the identity for a degree-$d$ design with $|P|\leq d$, repeatedly
apply its column equation to the columns of $S$.  At each application
the partial matching has fewer than $|P|$ edges, so the degree restriction
is satisfied.  An occupied row contributes zero by collision vanishing.
For a still unassigned column the sum over available rows in $U$ is
therefore the previous value minus the sum over rows in $B$.  Expanding
these successive differences gives
\[
 \sum_{J:S\to U}\psi(P_0\cup J)
  =\sum_{A'\subseteq S}(-1)^{|A'|}
       \sum_{J:A'\to B}\psi(P_0\cup J).
\]
Every surviving term is an injection, because collisions have value zero.
The right side contains each partial injection exactly once, indexed by
its column set.  Isolating the term $J=A$ on the left proves the stated
reduction.  The column step follows by interchanging rows and columns in
the design equations.  No assertion that the deterministic reduction
itself commutes with transposition is needed.

We next prove termination, including the equal-degree case.  For the
fixed grid $G=[N]\times[M]$, set $H=N+M$, $c=NM+1$, and define
\[
 w(i,j)=c^{H-i-j},\qquad W(P)=\sum_{(i,j)\in P}w(i,j),\qquad
 \rho(P)=|P|(W(G)+1)+W(P).
\]
Every lower term has fewer edges, since an injection into $B$ has at most
$|B|<|A|$ edges.  Thus its $\rho$ is smaller irrespective of its weight.
For a full injection $J\ne A$ into $U$, let $e$ be the first edge of $A$
not belonging to $J$, in increasing row order.  Every earlier chain edge
still belongs to $J$.  If $x\in J\setminus A$ had row smaller than that
of $e$, its row would be an earlier chain row already occupied by its
retained edge; this contradicts the matching condition.  A row from the
other part of $U$ is greater than $b$ and hence cannot be smaller.
Likewise, a column of $x$ smaller than the column of $e$ belongs to an
earlier retained chain edge, again giving a collision.  Consequently
\[
 e_1\leq x_1,\qquad e_2\leq x_2,\qquad e_1+e_2<x_1+x_2.
\]
At least one inequality is strict because $x\ne e$.  Each newly added
edge has weight at most $w(e)/c$, and there are at most $NM<c$ such
edges.  Their total weight is strictly smaller than $w(e)$.  Cancelling
the common edges of $A$ and $J$ shows $W(P_0\cup J)<W(P)$.
The identical argument applies to a column step, since $i+j$ is
unchanged by transposition.  Every child therefore has strictly smaller
nonnegative integer rank $\rho$.  The tree is finitely branching, and
every path has length at most its initial rank.  Hence the complete
signed reduction tree is finite.

For terminal $T$, let $C_{N,M}(P,T)\in\mathbb Z$ be the sum, over paths
from $P$ to $T$, of the products of their step signs; set it to zero for
colliding $P$.  The empty path gives
\[
 C_{N,M}(S,T)=\delta_{S,T}\quad(S\text{ terminal}),\qquad
 C_{N,M}(P,T)=0\quad(|T|>|P|).
\]
Induction on $\rho(P)$ using the reduction identity gives, for every
degree-$d$ design,
\begin{equation}\label{eq:terminal-expansion}
 \psi(P)=\sum_{\substack{T\text{ terminal}\\|T|\leq d}}
              C_{N,M}(P,T)\psi(T)\qquad(|P|\leq d).
\end{equation}
Thus the terminal values determine every design in the relevant range.
It remains to show that these coordinates themselves satisfy the required
totality equations and can be restricted from a square to a rectangle.

\subsection{Integral independence in a square}
The next argument uses signed sums of evaluations at actual square
permutations to isolate terminal coordinates. A diagonal value of
$1$ or $-1$ is invertible in every coefficient ring, unlike a factorial.
This is why the resulting independence remains valid after reduction
to any characteristic.

Let $T\subseteq[N]\times[N]$ be terminal.  For an edge $e\in T$, let
its height be the maximum chain length in the southwest downset
$\{f\in T:f_1\leq e_1,\ f_2\leq e_2\}$.  A longest such chain ends
at $e$: otherwise $e$ can be appended, because distinct edges of a
matching have distinct row and column coordinates.  If $e$ is southwest
of $f$ with both coordinates strictly smaller, appending $f$ raises the
height.  It follows that every equal-height level $T_t$ is decreasing:
as its row coordinate increases, its column coordinate decreases.
The levels partition $T$.

Let $h=h(T)$.  The terminal inequality at prefix $N$ implies that there
are at least $h$ unused rows and $h$ unused columns.  Let $a_t$ and $b_t$
be the $t$th smallest unused row and column, respectively, for
$1\leq t\leq h$.  These anchors precede every coordinate in $T_t$.
For example, if $e\in T_t$, a chain of length $t$ ending at $e$ belongs
to its row prefix.  Terminality requires at least $t$ unused rows in
that prefix; row $e_1$ is occupied, so the $t$th unused row is strictly
less than $e_1$.  The column argument is identical.  Form the paired blocks
\[
 A_t=\operatorname{rows}T_t\cup\{a_t\},\qquad
 B_t=\operatorname{cols}T_t\cup\{b_t\}.
\]
They are disjoint on each side, their paired sizes agree, and their
minima are $a_t,b_t$.  Their complements also have equal sizes.  Fix the
increasing bijection on the complements, and an increasing base bijection
$A_t\to B_t$ in each block.  Vary each base bijection by a permutation
$\sigma_t$ of $A_t$.  The union of all these bijections is an actual
permutation $\pi_\sigma$ of $[N]$.

For any edge set $P$, define the integer signed moment
\[
 M_T(P)=\sum_{\sigma}
     \left(\prod_{t=1}^{h}\operatorname{sgn}\sigma_t\right)
     \boldsymbol{1}_{P\subseteq\pi_\sigma}.
\]
If a block has two rows absent from $\operatorname{rows}P$, interchange
their images.  This is a fixed-point-free involution on the choices,
changes the sign, and preserves the containment condition.  Its terms
cancel over $\mathbb Z$.  In particular, $M_T(P)=0$ when $|P|<|T|$:
otherwise every block would use at least $|A_t|-1=|T_t|$ rows, forcing
$|P|\geq\sum_t|T_t|=|T|$.

The diagonal is a unit.  In each block the edges of $T_t$ prescribe the
bijection on every row except $a_t$; its remaining value is necessarily
$b_t$.  Thus exactly one choice $\sigma$ contains $T$, and
\[
 M_T(T)\in\{1,-1\}.
\]
This is a single signed contribution; no factorial or positive integer
multiplicity must be inverted.

For equal-degree triangularity, suppose $|P|=|T|$ and $M_T(P)\ne0$.
There is a choice containing $P$, so $P$ is a matching.  Cancellation
shows that at most one row in each block is unused.  Counting over the
disjoint blocks forces exactly $|T_t|$ edges in block $A_t\times B_t$,
one unused row in every block, and no edge in the complement.  Write
$P_t$ for this block matching.

Among all matchings of size $|T_t|$ in $A_t\times B_t$, the decreasing
matching $T_t$ has uniquely least weight $W$.  Here is a direct proof.
Choose a weight minimiser $Q$ from this finite set.  If it uses the
minimum row $a_t$, move that edge to the unique unused row.  This is a
legal matching, has the same cardinality, and strictly smaller weight,
because its row coordinate has increased.  Thus $a_t$ is unused.
The same argument shows that the minimum column $b_t$ is unused.
If $Q$ contains an increasing pair $(i,j),(i',j')$ with
$i<i'$ and $j<j'$, replace it by the crossed pair
$(i,j'),(i',j)$.  Both added edges have coordinate sum strictly larger
than that of $(i,j)$, and their combined weight is smaller than its
weight, since $c=N^2+1>2$ when such a pair exists.  The unchanged context
is disjoint from the crossed pair.  This again contradicts minimality.
Therefore $Q$ is decreasing, uses precisely $A_t\setminus\{a_t\}$
and $B_t\setminus\{b_t\}$, and must equal $T_t$: on two fixed ordered
sets there is exactly one order-reversing bijection.  Any distinct
matching with the same minimum weight would also be a minimiser and
would give the same conclusion.  Hence
\[
 W(P_t)\geq W(T_t),\quad\text{with equality only if }P_t=T_t.
\]
Summing over the disjoint blocks proves that
\[
 |P|=|T|,\quad P\ne T,\quad M_T(P)\ne0
 \quad\Longrightarrow\quad W(P)>W(T).
\]
Together with the lower-degree cancellation, this says that $M_T(P)=0$
for every $P\ne T$ with $\rho(P)\leq\rho(T)$.

Now suppose a finite linear combination of terminal matching indicators
vanishes at every permutation of $[N]$, with coefficients in any
commutative ring $R$.  Among its nonzero coefficients choose a terminal
$T$ of greatest rank $\rho$.  Apply the integer signed moment $M_T$ to
this relation, then map the integers to $R$.  All other nonzero terms
vanish by triangularity; the result is the coefficient of $T$ multiplied
by $1$ or $-1$.  This contradicts its being nonzero.  The terminal
indicators are therefore independent over every commutative ring,
including fields of characteristic two and rings with zero divisors.

Each square permutation satisfies the design equations in every degree.
Equation~\eqref{eq:terminal-expansion}, applied to its containment
indicator, expresses every matching indicator in terminal coordinates.
For a fixed row $i$, the permutation identity
\[
 \sum_{j=1}^{N}\boldsymbol{1}_{P\cup\{(i,j)\}\subseteq\pi}
       =\boldsymbol{1}_{P\subseteq\pi}
\]
holds because $\pi$ has exactly one edge in that row.  Expand all terms
in terminal coordinates and use the integral independence just proved.
For each terminal $T$ this yields
\[
 \sum_{j=1}^{N}C_{N,N}(P\cup\{(i,j)\},T)=C_{N,N}(P,T).
\]
The column identity follows in the same way from the unique edge in a
column.  Colliding sets have zero coordinates and zero permutation
indicators, so they are included.  Thus every square coordinate is an
unnormalised design, in every degree, over the integers.

\subsection{Whole-history compression and the marked binomial identity}
The support recorded below belongs to an entire reduction path, not
only its final matching. Grouping paths by this support will produce
binomial coefficients in the number of appended rows. Keeping the
initial mark ensures that the coefficient with support size zero is
absent, even if its edge is later deleted.

Fix $P,T\subseteq[r]\times[m]$, with $P$ a matching and $T$ terminal, and a column
$j\notin\operatorname{cols}P$.  Append $L$ rows, and consider all pairs
consisting of a marked initial row $i\in\{r+1,\ldots,r+L\}$ and a signed
reduction path from $P\cup\{(i,j)\}$ to $T$.  The weight of a pair is the
product of its path signs.  Its \emph{history support} is the set of
appended rows appearing anywhere on the path, including the initial
state and edges that are subsequently deleted.

If $|P|+1\leq D$ and $2D\leq r$, every such support is nonempty and has
size at most $|P|+1$.  To prove the latter assertion, give one token to
each initial edge.  In a row rewrite, transport a token by its column
to the unique child edge in that column, or delete it if that column
is absent.  The child uses only columns of the parent, so this describes
every child edge exactly once.  In a column rewrite transport by row,
with the analogous deletion rule.  No deleted token reappears, and no
two surviving tokens represent the same edge.

For any matching $Q$ of at most $D$ edges, failure at row prefix $b$
implies
\[
 b< h(Q_{\leq b})+
       |\operatorname{rows}(Q_{\leq b})|\leq 2|Q|\leq2D\leq r.
\]
Thus every chosen row chain lies in old rows.  An edge already in an
appended row is part of the unchanged context of a row rewrite and is
retained verbatim.  A column rewrite preserves its token's row or
deletes the token.  It follows that a token can encounter at most one
appended row throughout its lifetime.  There are $|P|+1$ initial tokens,
so the full history has at most that many appended rows.  The initial
mark belongs to it even when its token later disappears.

We next verify that histories of the same size carry the same total
integer weight.  Let $S$ be a prescribed $k$-element subset of the $L$
appended rows, and let $f$ be the increasing injection fixing $[r]$ and
sending $r+1,\ldots,r+k$ to the elements of $S$ in order.  For $t\geq1$, set $f(r+k+t)=r+L+t$.
This extends $f$ increasingly and makes the inverse image of the large
row interval $[r+L]$ exactly $[r+k]$.  Rename the row coordinate of every
edge by $f$ and retain its column and step sign.

First, this renaming preserves the deterministic selector.  Every failed
row prefix lies below $r$, as just proved; those prefixes and their
unused-row counts are unchanged.  Their chains are unchanged as labelled
edge sets.  For a column prefix, the number of occupied columns and all
chain lengths are preserved by the increasing row renaming.  The first
failed column prefix is therefore unchanged.  The lexicographic tie
choice is also preserved: a strictly increasing map preserves the
lexicographic order of sorted edge lists, including the transposed lists
used in a column step.  Orientation, first failure and chosen chain all
agree under renaming, and terminality is preserved.

Second, renaming gives a bijection on the branches whose child edges
remain in $[r]\cup S$.  In a row step, the inverse images of the sets
$B$ and $U$ are exactly the corresponding sets in the compressed
rectangle: order comparisons with the old endpoint and occupied-row
membership are preserved.  An image-supported replacement therefore
has a unique inverse injection.  Its column set, cardinality, exclusion
of the original chain, collision status and sign are all preserved.
For a column step the permissible column sets are unchanged; the same
argument, using row membership, applies.  A child cannot encode two
different replacement injections, since its replacement edges are
recovered by removing the fixed context: their columns are disjoint
from the context columns in a row step, and their rows are disjoint
in a column step.  The lower and upper branch families are disjoint
because their cardinalities differ.  Thus no multiplicity is lost.
There may be other branches using rows outside $S$, but these belong to
different history supports and are not part of this correspondence.

Applying the branch bijection successively gives a two-way, sign-preserving
bijection between marked paths with history exactly
$\{r+1,\ldots,r+k\}$ in the compressed rectangle and marked paths with
history exactly $S$ in the large rectangle.  Its inverse simply renames
every state and the initial mark back.  The endpoint $T$ is fixed because
it uses only old rows.  All intermediate states have at most $D$ edges,
so the selector argument applies at every step.

Define $c_k(P,T,j)$ to be the total weight of the marked paths in the
rectangle $[r+k]\times[m]$ whose full history is all $k$ appended rows.
Partitioning the marked paths by their history support, and then by its
cardinality, proves the integer identity
\begin{equation}\label{eq:marked-coordinate}
 \sum_{i=r+1}^{r+L}
       C_{r+L,m}(P\cup\{(i,j)\},T)
    =\sum_{k=1}^{D}\binom{L}{k}c_k(P,T,j).
\end{equation}
The coefficients are independent of $L$.  Its constant term is absent
because each path retains its initial mark in the \emph{history}, not
necessarily in its final state.  This distinction is essential.

\subsection{Rectangular designs and their extensions}

Assume now the hypotheses of Lemma~\ref{lem:rectangle-obstruction}, with
$D\geq1$, and put $q=p^a$ and $L=m-r$.  For $1\leq k\leq D<q$ the
integer $\binom{L}{k}$ is zero in $K$.  Indeed, $q\mid L$ and in
characteristic $p$,
\[
 (1+z)^L=(1+z^q)^{L/q};
\]
the right side has no coefficient in these degrees.  Take the square
of size $m$ and an old terminal matching $T\subseteq[r]\times[m]$.
The function
\[
 \theta_T(P)=C_{m,m}(P,T)\quad\text{in }K,
       \qquad P\subseteq[r]\times[m],
\]
is a degree-$D$ rectangular design.  Its row equations are the square
row equations, with the same column set.  Its column equations follow
from the square column equation after deleting the appended-row sum,
which vanishes by~\eqref{eq:marked-coordinate} for $|P|<D$ and an
unused column.  If the column is occupied, every appended insertion
collides, so the omitted sum is already zero.  If $P$ collides, every
insertion still collides and both sides are zero.  This covers all
design equations.

On old terminal inputs, the square reduction is still terminal and
\[
 \theta_T(S)=\delta_{S,T}.
\]
In particular, $\theta_{\varnothing}$ is normalised.  For an arbitrary
$\psi\in E_{D-1}(r,m)$ define explicitly
\[
 \widetilde\psi(P)=
    \sum_{\substack{T\text{ terminal in }[r]\times[m]\\|T|\leq D-1}}
                     \psi(T)\theta_T(P).
\]
This is a member of $E_D(r,m)$ and agrees with $\psi$ on all terminal
inputs of size at most $D-1$.  Apply~\eqref{eq:terminal-expansion} in the
old rectangle to both designs, after restricting the new design to
degree $D-1$.  They agree on every edge set of size at most $D-1$.
Thus the extension exists for every unnormalised design, with no
choice of an unspecified lift.

\subsection{The invariant through ordinary-PC degree falls}
For $D\ge1$, the extension property is stronger than existence of one normalised
design. Multiplying a polynomial by an edge variable shifts its test
functional to the preceding degree window. Extending that shifted
design back to degree $D$ lets the induction test every PC line,
including lines whose leading terms have cancelled.

For a design $\psi$ define a linear functional on the original polynomial
ring by
\[
 \Lambda_\psi\!\left(\sum_\alpha b_\alpha x^\alpha\right)
      =\sum_\alpha b_\alpha\psi(\operatorname{supp}\alpha),
\]
where $\operatorname{supp}\alpha$ is the set of edges with positive
exponent.  This formula is a functional on ordinary polynomials; it
does not replace a PC proof by a proof in a Boolean quotient.
Every degree-$D$ design annihilates each axiom usable in a degree-$D$
proof.  Totalities use~\eqref{eq:matching-design} at $P=\varnothing$;
collisions use vanishing on nonmatchings; and the Boolean equations
use the equality of the supports of $x_e^2$ and $x_e$.

For an edge $e$, put $(s_e\psi)(P)=\psi(P\cup\{e\})$.
If $\psi\in E_D(r,m)$, then $s_e\psi\in E_{D-1}(r,m)$.
Collision vanishing is immediate.  In either totality equation at
$|P|<D-1$, apply the degree-$D$ equation to $P\cup\{e\}$, whose
cardinality is at most $D-1$.  The insertion operations commute,
including repeated insertion of the same edge.  Also, directly from
the monomial formula,
\[
 \Lambda_\psi(x_ef)=\Lambda_{s_e\psi}(f).
\]

Induct through a degree-$D$ ordinary-PC derivation, asserting that every
line is annihilated by \emph{every} member of $E_D(r,m)$.  Input axioms
were handled above, and linear combinations preserve the assertion
even when their highest-degree terms cancel.  Consider multiplication
of an earlier line $f$ by $x_e$.  If $f=0$ there is nothing to prove.
Otherwise ordinary polynomial degree gives
$\deg(x_ef)=\deg f+1$, so $\deg f\leq D-1$.
For a given $\psi\in E_D(r,m)$, extend $s_e\psi$ to a design
$\nu\in E_D(r,m)$ using the preceding construction.  The earlier line
is annihilated by $\nu$ by the inductive assertion.  Since the two
designs agree on all edge sets of size at most $D-1$, and every monomial
of $f$ has support of that size or smaller,
\[
 \Lambda_\psi(x_ef)=\Lambda_{s_e\psi}(f)=\Lambda_\nu(f)=0.
\]
This proves the invariant for multiplication as well.  Finally the
normalised design $\theta_{\varnothing}$ sends the constant $1$ to $1$,
so that constant cannot be the last line of the derivation.

For $D=0$, all displayed axioms have positive degree because both sides
of the rectangle are nonempty.  No nonzero constant axiom is available,
and multiplication of a nonzero polynomial cannot produce a degree-zero
line.  Linear combinations of the available zero lines remain zero.
Thus a degree-zero refutation is impossible directly.  This completes
the ordinary-PC proof in every case.  The NS conclusion follows from
the degree-preserving simulation of a total-product NS certificate by
ordinary PC.

\section{No algorithm can distinguish the types}\label{sec:undecidability}
Already in the special case of the fixed target $G^{(n)}=1$, the
classification cannot be turned into an algorithm
which determines the case from a sentence. This remains true under the
promise that every finite instance is inconsistent. Throughout this section,
the input is a well-formed normalised sentence and the request is a
refutation of its translated axioms, with the translation of
Section~\ref{sec:systems}; the characteristic $p$ is fixed.

The promise is already part of the every-union undecidability theorem
in LICS~2008~\cite[Theorem~1]{Riis}. Its Lemma~8 switches between
$\mathrm T_1$ and a prescribed other type. Here the controlled-join
construction gives explicit reductions for every pair, including
pairs wholly inside $\{\mathrm T_2,\mathrm T_3,\mathrm T_4\}$.
This refines the published formulation and the inspected earlier
construction; it is not a claim of exhaustive historical priority.

\begin{theorem}\label{thm:undecidability}
Let $A,B$ be disjoint nonempty subsets of $\{\mathrm T_1,\mathrm T_2,
\mathrm T_3,\mathrm T_4\}$. There is no partial computable algorithm which,
on every normalised sentence having no finite nonempty model and having
type in $A\cup B$, halts and correctly distinguishes membership in $A$
from membership in $B$. This holds for NS and for ordinary PC, over every
field of characteristic $p$.
\end{theorem}

In particular, each of the six pairs is undecidable, and so is membership
in any nonempty proper union of types. The last conclusion is the scope
of the historical every-union theorem. The pairwise statement requires the
additional construction below; it does not follow merely by listing all
the undecidable union problems.

\subsection{A sentence that records a computation prefix}
A reduction is an effective transformation of finite input descriptions;
it must not first decide whether its input computation halts. Register
programs and the standard halting theorem are discussed in \cite{Cutland}.
The normalised sentence and its degree bounds are constructed below.

A counter program has finitely many natural-number registers and instructions
for incrementing a register, branching on zero and decrementing, and halting.
All registers initially contain zero. The number of registers is finite
but need not be two. Write $c_M$ for one more than the largest register index
occurring in $M$, so $c_M\ge1$.

From a program $M$, form a sentence $\theta_M$ with a strict linear order,
least and greatest elements $o,e$, a successor function $S$, counter-value
functions $C_r$, and state predicates $Q_\ell$. The finite vocabulary may
grow with the input program, but is fixed independently of the universe
size. Its clauses have four roles:
\begin{enumerate}
\item \emph{Time order.} Require $S(e)=e$ and require $S(t)$ to be
the immediate successor of $t$ when $t\ne e$.
\item \emph{Initial and allowed configurations.} At $o$, all counters
equal $o$ and the state is the initial one. Exactly one state holds
at each element, and halting states are forbidden.
\item \emph{Increment transitions.} At $t\ne e$, the current state
determines the instruction and the next state at $S(t)$. Increment
replaces $C_r(t)$ by $S(C_r(t))$, with the guard $C_r(t)\ne e$.
\item \emph{Zero tests and decrement transitions.} Decrement branches
according as $C_r(t)=o$; on the positive branch, the new value $v$
satisfies $v<C_r(t)$ and $S(v)=C_r(t)$.
\end{enumerate}
In every transition, registers not changed by the instruction retain
their values. These are finite universal clauses with flattened function graphs,
and $M\mapsto\theta_M$ is primitive recursive.

Primitive recursion permits composition and loops with iteration counts
specified by their inputs. It describes the finite procedure producing
the sentence, not a bounded running time for the program encoded by it.
The latter may perform unbounded search or never halt.

\begin{lemma}[Computation switch]\label{lem:switch}
If $M$ never halts, $\theta_M$ has a model of every size $n\ge1$.
If the run of $M$ first reaches a halting state at time $T$, the translated system has an NS refutation of
degree bounded by a constant $C_M$ at every size $n\ge T+1$.
\end{lemma}
\begin{proof}
In the first case, use $[n]=\{0,\ldots,n-1\}$ in its natural order and
record the first $n$ configurations. Every counter value at time $j$ is at
most $j$, so no guarded increment overflows before the last element.
This includes $n=1$.

For the second case put $a_j=S^j(o)$. If the structure has $T+1$ distinct
elements, then $a_j\ne e$ for $j<T$: otherwise immediacy of successor and
linearity of the order leave only $a_0,\ldots,a_j$. Induction along this
finite prefix forces the actual state at $a_j$ and forces
$C_r(a_j)=a_{c_r(j)}$, where $c_r(j)$ is the numerical counter value.
At $a_T$ this gives a forbidden halting state. The argument needs neither
well-ordering nor an induction axiom in the structure.

This local contradiction gives a degree bound by the finite-term
certificate in Lemma~\ref{lem:finite-certificate}. More explicitly, with $c_M$
counters one may take
\[
 m=T+2+2c_M(T+1),\qquad
 C_M=m+\sum_R(m+T+1)^{\operatorname{arity}(R)}
                  +\max_{C\text{ a clause}}|C|.
\]
The term nodes are $o,e,a_1,\ldots,a_T,C_r(a_j),S(C_r(a_j))$;
$T+1$ distinct additional labels supply the size condition. The constant
depends on the program and its halting time, never on $n$.
\end{proof}

For completeness, the reduction needs effective universality, not a
decision procedure for the halting time. Every primitive recursive function
can be compiled into finite counter code: composition uses fresh registers,
and primitive recursion uses a counted loop, preserving the input registers.
Let $H(k,c)$ equal one when the program with code $c$, on its fixed
encoded input, halts within $k$ simulated steps, and zero otherwise.
Bounded simulation computes $H$ by a fixed number of primitive-recursive
transition updates, so this predicate is primitive recursive. A counter program that successively tests
$H(0,c),H(1,c),\ldots$ therefore halts exactly when the coded program does.
Loading $c$ into this fixed search program is primitive recursive. This
reduces the usual undecidable halting problem to the programs just used.
Only effective compilation and the standard undecidability of halting are used; the number of counters is unrestricted but finite.

\subsection{Finite local contradictions and their certificates}
\label{app:finite-certificate}
The halting switch uses a fixed finite computation prefix even though
function totality is expressed by sums with $n$ summands. The following
elementary certificate construction separates these two sizes.

The degree bound counts a fixed list of term evaluations and relation
atoms. The number of possible labellings, and hence the number of
certificate summands, may still grow with $n$. Bounded degree here
does not assert bounded certificate size.

\begin{lemma}[Finite-term certificate]\label{lem:finite-certificate}
Let $n\ge1$ and let a finite acyclic list of $m$ function evaluations be given, with every
argument supplied by an earlier node. Fix $b$ distinct labels. Suppose that
every assignment of labels to the nodes, and every Boolean assignment to
the relation atoms on the named and fixed labels, either has inconsistent
function values or falsifies a clause instance whose variables lie among the named and fixed
labels and all of whose function evaluations occur in the list. If clauses have length at most $L$, the
graph-and-sum translation has an NS refutation of degree at most
\[
                 m+\sum_R(m+b)^{\operatorname{arity}(R)}+L.
\]
The hypothesis and conclusion concern sizes for which the $b$ distinct
labels exist.
\end{lemma}
\begin{proof}
For a labelling $\alpha$ of the nodes, let $q_\alpha$ be the product of the
$m$ graph variables asserting those function evaluations. Expanding the
totality sum for each node, in the order of the acyclic list, gives
\[
                         \sum_\alpha q_\alpha-1
\]
in the degree-$m$ NS span. This is a telescoping identity: at step $j$, the
product for the first $j-1$ evaluations multiplies the totality equation
for evaluation $j$. Summing over the possible earlier values gives the
next partial sum. Repeated graph variables cause no problem.

If $\alpha$ gives different values to the same function on the same input,
$q_\alpha$ contains a functionality collision and already belongs to the
span within degree $m$. Otherwise, list the distinct relation variables
on the at most $m+b$ named labels. Write this actual set as
$Y_\alpha=\{y_1,\ldots,y_{a_\alpha}\}$, where
$a_\alpha\leq a_{\mathrm{rel}}=\sum_R(m+b)^{\operatorname{arity}(R)}$.
Expand over these distinct variables:
\[
             1=\prod_{i=1}^{a_\alpha}(y_i+(1-y_i))
              =\sum_\beta u_\beta.
\]
Each $u_\beta$ is the Boolean assignment monomial for one partial relational
structure. By hypothesis that structure falsifies an available clause.
Its clause polynomial is a product of at most $L$ falsity indicators.
Write $P=q_\alpha u_\beta$ and the clause polynomial as
$F_1\cdots F_\ell$, where each $F_j$ is the falsity indicator of a
literal false in the chosen partial structure. For each literal,
$P(1-F_j)$ has an explicit expression in the structural NS span:
\begin{itemize}
\item For a positive relation literal, $F_j=1-x$ and $u_\beta=(1-x)u'$.
Thus $P(1-F_j)=q_\alpha u'x(1-x)=-q_\alpha u'(x^2-x)$.
For a negative relation literal, $F_j=x$ and $u_\beta=xu'$; the same
identity with $x(1-x)$ applies.
\item For a negative graph literal false in the partial structure,
its graph variable $x$ is one of the chosen evaluations, so
$q_\alpha=xq'$. Its falsity indicator is $F_j=x$ and
$P(1-F_j)=-q'u_\beta(x^2-x)$.
\item For a positive graph literal false there, the chosen evaluation
has a different output variable $x'$ on the same function input.
Here $q_\alpha=x'q'$, $F_j=1-x$, and
$P(1-F_j)=q'u_\beta x'x$, an explicit functionality-axiom multiple.
\item A false equality or disequality has the constant falsity
indicator $F_j=1$, so $P(1-F_j)=0$.
\end{itemize}
These are all cases of the normalised literals. Each identity has
product degree at most $m+a_{\mathrm{rel}}+1$.
The exact telescoping identity
\[
 P(1-F_1\cdots F_\ell)
   =\sum_{j=1}^{\ell}P(1-F_j)\prod_{i<j}F_i
\]
therefore puts its left side in the NS span within degree
$m+a_{\mathrm{rel}}+\ell$. Since $PF_1\cdots F_\ell$ is a
multiple of the falsified clause axiom, $P$ is in that same span.
Sum first over $\beta$, then over $\alpha$, and use the totality identity
to obtain one.
\end{proof}

A \emph{partial structure} for $\theta_M$ has a domain of labels, all relations defined on it, and
function values defined only on some arguments. It is \emph{locally consistent} if every clause
instance whose variables take values in the domain, and whose function atoms are all defined,
has a true literal.

\begin{lemma}[Local switch]\label{lem:local-switch}
Let the run of $M$ first reach a halting state at time $T$, and let $M$ have $c_M$ counters.
Suppose a locally consistent partial structure defines $o$, $e$, $a_{j+1}=S(a_j)$ for $j<T$
\textup{(}with $a_0=o$\textup{)}, $C_r(a_j)$ and $S(C_r(a_j))$ for $j\le T$, $r<c_M$, and that all these
elements lie in its domain. Then its domain has at most $T$ distinct elements.
\end{lemma}
\begin{proof}
Suppose it contains $T+1$ distinct elements. If $a_j=e$ for some $j<T$, take $j$ least. Every
domain element $x$ satisfies $x=a_0$ or $a_0<x$ (least element), and, if $a_i<x$ with $i<j$, then
$x=a_{i+1}$ or $a_{i+1}<x$ (no element strictly between $a_i$ and $S(a_i)$, and linearity). As
$a_j=e$ is greatest, the domain is $\{a_0,\ldots,a_j\}$, too small. So $a_j\ne e$ for $j<T$, and
the order clauses make $a_0<a_1<\cdots<a_T$. For the counter induction, the initial clauses give the initial
state and $C_r(a_0)=a_0$ for every counter. A counter value at time $j$
is at most $j$, so the notation $a_v$ uses only this named prefix.
Suppose its states and counters agree through time $j<T$.
An increment of value $v$ gives $S(a_v)=a_{v+1}$; because $v\le j<T$,
this successor is among the defined evaluations and its guard is valid.
The zero branch of a decrement is selected exactly when $v=0$.
On a positive decrement, the transition specifies a value $y<a_v$
with $S(y)=a_v$. The immediate-predecessor property forces
$y=a_{v-1}$: if $y<a_{v-1}$ then $a_{v-1}$ lies strictly between
$y$ and $S(y)$, and if $a_{v-1}<y<a_v$ then $y$ lies strictly
between $a_{v-1}$ and $S(a_{v-1})$. Both alternatives contradict
immediacy. Unchanged counters retain their named values. State
uniqueness and the chosen transition identify the next state.
Every clause used has its variables in the domain and its function
atoms in the specified finite evaluation list. This proves the
induction through time $T$. At $j=T$ the state is halting, which a clause forbids.
\end{proof}

\begin{corollary}\label{cor:switch-degree}
In that case, for every $n\ge T+1$ the translation of $\theta_M$ at size $n$ has an NS refutation
of degree at most
\[
 C_M=m+\sum_R(m+T+1)^{\operatorname{arity}(R)}+\max_{C}|C|,\qquad m=T+2+2c_M(T+1).
\]
\end{corollary}
\begin{proof}
Apply Lemma~\ref{lem:finite-certificate} to the $m$ evaluations $o$, $e$, $a_1,\ldots,a_T$,
$C_r(a_j)$, $S(C_r(a_j))$ ($j\le T$, $r<c_M$), each argument supplied by an earlier node, with
$b=T+1$ fixed labels. A consistent labelling and a Boolean assignment give a partial structure of
the kind in Lemma~\ref{lem:local-switch}, whose domain contains the $T+1$ fixed labels; so it is not
locally consistent, which is the hypothesis of Lemma~\ref{lem:finite-certificate}.
\end{proof}

Here $c_M$ is one more than the largest counter index occurring in $M$ (so $c_M\ge1$), and $T$ must be
the first halting time. Sizes $n\le T$ are genuine exceptions: $\theta_M$ can have models there.

\subsection{A disjunction with controlled degree}
Conjunction is easy to control when one block is satisfiable. In disjoint
signatures, if $\beta$ has a root at size $n$, then
\begin{equation}\label{eq:satisfiable-block}
 d_*(\alpha\wedge\beta,n)=d_*(\alpha,n),
 \qquad *=\NS,\PC.
\end{equation}
One inequality follows by using the refutation of $\alpha$; the other
follows by substituting the root of $\beta$ into every proof line.

We also need a disjunction which does not accidentally simplify either
branch. Introduce a fresh unary relation $Z$. Keep both branches' Boolean,
totality and functionality equations. Replace every clause polynomial $f$
of $\alpha$ by $z_t f$, and every clause polynomial $g$ of $\beta$ by
$(1-z_t)g$, for a fresh universally quantified label $t$. Denote this
normalised sentence by $\alpha\join\beta$. At the clause level, these
replacements are $\neg Z(t)\vee C$ for clauses $C$ of $\alpha$ and
$Z(t)\vee C'$ for clauses $C'$ of $\beta$.

A guarded branch is simulated using the guard times each of its
original proof lines; the unguarded line is not assumed derivable
from the joined system. The same guard passes through linear
combinations and variable multiplication. Structural axioms remain
available, and can acquire the guard using one variable multiplication
and linear combinations.

\begin{lemma}[Controlled join]\label{lem:join}
For $n\ge1$, with $\infty$ allowed, each of NS and PC satisfies
\[
 \max\{d_*(\alpha,n),d_*(\beta,n)\}
 \le d_*(\alpha\join\beta,n)
 \le\max\{d_*(\alpha,n),d_*(\beta,n)\}+1.
\]
\end{lemma}
\begin{proof}
Set all selectors to one, and give the inactive block a structural root:
all its functions take one fixed value and all its relations are false.
Its clauses are guarded away, while its structural equations vanish.
The resulting proof refutes $\alpha$ without increasing degree. Selectors
zero give the other lower inequality.

For the upper bound fix a label $t_0$. Multiply a refutation of $\alpha$
line by line by $z_{t_0}$ and a refutation of $\beta$ by $1-z_{t_0}$.
The guarded clauses are axioms; structural axioms can be multiplied by
the guard at a cost of one degree. Linear combinations and variable
multiplication remain valid, including after cancellations. The two
derivations end in $z_{t_0}$ and $1-z_{t_0}$, whose sum is one. For NS
the same construction multiplies the two certificates by the guards.
\end{proof}

\subsection{The reduction}
Let $\psi_1,\psi_2,\psi_3,\psi_4$ be the four sentences of
Section~\ref{sec:examples}. Each has no finite nonempty model. For these particular
witnesses, the dominant type under the maximum operation is ordered by
\[
                 \mathrm T_1\prec\mathrm T_4\prec\mathrm T_3\prec\mathrm T_2.
\]
This is a property of the chosen witnesses, not an ordering assertion about
arbitrary profiles. A bounded profile has no effect on the type of a maximum;
the all-size linear lower bound for $\psi_2$ makes it dominant; and the
maximum of $\psi_4$ and $\psi_3$ is bounded off the multiples of $p$ and
linear on them. An additive error of at most one preserves these conclusions.
More generally, adjoining an eventually bounded profile and allowing this
error preserves each of the four intrinsic types: recurrence transfers
through finite intervals of degrees, and unboundedness transfers directly.

Take $a\prec b$ and define, using disjoint signatures,
\[
             \Phi_M=\psi_a\join(\psi_b\wedge\theta_M).
\]
If $M$ does not halt, \eqref{eq:satisfiable-block} makes the second branch
have exactly the profile of $\psi_b$. The join has type $b$. If $M$ halts,
the second branch is eventually bounded by Lemma~\ref{lem:switch}: a
refutation of $\theta_M$ is also a refutation of
$\psi_b\wedge\theta_M$. The join has type $a$. Every $\Phi_M$ is within the all-finite-inconsistent
promise, whether or not $M$ halts. The map is primitive recursive and
uses no information about that alternative. A separator of types $a,b$
would decide halting. Choosing $a$ and $b$ on opposite sides of any
disjoint nonempty $A,B$ proves Theorem~\ref{thm:undecidability}; if the
smaller type lies in $B$, complement the proposed separator.

\section{Conclusion}\label{sec:conclusion}
The linear bounds arise from a way of comparing bounded-degree algebraic
calculations across universe sizes. For a fixed uniform request over a
field of positive characteristic, the construction transports the actual
axiom, multiplication and target maps. It preserves the linear-space
operations defining the consequence spaces, including the cancellations
which make ordinary PC stronger at a given degree. A separate detection
argument reflects whether the prescribed target belongs to that space.
Together these steps make the construction apply to uniform consequences,
with refutation as the target-$1$ special case.

The quantitative gain is a linear onset and a linearly bounded prime-power
period for each degree threshold. Repeated lifting keeps the numerical
degree fixed while the universe grows. A nonrecurring degree can therefore
occur only before its own onset; the period bound also locates recurring
large degrees at proportionate universe sizes. This gives bounded degree
in $\mathrm T_1$, linear hardness at finite values in $\mathrm T_2$, linear
hardness on the hard residues in $\mathrm T_3$, and finite linear peaks
with a finite liminf in $\mathrm T_4$. Underivable requests retain degree
$\infty$, and the stated initial exceptions remain part of the results.

The linear degree lower bound is sharp in general. The injection example has matching
$\Theta(n)$ degrees for both systems, and the weak-bijection example has
finite $\Theta(n)$ peaks; its degree-one target extension has the same
peak behaviour. These witnesses establish optimality of the general
degree lower bound, while the theorem leaves the individual upper bounds to the
particular uniform request.

The uniform NS--PC comparison shows that these patterns belong to the
request itself: both systems assign it the same type, even when their
minimum degrees differ. Their common hardness spectrum gives the same
four alternatives as topological shapes in $\mathbb Z_p$, for all
uniform requests. With full explicit coefficient lists, including
zeros, the degree results give the stated polynomial upper bounds and
exponential size lower bounds. The first-order refutation subcase connects
this algebraic classification to logic: explicit reductions separate
any two types under the no-finite-model promise already stated in
LICS~2008~\cite[Theorem~1]{Riis}, refining its every-union result.
Thus the stronger quantitative bounds preserve the four-way
structure of the earlier classification while extending its algebraic
scope.

\paragraph{Proof sources and verification.}
The supporting archive~\cite{RiisSupporting} supplies the
curated proof sources, pinned Lean~4.26.0 and mathlib dependency data,
and the source, statement, axiom and import-closure audit commands
and their recorded execution provenance. The core degree, comparison,
classification and undecidability endpoints have Lean proofs; the
archive identifies their exact types and assumptions. The injection
upper certificate, the optimality deductions, the Boolean completeness
bound in Section~2.4 and the size presentation are written arguments. The topological corollary in
Section~\ref{sec:model-theory} is likewise a written deduction from
periodicity, comparison and compactness of $\mathbb Z_p$; it is not
an addition to the Lean formalisation. Recorded fresh replay covers the named import
closures, with separately compiled scope adapters distinguished in the
evidence; it is not a fresh replay of every package module.

\paragraph{Acknowledgements.}
I used the AI systems Astra and Claude for sustained proof search,
formalisation, verification support and development of the exposition,
under my mathematical direction and editorial supervision. The proof
search built on my 53-page 2009 manuscript~\cite{RiisLong} and my
unpublished notes on the problem. I supplied these sources, set the
research programme, and checked and revised the arguments and their
presentation. The human work in proof complexity,
model theory, modular representation theory and algebraic combinatorics
cited throughout the paper provides important antecedents and connections,
including~\cite{NSFoundations,PCFoundations,Ajtai,Krajicek,KrajEuler,
BussDesign,BeameRiis,RiisSitharamTautologies,RiisSitharamModules,
CEF,Harman,HKPSYZ,BDM}. Responsibility
for the statements, interpretation, attribution and final text remains mine.

\end{document}